# Exceptional Alkaline Methanol Electrooxidation on Bi-modified $Pt_3M$ Intermetallics: Kinetic Origins and an OH Binding Energy Descriptor

Lecheng Liang,[1,3,6] Hengyu Li,[2,6] Shao Ye,[1,4,6] Peng Li,[3] Kaiyang Xu,[5]
Jinhui Liang,[1] Binwen Zeng,[1] Bo Shen,[4] Taisuke Ozaki,[2] Zhiming Cui[1,7,*]

[1]Guangdong Provincial Key Laboratory of Fuel Cell Technology, School of Chemistry and Chemical Engineering, South China University of Technology, 510641 Guangzhou, China.
[2]Institute for Solid State Physics, The University of Tokyo, 277-8581 Kashiwa, Japan.
[3]College of Chemistry and Molecular Sciences, Wuhan University, 430072 Wuhan, China.
[4]Department of Materials Science and Engineering, City University of Hong Kong, Hong Kong SAR, China.
[5]Songshan Lake Materials Laboratory, Dongguan 523808, China.
[6]These authors contributed equally
[7]Lead contact
*Correspondence: zmcui@scut.edu.cn

**SUMMARY**

The exploration of advanced CO-free catalysts and clarifying the ambiguous kinetic origins and governing factors would undoubtedly open up opportunities to overcome the sluggish kinetics of methanol electrooxidation and promote the development of direct methanol fuel cells. Herein, we constructed a family of Bi-modified $Pt_3M$ intermetallic catalysts (Bi-$Pt_3M$/C, M=Cr, Mn, Co, Zn, In, Ga, and Sn) that follow CO-free dominated pathway and exhibit exceptional catalytic activity. More significantly, leveraging this platform, we have identified the pivotal factor governing the reaction kinetics in CO-free pathway, namely OH binding energy (OHBE). This arises because the rate-determining step (RDS) encompasses both C-H bond activation and water dissociation, whose respective barriers can be reflected by the OHBE. Accordingly, OHBE can act as an activity descriptor. Specifically, Bi-$Pt_3In$/C stands out from other Bi-$Pt_3M$/C and delivers the unprecedented mass activity of 36.7 A $mg_{Pt}^{-1}$ at peak potential, far exceeding state-of-the-art Pt-based catalysts reported to date. Taking Bi-$Pt_3In$/C as a proof of concept, we clearly elucidate the origin of enhanced MOR activity by combining theoretical calculations, kinetic isotope effects, and formaldehyde electrooxidation. Moreover, there exhibits a volcano-type trend between OHBE and the activity of Bi-$Pt_3M$/C. Beyond the discovery of ultrahigh-performance catalysts, these findings provide a detailed mechanistic picture of RDS and offer an innovative design principle for advanced catalysts.

## INTRODUCTION

Methanol electrooxidation pertains to one of the most important model reactions in the study of electrocatalysis of C1 compounds and the development of direct methanol fuel cells (DMFCs) for over 60 years.[1,2] Revisiting the mechanism of methanol electrooxidation at Pt electrodes, the dual-pathway scheme proposed by Breiter[3,4] has been well-established by the community. The two parallel pathways typically present in scheme while almost Pt-based catalysts follow a CO-dominated pathway, regardless of whether in acidic or alkaline electrolytes. The strong adsorption of CO on the Pt sites induces a serious poisoning effect and impedes further electrocatalytic reaction, resulting in substantial activity decay.[5] Besides, methanol electrooxidation is a complex six-electron reaction with sluggish kinetics, several orders of magnitude slower than hydrogen oxidation, which significantly limits the widespread application of DMFCs. Therefore, it is highly desirable yet challenging to develop a highly efficient and robust MOR catalytic system with a CO-free pathway, simultaneously elucidating its intrinsic mechanism at the atomic-level.

Over the past decades, numerous studies have focused on alleviating the CO poisoning effect on Pt-based catalysts. Since the key reaction step in the CO pathway involves the interaction between adsorbed CO (CO*) and OH* to form COOH*, various strategies have been proposed to improve the CO tolerance of Pt-based catalysts either by weakening the CO binding energy (COBE)[6, 7] or strengthening the OHBE[8, 9]. Despite the fruitful progress in the CO tolerance of catalysts, the CO pathway has not been fully eliminated, inevitably leading to catalyst deactivation. Consequently, researchers are dedicated to exploring emerging catalysts that can achieve a CO-free dominated pathway to overcome the above-mentioned dilemma. Unexpectedly, Pt-Bi alloys are found to exhibit exceptional CO tolerance when switching from acidic to alkaline, and are subsequently speculated to follow a CO-free dominated pathway based on spectral evidence.[10-13] This distinctive catalytic property makes Pt-Bi alloys a highly promising new class of catalysts for alkaline methanol electrooxidation. The excellent anti-CO poisoning ability is generally inferred to result from the disruption of contiguous Pt ensembles by Bi atoms, thereby suppressing the CO formation pathway. Regrettably, the classical ensemble effect theory does not provide a satisfactory explanation for the mechanistic origin of the pH-dependent anti-CO ability on Bi-Pt surfaces.[13, 14] Taking Bi-modified Pt/C as a model catalyst, our group recently presented a complete reaction landscape and identified that the pathway selectivity depends on the formaldehyde (HCHO) intermediate.[15] The true role of Bi adatoms is to facilitate the HCHO* desorption and enable the subsequent conversion of HCHO to the $H_2COOH$ anion at the alkaline interface, thereby circumventing CO formation.

In addition to pathway selectivity, understanding the intrinsic factors that affect reaction kinetics is crucial for the rational design of advanced Pt-Bi catalysts. Notably, recent experimental studies have demonstrated that optimizing the surface structure and composition of the host matrix can reduce the energy barrier of the rate-determining step (RDS) in proton-coupled electron transfer (PCET) processes, thereby leading to a positive effect on methanol electrocatalytic activity.[16-19] Such results imply an underlying correlation between host matrix and activity, however, there is still no consensus on identifying the RDS in methanol electrocatalysis. Up to now, there are two main schools of thought regarding the RDS: one suggests that the RDS involves the coupling between specific carbon-containing and oxygen-containing intermediates.[20] For instance, the coupling of HCO* and OH* is regarded as the RDS in the CO-free pathway, but it lacks sufficient experimental verification;[21, 22] the other argues that C-H bond activation is the RDS,[23, 24] while this conclusion is generally based on studies using Pt electrodes. Undoubtedly, it is imperative to explore an ideal materials platform for elucidating the correlation between structures and performance and the intrinsic factors governing reaction kinetics. Distinct from traditional disordered alloys, intermetallic compounds (IMCs) feature ordered crystal structures and precise stoichiometry, which offer predictable control over the structure and electronic effects, resulting in better consistency between experimental results and theoretical modeling.[25, 26 27] Such unparalleled merits make IMCs an excellent host platform for systematic structure–property studies.

In this contribution, we developed a family of Bi-modified $Pt_3M$ intermetallic catalysts (Bi-$Pt_3M$/C) (M = Cr, Mn, Co, Zn, In, Ga, and Sn) as a research platform and revealed that the RDS both involves water dissociation as well as C-H bond cleavage of $CH_3O$*. Moreover, according to the proposed mechanism, OHBE is identified as the critical factor governing reaction kinetics and can therefore serve as a reliable descriptor for alkaline methanol electrooxidation activity. Concretely, all synthesized Bi-$Pt_3M$/C catalysts exhibit extraordinary performance for alkaline methanol electrooxidation, far superior to those of Bi-Pt/C (14.15 A $mg_{Pt}^{-1}$) and Pt/C (3.26 A $mg_{Pt}^{-1}$). Among them, Bi-$Pt_3In$/C achieves an impressive mass activity as high as 36.66 A $mg_{Pt}^{-1}$, representing the best-performing Pt-based electrocatalyst reported to date. Taking Bi-$Pt_3In$/C sample as a proof of concept, it unambiguously demonstrates that the origin of remarkably enhanced activity is attributed to the rapid cleavage of C-H bonds in the CO-free pathway, based on a comprehensive approach encompassing ab initio molecular dynamics (AIMD) simulation, density functional theory (DFT) calculations, kinetic isotope effect (KIE) tests, and formaldehyde electrooxidation (Scheme 1). Furthermore, an observation was made that the activity of Bi-$Pt_3M$/C catalysts exhibited a volcano-like trend in relation to the OHBE change. It is evident from the comprehensive reaction landscape that the OH* generated from water dissociation performs a pivotal function in the activation of methanol. Specifically, hydrogen from the hydroxyl group of methanol can be rapidly transferred to OH* via hydrogen bonding, forming adsorbed methoxy ($CH_3O$*) and subsequently converting back to $H_2O$*. During this process, OH* and $H_2O$* exhibit dynamic conversion characteristics. In addition to the C-H bond activation, this dynamic behavior is identified as another critical factor that dominates the reaction kinetics, which can be manifested by the OHBE.

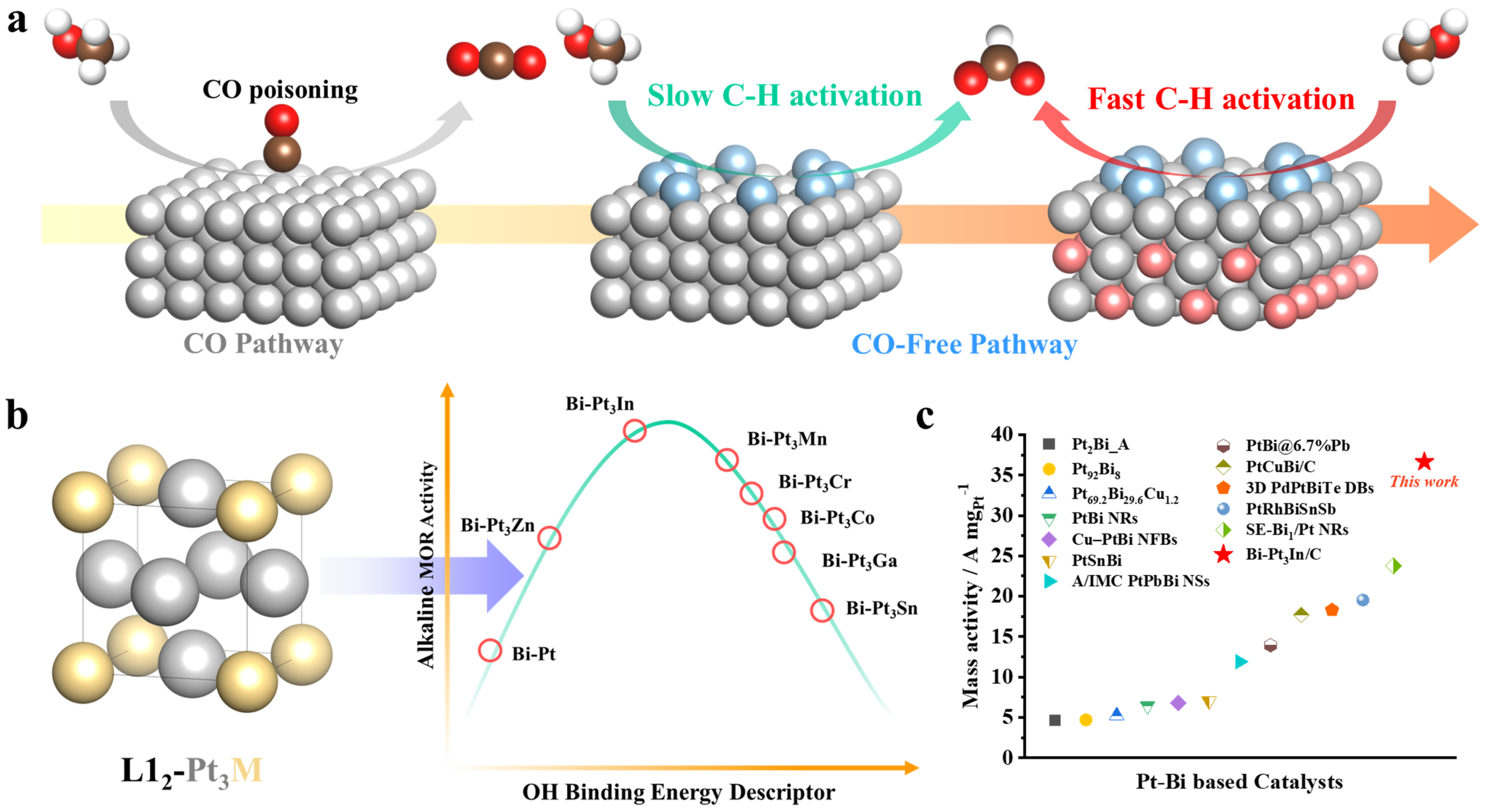

**Scheme 1**. The kinetic origin of exceptional alkaline MOR performance of Bi-$Pt_3M$ intermetallics. (a) Schematic illustration of the proposed mechanism and the key reaction step on Pt, Bi-Pt, and Bi-$Pt_3M$ surfaces. (b) The volcano-type trend based on the relationship between OH binding energy of Bi-$Pt_3M$ and their alkaline MOR activities. (c) Mass activity of Bi-$Pt_3In$/C in this work and Pt-Bi based catalysts reported in recent years.

## RESULTS

**Bi Modified $Pt_3In$/C as a Proof of Concept for Efficient Alkaline Methanol Electrooxidation Catalyst.** Carbon-supported ordered $Pt_3In$ nanoparticles (denoted $Pt_3In$/C) were synthesized through a two-step protocol comprising freeze-drying impregnation and annealing reduction (see Experimental Section for details). The crystal structure of $Pt_3In$/C was characterized using X-ray diffraction (XRD). As shown in Figure 1a, the XRD pattern of the synthesized $Pt_3In$/C exhibits not only the five characteristic diffraction peaks of the face-centered cubic (FCC) structure, but also a well-defined superlattice peak at 31.7° that corresponds precisely to the (110) reflection of ordered $Pt_3In$. The observed superlattice reflection indicates the successful formation of an $L1_2$-type ordered intermetallic phase with long-range atomic ordering. The Pt loading in $Pt_3In$/C was quantified using inductively coupled plasma optical emission spectroscopy (ICP-OES), provided in Table S1. Subsequently, the construction of $Pt_3In$ NPs with well-defined core–shell structure and formation of Pt-rich shells was achieved through an electrochemical dealloying method. Electrochemical dealloying was performed via cyclic voltammetry (CV) between 0.05 and 1.2 V vs. RHE at 250 mV $s^{-1}$. During this process, we observed a progressive increase in current density within H desorption/adsorption region with successive CV cycles, providing compelling evidence for the selective dissolution of surface/subsurface In atoms into the electrolyte and consequent enrichment of Pt sites (Figure S1). The dealloying process was considered complete when both the voltammetric profile and corresponding current densities stabilized, indicating the formation of a structurally stable Pt-enriched shell. Building on our previously established method for Bi-modified Pt/C catalyst, we applied underpotential deposition to decorate Pt-enriched surfaces with Bi adatoms. Specifically, Bi- $Pt_3In$/C was prepared via electrodeposition in 0.5 M $H_2SO_4$ containing 1 mM $Bi^{3+}$ at -0.16 V vs. Ag/AgCl (saturated KCl) for 120 s.

Transmission electron microscopy (TEM) images of $Pt_3In$ NPs in Figure 1b reveal uniformly dispersed nanoparticles with an average diameter of 4.8 ± 1 nm on the Ketjen Black-300J carbon support with no apparent aggregation. Following Bi modification, the resulting Bi-$Pt_3In$ NPs exhibited similar morphological characteristics, with the particle size remaining largely unchanged (5.1 ± 0.8 nm), while maintaining excellent dispersion (Figure 1c). The atomic-level structural ordering of Pt and In was characterized by high-angle annular dark-field scanning transmission electron microscopy (HAADF-STEM). Figure 1d presents the HAADF-STEM of an individual nanoparticle. The alternating bright and dark columns corresponding to Pt and In atoms, respectively, which unambiguously confirms the formation of the ordered intermetallic phase. Energy dispersive spectroscopy (EDS) elemental mapping further verifies the homogeneous spatial distribution of Pt, In, and Bi atoms throughout the nanoparticles (NPs) in Figures 1c and S2. Notably, EDS line scanning of a single Bi-$Pt_3In$ NP (Figure 1d) provides direct evidence for the formation of the core−shell structure featuring a thin Pt shell (ca. 0.48 nm, approximately 2 atomic layers) decorated with Bi adatoms.

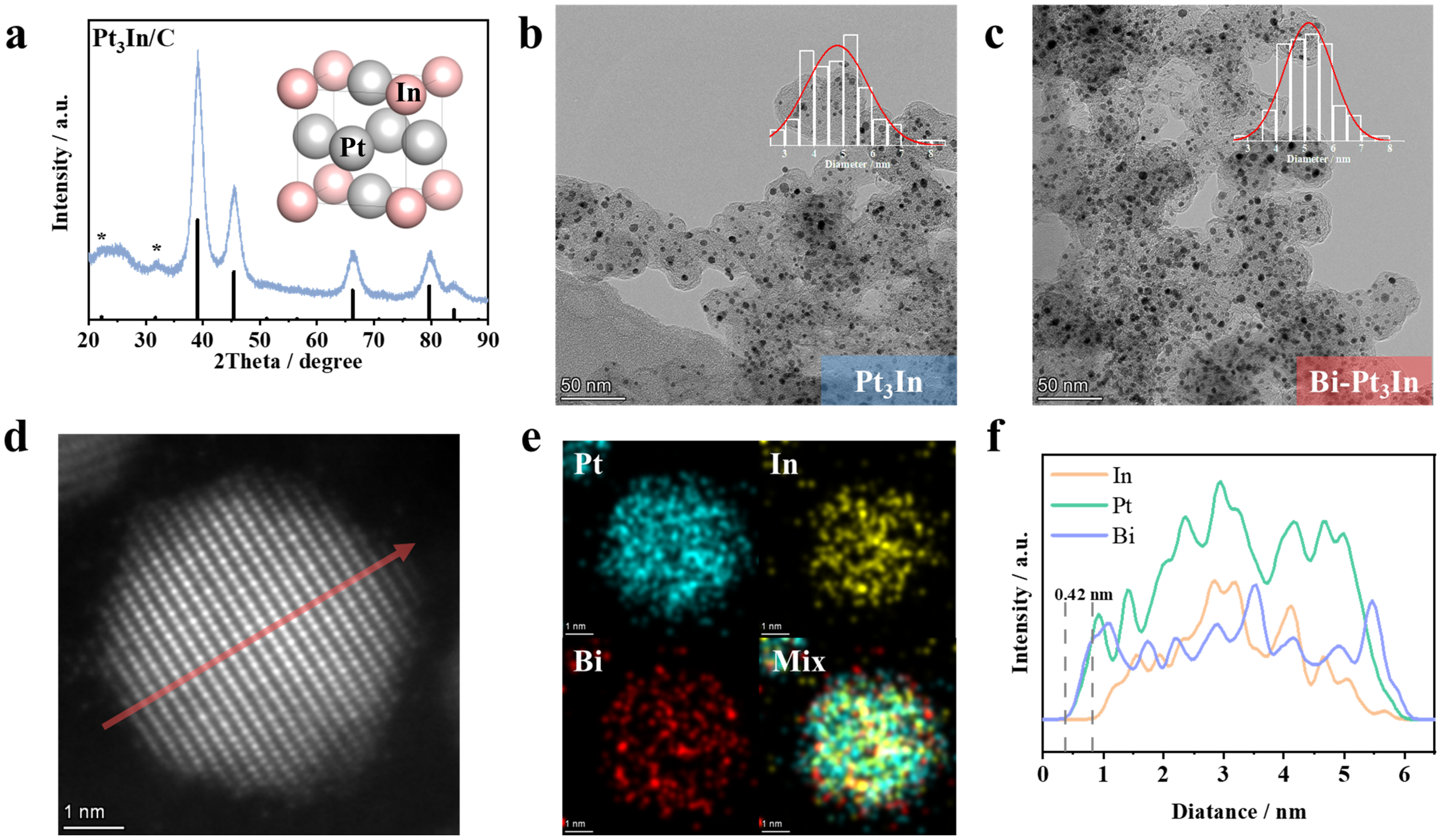

**Figure 1. Structure characterization.** (a) XRD patterns of $Pt_3In/C$. The inset displays the crystal structure of $L1_2$ type ordered $Pt_3In$ intermetallic. STEM images of (b) $Pt_3In/C$ and (c) Bi-$Pt_3In/C$. (d) HAADF-STEM image of a Bi-$Pt_3In$ NP and (e) EDS-elemental mappings of In, Pt, and Bi of the NP. (f) EDS line scanning along the red line in (d).

Such Bi adatoms profoundly altered the electrochemical behavior of Pt surfaces. Figures 2a and S3 display the stable cyclic voltammograms (CV) curves of Bi-$Pt_3In/C$, $Pt_3In/C$, Bi-Pt/C, and Pt/C in $N_2$-saturated 1M KOH. As demonstrated, the incorporation of Bi adatoms induced a significant reduction of current within the hydrogen adsorption/desorption region, which spans from 0.05 to 0.45 $V_{RHE}$. Specially, the disappearance of the sharp peak current at 0.25 $V_{RHE}$ results from Bi adatoms inhibiting the adsorption of hydrogen and hydroxide ions at Pt (110) step sites.[28-32] Besides, Bi-$Pt_3In/C$ exhibits even lower hydrogen region currents compared to Bi-Pt/C, suggesting that the underlying In atoms in the intermetallic core effectively modulate the electronic structure of the outermost Pt atoms, further attenuating the binding strength with adsorbed hydrogen species (Figure S4).

The electrocatalytic performance toward methanol oxidation reaction (MOR) of the above samples were evaluated in $N_2$-saturated 1.0 M KOH+1.0 M $CH_3OH$ electrolyte. Exhilaratingly, as depicted in Figures 2b and 2c, Bi-$Pt_3In/C$ exhibits an exceptional mass activity (MA) of 36.66 A $mg_{Pt}^{-1}$ at peak potential. This performance represents a 2.59-fold enhancement compare to Bi-Pt/C (14.15 A $mg_{Pt}^{-1}$), and much superior to $Pt_3In/C$ (5.22 A $mg_{Pt}^{-1}$, 7.03-fold) and commercial Pt/C (3.26 A $mg_{Pt}^{-1}$, 12.6-fold). In addition, the current density was further normalized by electrochemical surface areas (ECSAs) to assess changes in the intrinsic activity of the Pt sites. The ECSAs of the catalysts were measured by the CO-stripping voltammetry. Specifically, Bi-$Pt_3In/C$ also achieves a high specific activity of 50.58 mA $cm_{Pt}^{-2}$, which is 2.54 times higher than that of Bi-Pt/C (19.89 mA $cm_{Pt}^{-2}$) and far outperforms $Pt_3In/C$ (4.68 mA $cm_{Pt}^{-2}$) and Pt/C (4.01 mA $cm_{Pt}^{-2}$). The activity enhancement exhibits potential-dependent behavior, with the MA ratio between Bi- $Pt_3In/C$ and $Pt_3In/C$ reaching a maximum value of 12.6 at 0.6 $V_{RHE}$, indicating substantially enhanced reaction kinetics at suitable overpotentials (Figure S5). To gain deeper insights into the charge transfer kinetics, we conducted electrochemical impedance spectroscopy (EIS). The Nyquist plots in Figure 2d displays that Bi-$Pt_3In/C$ possesses the smallest charge transfer resistance ($R_{ct}$) of merely 5.6 Ω, much lower than those of Bi-Pt/C (14.5 Ω), $Pt_3In/C$ (65 Ω), and Pt/C (124 Ω). This dramatic reduction in charge transfer resistance quantitatively confirms the superior electron transfer capabilities of the Bi-$Pt_3In/C$ catalyst. Beyond its outstanding intrinsic MOR activity, Bi-$Pt_3In/C$ also exhibits superior electrochemical durability. The chronoamperometry (CA) measurements of Bi-$Pt_3In/C$, $Pt_3In/C$, Bi-Pt/C, and Pt/C were executed at 0.45 $V_{RHE}$ for 3600s (Figure 2e). As demonstrated in Figure S6, Bi-$Pt_3In/C$ not only maintained the highest MA throughout the whole duration but also reserved 89% of its initial catalytic activity after CA test, significantly outperforming Bi-Pt/C (76% retention), underscoring the robust stability of Bi-$Pt_3In/C$. As plotted in Figure 2f, chronopotentiometry at a constant current of 50 mA $cm^{-2}$ is performed to estimate the durability of these catalysts. It has been demonstrated that the initial potential on Bi-$Pt_3In/C$ is more negative than those of Bi-Pt/C, $Pt_3In/C$, and Pt/C. Furthermore, Bi-$Pt_3In/C$ also exhibits lower voltage degradation rates, also indicating its outstanding durability. The improved durability can be attributed to the heteroatomic bonding in intermetallic compounds, which typically exhibits a more negative formation enthalpy compared to random distributed alloys, thereby providing greater resistance to structural degradation during electrocatalysis.[33, 34] Comparative analysis with state-of-the-art catalysts reported in the literature (Figure 2g and Table S2) demonstrates that the mass activity of Bi-$Pt_3In/C$ surpasses most advanced Pt-based catalysts documented to date, representing a major leap for MOR.

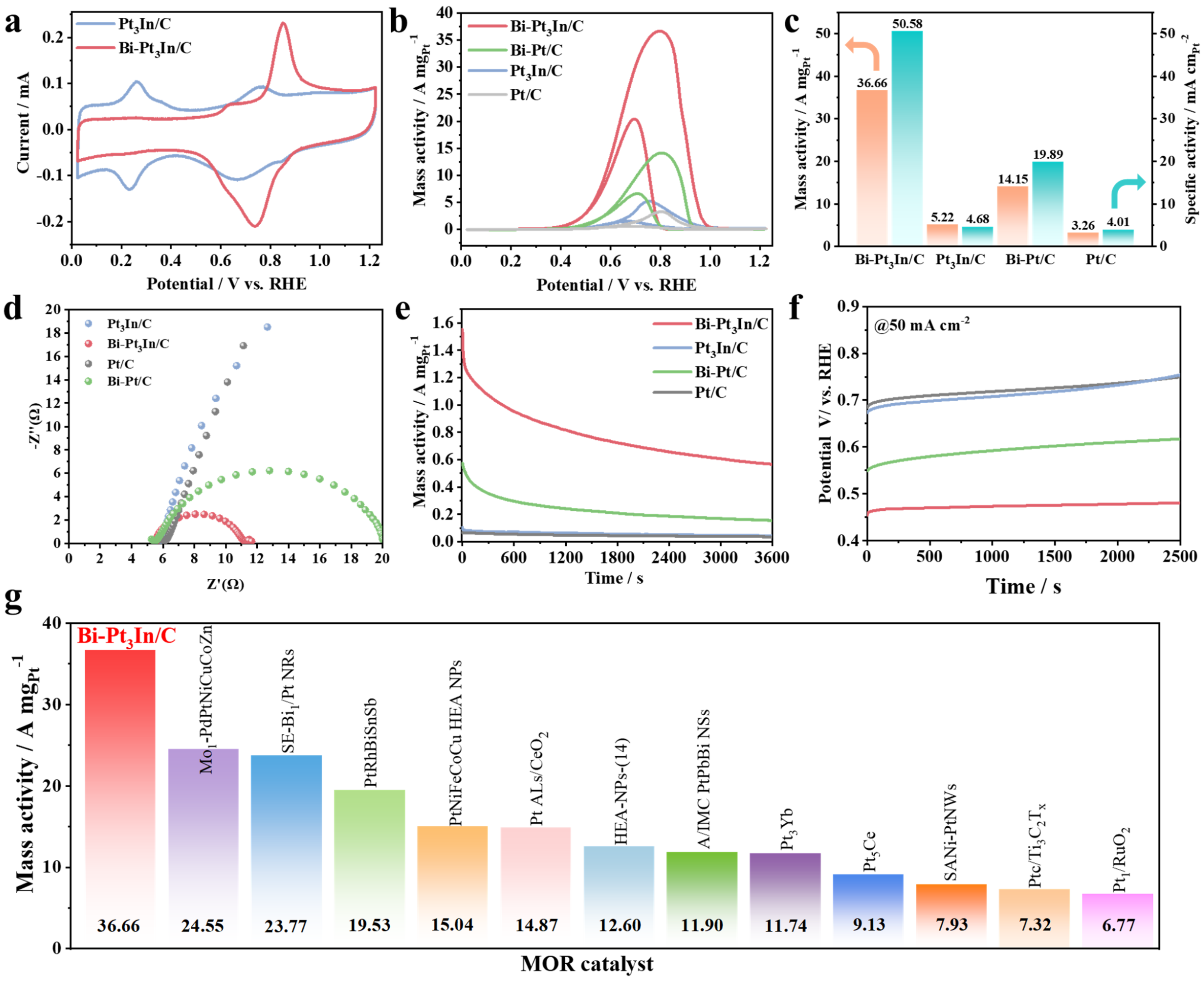

**Figure 2. Alkaline methanol electrooxidation performance.** (a) CV curves of $Pt_3In/C$ and Bi-$Pt_3In/C$ in $N_2$-saturated 1M KOH. (b) CV curves of Bi-$Pt_3In/C$, Bi-Pt/C, and $Pt_3In/C$ recorded at a scan rate of 50 mV $s^{-1}$ in 1M $CH_3OH$/1M KOH. (c) Peak current density of specific activity and mass activity. (d) EIS plots of these catalysts at 0.53 $V_{RHE}$. (e) Mass-normalized i-t curves of these catalysts recorded at 0.45 $V_{RHE}$. (f) Potential versus time plots measured by chronopotentiometry at a constant load of 50 mA $cm^{-2}$. (g) A comparison of mass activity between Bi-$Pt_3In/C$ and other recently reported Pt-based catalysts.

**Kinetic Origin of the Activity Enhancement.** To uncover the fundamental origin of the exceptional activity observed for Bi-$Pt_3In/C$, we performed mechanistic studies using kinetic isotope effects (KIEs) to elucidate reaction mechanisms. The KIEs factor is crucial for identifying the rate-determining step (RDS), with an H/D KIE value exceeding 1.4 typically indicating that the C-H/D (or O-H/D) bond cleavage constitutes the RDS for the specific reaction.[35, 36] Figures 3a-b and S7 show the profile of deuterium KIE experiments in solution containing 1 M $CD_3OD$/1 M KOH and 1M $CH_3OD$/1M KOH, respectively. Replacing $CH_3OH$ with $CD_3OD$ in the electrolyte caused significant MA reductions at peak potentials, exhibiting KIE values of 1.86 and 1.53 for Bi-$Pt_3In/C$ and $Pt_3In/C$, respectively. In striking contrast, MA values showed no significant change when using $CH_3OD$ compared to $CH_3OH$. These results align precisely with our previously fundings, underscoring the considerable influence of C−H bond cleavage on the reaction kinetics, whereas the effect of O−H bond cleavage appears to be insignificant. Moreover, we conducted comparative analysis of the electrocatalytic activities of Bi-$Pt_3In/C$ and Bi-Pt/C catalysts for the formaldehyde (HCHO) oxidation reaction (FOR) to gain valuable insights into the MOR mechanism.[37-39] As illustrated in Figure 3c, Bi-$Pt_3In/C$ demonstrates markedly superior activity for alkaline FOR compared to Bi-Pt/C. Bi-$Pt_3In/C$ required only 209 mV overpotential to achieve 1 A $mg^{-1}$ mass activity, demonstrating enhanced reaction kinetics compared to Bi-Pt/C (380 mV). The results from KIEs and FOR tests provide compelling evidence that the boosted MOR activity of Bi-$Pt_3In/C$ is attributable to the rapid activation of C-H bonds. As the FOR mechanism illustrated in Figure 3d, initially, HCHO is converted to $H_2COOH^-$ anion in alkaline solution via reversible hydration and deprotonation process. Subsequently, the $H_2COOH^-$ anion adsorb onto the catalyst surface from the bulk solution, where it interacts preferentially with Pt atoms through its oxygen end, accompanied by electron transformation. The $H_2COOH^*$ intermediate undergoes a proton-coupled electron transfer step via the cleavage of C-H bond to yield HCOOH and $H^+$. Compared to ideal Pt bulk, the $Pt_3In$ intermetallic core optimizes the electronic structure of surface Pt atoms, thereby lowering the activation barrier for this rate-determining C-H bond scission step.

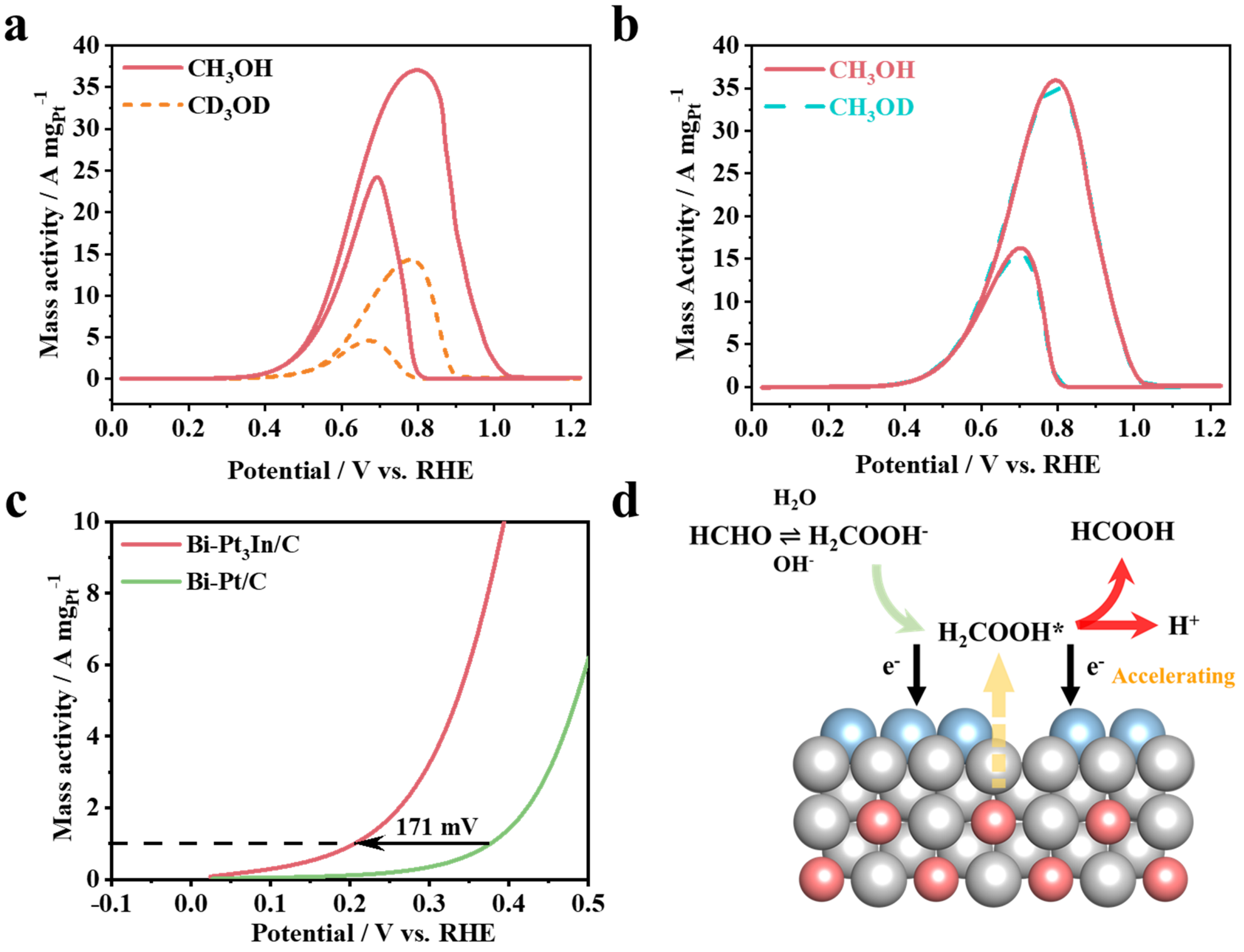

**Figure 3. Kinetic isotope effects and alkaline formaldehyde electrooxidation.** (a) A comparison of CV curves of Bi-$Pt_3In$/C between 1M $CH_3OH$/1M KOH and 1M $CD_3OD$/1M KOH. (b) A comparison of CV curves of Bi-$Pt_3In$/C between 1M $CH_3OH$/1M KOH and 1M $CH_3OD$/1M KOH. (c) positive-going polarization curves Bi-$Pt_3In$/C and Bi-Pt/C recorded at a scan rate of 50 mV $s^{-1}$ in 1M HCHO/1M KOH. (d) Schematic illustration of the alkaline formaldehyde electrooxidation mechanism on Bi-$Pt_3In$ surface.

To reveal the underlying mechanism at level atomic-molecular level, we employed an integrated methodology combining CO-stripping experiments with density functional theory (DFT) calculations, and ab initio molecular dynamics (AIMD) simulations. The CO-stripping experiments depicted in Figures 4a and 4b confirm a substantial decrease in CO oxidation current for Bi-$Pt_3In$/C compared to $Pt_3In$/C, which is attributed to the effective inhibition of CO poisoning by Bi adatoms, in good consistency with previous findings in the literatures.[10, 12, 40] However, the evident CO oxidation current on Bi-$Pt_3In$/C indicates the presence of a heterogeneous surface structure where both CO-poisoned and CO-resistant Pt sites coexist due to the specific arrangement of Bi adatoms. Besides, both Bi-$Pt_3In$/C and $Pt_3In$/C catalysts display lower CO oxidation peak potentials compared to the Bi-Pt/C and Pt/C catalysts, implying that In atoms promote the removal of CO by weakening CO adsorption strength (Figure S8). According to the results from physical and electrochemical characterization, we constructed atomically precise computational models using $Pt_3In$(110) as the base surface with a core-shell structure comprising an intermetallic core covered by two atomic layers of Pt. To obtain a reasonable coverage of Bi atoms on the (110) surface, a truncated octahedral model was used to calculate the surface-to-core atomic ratio (0.22) of a 5 nm Pt nanoparticle.[41] The molar ratio of 12.2:3:1 for Pt/In/Bi determined by X-ray absorption ICP-OES data, the surface coverage of Bi atoms was estimated to be approximately 0.33 monolayers (ML). Figures 4c and S9 illustrate four distinct Pt adsorption sites on the Bi-$Pt_3In$(110) surface, labeled as sites 1, 2, 3, and 4. These sites are defined by their unique local coordination environments of Pt and Bi atoms: Site 1, referred to as the Pt (P) site, is composed exclusively of Pt atoms; Site 2, known as the Pt-Bi-$Pt_V$ trio ($PBP_V$) site, features a Bi atom bridging two Pt atoms in a V-shaped configuration; Site 3, termed the Bi-Pt-$Bi_L$ trio ($BPB_L$) site, consists of a Pt atom linking two Bi atoms in a linear arrangement; Lastly, Site 4, designated the Bi-Pt-$Bi_V$ trio ($BPB_V$) site, exhibits a configuration similar to that of Site 2. Bader charge analysis in Figure S10 indicates a considerable charge transfer from Bi to Pt atoms, with each Bi adatom donating approximately 0.7 electrons, which aligns with the observations from X-ray photoelectron spectroscopy (XPS) results (Figure S11).

The CO binding energies (COBEs) at potential active sites were calculated. As illustrated in Figures 4d and S12, the CO binding energy (COBE) at site 1 is -2.14 eV, differing by a mere 0.12 eV from the top site of $Pt_3In$(110) (-2.26 eV). Similarly, the calculated COBE values for various bridge sites closely approximate to that of the bridge site on $Pt_3In$(110), with the maximum deviations not exceeding 0.24 eV (Figure S13). As a result, sites 1, 2, and 4 maintain strong CO adsorption characteristics and are consequently susceptible to CO poisoning during the reaction. In contrast, site 3 shows the weakest COBE, with a remarkably weaker value of -0.86 eV, signifying a highly unfavorable interaction with CO and thus conferring inherent resistance to CO poisoning. Collectively, the COBE results demonstrate that the multi-site effect on the Bi-Pt surface leads to the coexistence of both CO-poisoned and CO-resistant Pt sites, in agreement with experimental observations from CO-stripping tests. Further, we investigated hydroxyl binding energies (OHBEs) across the various sites on the Bi-$Pt_3In$ surface (Figures 4e and S14), given their critical role in methanol oxidation kinetics. The OHBE at site 1 is consistent with that of Pt site on $Pt_3In$(110), with a value of 0.54 eV. However, the Pt atoms adjacent to Bi exhibit weaker OHBEs. Especially, Bi sites exhibit the weakest OHBE, with a value of 1.11 eV, supporting our conclusion that Bi modification weakens OHBE in recent study. Therefore, different from the conclusions in other previous studies, Bi atoms do not play a role in facilitating water dissociation.

According to our proposed alkaline MOR mechanism on Bi-Pt surface, Bi modification fundamentally alters the dynamic behavior of the HCHO intermediate, thereby effectively directing the reaction toward a CO-free pathway. Concretely, as illustrated in Figure 4f, Bi adatoms induce the formation of ensemble sites characterized by a $BPB_V$ configuration, which suppresses the cleavage of the C-H bond and subsequently promotes the desorption of HCHO. Following this, the conversion of HCHO into the $H_2COOH^-$ anion occurs through hydration and deprotonation at alkaline interfaces, inherently prevent the formation of CO. Figure S15 shows that Bi adatoms weaken the binding energy of HCHO on $Pt_3In$(110) from -0.41 eV to 0.04 eV, thereby effectively promoting the desorption of HCHO.

To model the electrified solid-liquid interface under realistic electrochemical conditions, we performed extensive AIMD simulations on the Bi-$Pt_3In$(110) electrolyte interface system incorporating explicit water molecules. We first investigate the potential of zero charge (PZC) of the Bi-$Pt_3In$(110)/water interface. As shown in Figure S16, the computed PZC of the Bi-$Pt_3In$(110)/water interface was evaluated to ca 0.37 $V_{SHE}$, which is higher than that of the Pt(110)/water interface. This increase in PZC is caused by the Bi modification weakens the interaction between the surface and water molecules, thereby reducing the charge redistribution at the interface.[42] To simulate the MOR process under alkaline conditions, a $CH_3OH$ and K ions were introduced into the water layer, and the electrode potential was adjusted to approximately 0.5 $V_{RHE}$, corresponding to a pH of 14. Our simulations revealed that when two K ions were present in the system, the electrode potential was around 0.23 $V_{RHE}$, whereas with only one K ion, the electrode potential was about 0.79 $V_{RHE}$ (Figure S17). Besides, $CH_3OH$ cannot stably adsorb on the surface at either low or high potentials, but rather preferentially coordinates with K ions at the interface. Based on the findings of Herrero et al.,[43, 44] which suggest that adsorbed OH* may play a crucial role in the initial PECT step, we introduced an additional OH* onto the Bi-$Pt_3In$ surface. The system containing one K ions and OH* was ultimately selected for subsequent simulations. As shown in Figure 4f, when OH* occupies site 1 and $CH_3OH$ approaches an adjacent site-1 position, the hydrogen of the methanol hydroxyl group is spontaneously transferred via hydrogen bonding to OH* and $CH_3O$* is yielded, in agreement with our previous simulations on Bi–Pt surfaces. The schematic diagram of the alkaline MOR pathway is depicted in Figure 4g. Specifically, the presence of OH* at site 1 is required to promote the activation of $CH_3OH$. Following C–H bond scission, the key intermediate HCHO* is formed and subsequently desorbs from the surface. In alkaline media, solvated HCHO(aq) undergoes hydration and deprotonation processes at the interface to yield the $H_2COOH^-$ anion, which can then reabsorb and be further transformed on the surface to HCOOH. Within this reaction pathway, C–H bond cleavage of $CH_3O$ is identified as the RDS. To accurately determine activation barriers under electrochemical conditions, we implemented the "constant-potential hybrid-solvation dynamic model" (CP-HS-DM) developed by Liu et al.[45] We constructed models of $CH_3O$* intermediates on both Bi-$Pt_3In$(110) and Bi-Pt(110) surfaces. Figure 4h presents the representative CP-HS-DM simulation snapshot of $CH_3O$*-to-HCHO* conversion on Bi-$Pt_3In$(110). The oxygen ends of the intermediate bind to site 1, and C−H bond length was employed as the collective variable (CV) in dehydrogenation step, as illustrated in Figure S18. The C-H bond cleavage on Bi-$Pt_3In$(110) requires overcoming a free energy barrier of 0.58 eV, which is significantly lower than the energy barrier of 0.95 eV required on Bi-Pt(110) (Figures 4i, S19, and S20). These results provide strong evidence that the introduction of In optimizes the electronic structure of surface Pt atoms, facilitating C-H bond cleavage and thus enhancing the alkaline MOR kinetics.

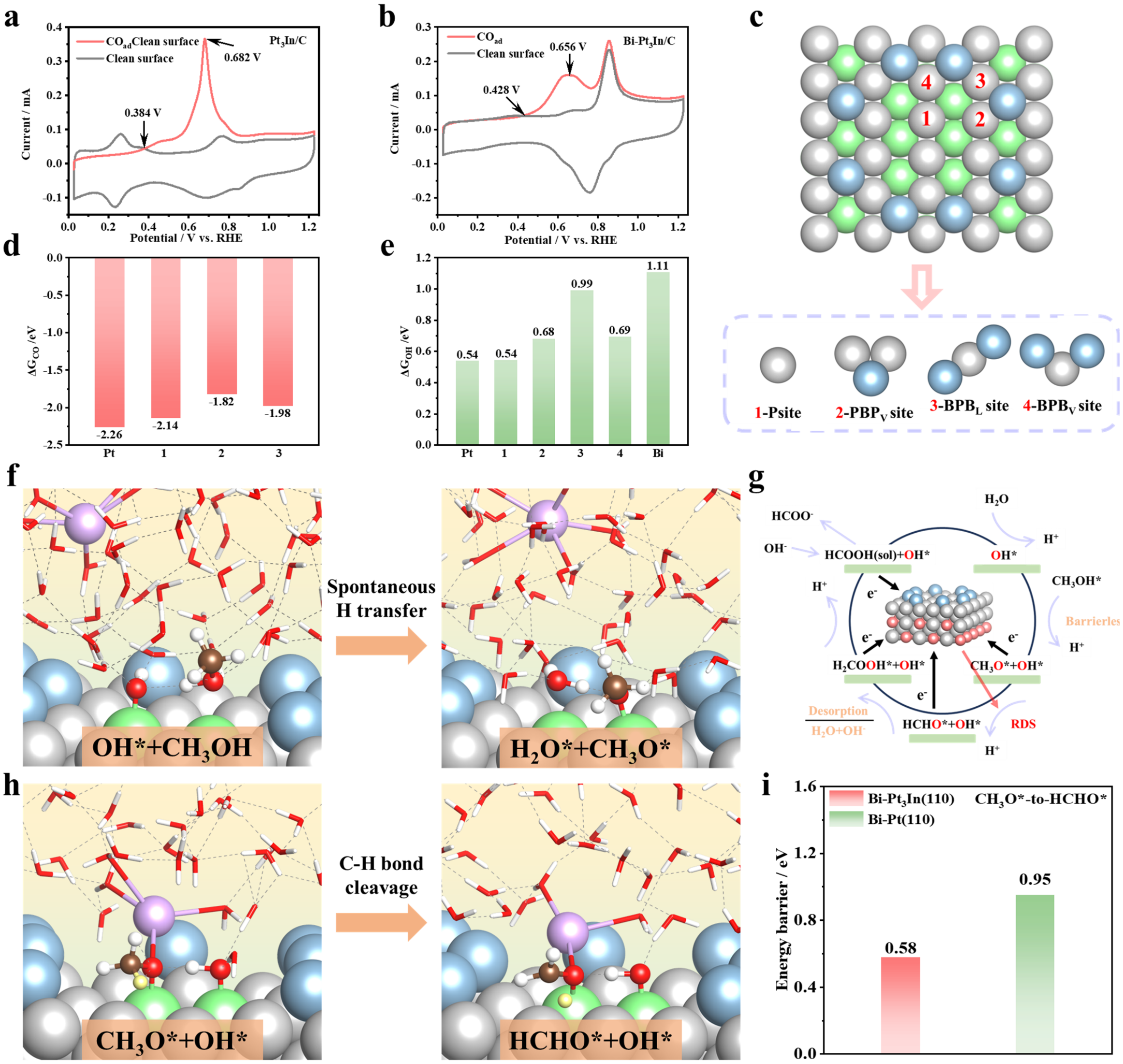

**Figure 4. The underlying reaction mechanism on Bi Modified Pt$_3$In surface.** CO-stripping curves of (a) Pt$_3$In/C and (b) Bi-Pt$_3$In/C in 1M KOH solution. (c) Surface structure of Bi-Pt$_3$In model with multiple sites and the definition of active sites depends on the local coordination environment of Pt and Bi atoms. The binding energy comparison of (d) CO and (e) OH at top sites. "Pt", denoted as the top site of Pt$_3$In(110). (f) Representative snapshots of from AIMD trajectory when OH adsorbs at site 1 and $CH_3OH$ is close to top of adjacent site 1. (g) Schematic diagrams of the alkaline MOR mechanism. (h) Representative snapshots of $CH_3O$*-to-HCHO* conversion. (i) A comparison of dehydrogenation barrier of $CH_3O$* between Bi-Pt$_3$In(110) and Bi-Pt(110) at 0.6 $V_{RHE}$. Color code: Pt, gray; site 1, green; K, purple; Bi, blue; In, light red; O, red; C, brown; H, white; and the cleavable H involved in the C–H bond, yellow.

**Expansion of the Bi-Modified Pt$_3$M/C Intermetallic Series.** To establish the structure-activity relationships during alkaline MOR, we systematically expanded our investigation to include six other L1$_2$-type Pt$_3$M intermetallic compounds (M = Cr, Mn, Co, Zn, Ga, and Sn). Following our established protocol, these materials were subjected to sequential electrochemical dealloying and underpotential deposition of Bi to create a family of core-shell electrocatalysts with tunable properties (Figure 5a). The crystallographic structure of each Pt$_3$M/C material was characterized by X-ray diffraction. As shown in Figure 5b, all synthesized materials exhibit not only major diffraction peaks but also small superlattice peaks, indicating the successful achievement of the ordered structure. The precise Pt loadings of these Pt$_3$M/C catalysts were also determined by ICP-OES, with detailed compositional data compiled in Table S1. Similar to the synthesis of Bi-Pt$_3$In/C, a series of other Bi-modified Pt$_3$M/C catalysts were obtained (denoted Bi-Pt$_3$M/C). The good dispersion of these Bi-Pt$_3$M intermetallic NPs can be further confirmed by TEM images (Figure S21). Apart from the slight large particle size observed in the TEM image of $Pt_3Sn$/C, other Bi-$Pt_3M$/C (M=Cr, Mn, Co, Zn, and Ga) intermetallic NPs exhibit small particle size. That is because the formation of ordered $Pt_3Sn$ phase requires higher annealing temperature, resulting in overgrowth of nanocrystals. The elemental distribution within individual nanoparticles was characterized by energy dispersive spectroscopy (EDS) mapping, confirming the homogeneous spatial distribution of Pt, M, and Bi throughout the nanostructures (Figures 5c, S22-S27).

Figure S28 shows the CV curves of Bi-$Pt_3M$/C and $Pt_3M$/C in 1 M KOH solution. It is evident that the hydrogen adsorption/desorption currents of Bi-$Pt_3M$/C significantly decrease, while their oxygen currents increase compared to their unmodified counterparts. According to the CO stripping experiments, the changes in the CO oxidation current curve indicate that there are still bits of Pt sites susceptible to CO poisoning after Bi modification, indicating that the surface structure of Bi-$Pt_3M$/C is similar to that of Bi-$Pt_3In$/C (Figure S29). The MOR activity of these catalysts was recorded in 1 M $CH_3OH$/1 M KOH electrolyte (Figure S30). All Bi-$Pt_3M$/C catalysts show significantly improved enhanced activity, much higher than those of unmodified $Pt_3M$/C and Bi-Pt/C (Figure 5c). This universal enhancement can be attributed to two synergistic effects: (1) the Bi modification redirects the reaction through a CO-free pathway, and (2) the incorporation of M atoms in the intermetallic core promotes C-H bond activation, resulting in dramatically accelerated reaction kinetics. As shown in Figures S30 and S31, the order of the MOR activities of the catalysts at peak potential and 0.5 $V_{RHE}$ is as follows: Bi-$Pt_3In$/C > Bi-$Pt_3Mn$/C > Bi-$Pt_3Co$/C > Bi-$Pt_3Cr$/C > Bi-$Pt_3Ga$/C > Bi-$Pt_3Zn$/C > Bi-$Pt_3Sn$/C > Bi-Pt/C. This well-defined activity sequence provides valuable insights into the electronic and geometric factors governing catalytic performance, establishing a foundation for elucidating quantitative structure-activity relationships.

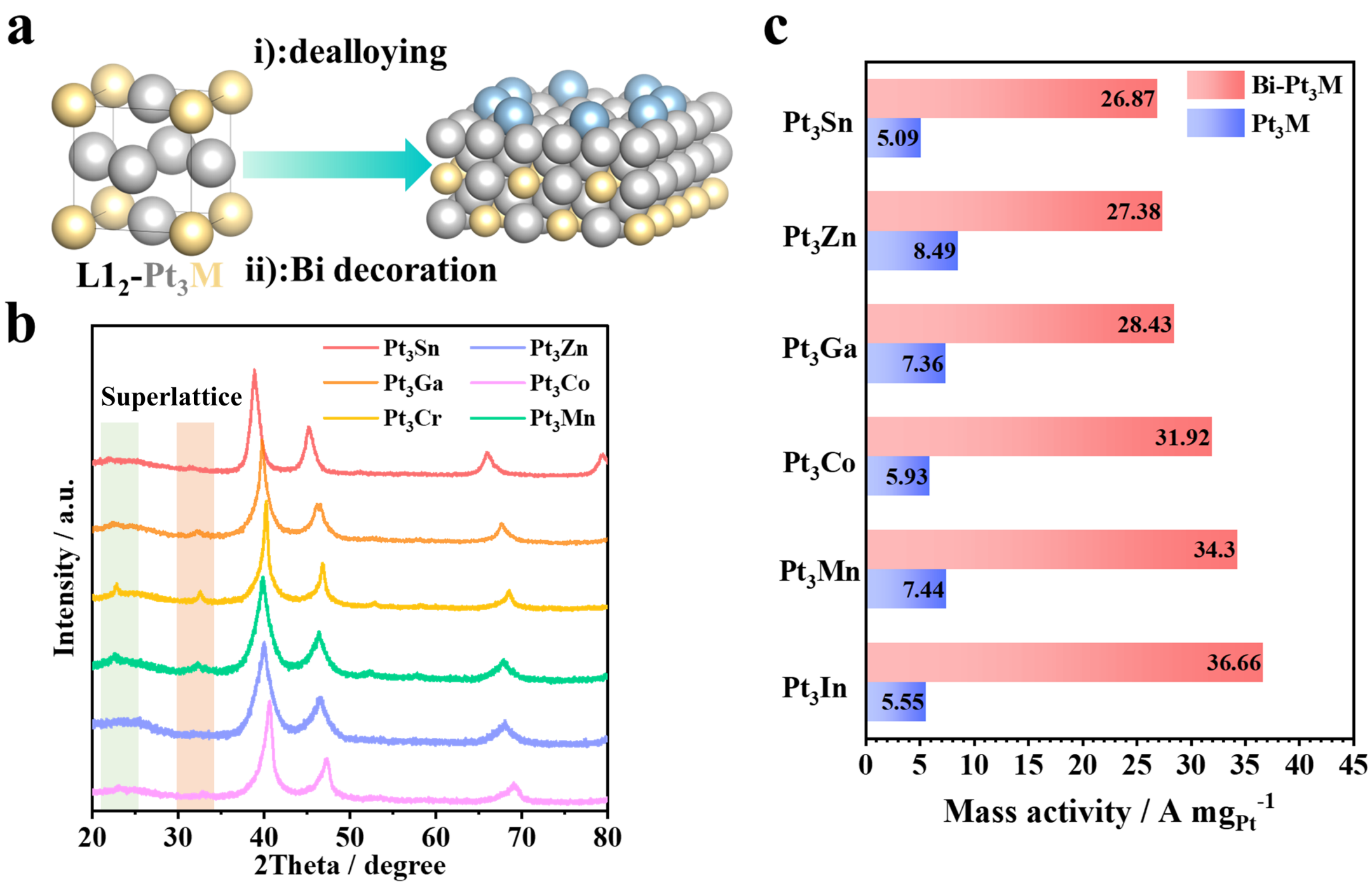

**Figure 5. A family of Bi Modified $Pt_3M$/C.** (a) Schematic illustrating the synthesis of a series of Bi-$Pt_3M$/C (M=Cr, Mn, Co, Zn, Ga, and Sn). (b) XRD patterns of $Pt_3M$/C. (c) Mass activity of $Pt_3M$/C and Bi-$Pt_3M$/C at peak potential.

**Kinetically Controlled Step and OH Binding Energy Descriptor for Bi-$Pt_3M$ Catalysts.** Leveraging our systematic series of Bi-$Pt_3M$/C catalysts with precisely controlled compositions, we aimed to identify a universal descriptor capable of quantitatively predicting catalytic activity and offering fundamental insights into the governing principles of alkaline methanol electrooxidation. Given the isotropic strain inherent to the $L1_2$-type $Pt_3M$ intermetallic, lattice strain was first considered as a possible descriptor to describe activity of Bi-$Pt_3M$/C. As shown in Figure S32, the correlation between lattice strain and activity of Bi-$Pt_3M$/C correlations produces an M-shaped curve with lattice strain changes, indicating that both excessive tensile and compressive strain leads to suboptimal performance. In parallel, the d-band center theory-widely regarded and extensively applied to rationalize the interaction between active sites and key intermediates. Thus, we evaluated the d-band center of Pt atoms in Bi-$Pt_3M$(110). Similarly, the activity plotted against the d-band center similarly followed an M-shaped curve (Figure S35b) rather than the anticipated volcano-type correlation. These phenomena are presumably attributable to the multi-site effect on the Bi-Pt surface, which complicates the use of lattice strain or the d-band center as reliable descriptors for capturing the structure–activity relationships of Bi-$Pt_3M$/C.

It is therefore essential to elucidate the origin of the structure-activity relationship by considering both the reaction mechanism and the nature of the active sites. According to our proposed mechanism, the principle adsorbed species throughout the reaction process are OH*, $CH_3O$*, and $H_2COOH$*, all of which interact with the surface through their oxygen atoms (Figure S33). Also, these oxygen-containing intermediates predominantly adsorb at sites 1. Guided by this insight, we examined the correlations between mass activity and the binding energies of OH* and $CH_3O$* at site 1. As shown in Figures 6a and S34-S35, both the OHBE and the $CH_3O$* binding energy exhibit a pronounced volcano-type relationship with activity, underscoring that excessively strong or weak binding diverts the activity from the optimal performance. Based on the above - mentioned mechanistic analysis, the C–H bond cleavage of $CH_3O$* is confirmed to be the RDS, with its binding strength exerting a decisive influence on C–H bond activation. In particular, weaker binding promotes electronic redistribution within the C–O bond, thereby enhancing the electron-withdrawing effect of the oxygen atom on the methyl group and facilitating polarization of the C–H bond. Bader charge analysis shows that the positive charge on the carbon atom of $CH_3O$* is 0.42 e, 0.44 e, and 0.45 e on Bi-Pt(110), Bi-$Pt_3In$(110), and Bi-$Pt_3Sn$(110), respectively (versus 0.47 e for the free $CH_3O$ moiety). In addition, the free energy barrier for C-H bond cleavage of $CH_3O$* on Bi-$Pt_3Sn$(110) is 0.45 eV-lower than that on Bi-$Pt_3In$(110) and Bi-Pt(110), further confirming that weak adsorption indeed facilitates C–H bond activation.

However, excessively weak adsorption results in diminished activity, which can be rationalized in terms of OHBE: weak OHBE is unfavorable for water dissociation and hence for initiating $CH_3OH$ activation. To validate this hypothesis, KIE measurements were conducted in a $CH_3OD$+ NaOD /$D_2O$ electrolyte (all hydroxyl hydrogens replaced by deuterium). As shown in Figures 6b–c, relative to $CH_3OH$+ NaOH /$H_2O$, the MOR activities of Bi-$Pt_3In$/C and Bi-Pt/C decrease significantly in the deuterated solution, demonstrating that water dissociation contributes to the RDS. Complementarily, we employed the CP-HS-DM approach to calculate the water-dissociation barrier at 0.6 $V_{RHE}$ for Bi-Pt(110), Bi-$Pt_3In$(110), and Bi-$Pt_3Sn$(110). Figure 6d presents the representative snapshots from the water dissociation process on the Bi-$Pt_3In$(110) surface with one $K^+$ and one $OH^-$ introduced into the water layer. The CV is defined as a combination of O–H distances associated with the corresponding $H_2O$*. In the initial state, $H_2O$ is stably adsorbed at site 1. The water-dissociation pathway involves dissociation of $H_2O$* to form OH* and a proton, the latter relayed through the interfacial hydrogen-bond network and subsequently consumed by $OH^-$ in the double layer, regenerating $H_2O$. As summarized in Figures 6e and S37–S39, the free energy barrier increases from 0.31 eV (Bi-Pt) to 0.41 eV (Bi–$Pt_3In$) and 0.68 eV (Bi–$Pt_3Sn$), consistent with the trend in OHBE. Therefore, when the OHBE is too weak, the high water-dissociation barrier hinders the activation of $CH_3OH$.

Collectively, OH/$H_2O$ interconversion dynamics, alongside C–H activation, constitute a second major determinant of the overall reaction kinetics. As depicted in Figure 6f, alkaline MOR on Bi–$Pt_3M$ and Bi-Pt is kinetically governed by both water dissociation and C–H bond cleavage within the RDS. OH* produced from water dissociation initiates $CH_3OH$ activation to yield $CH_3O$* via hydrogen bonding, while being simultaneously regenerated as $H_2O$; The $CH_3O$* intermediate then undergoes C–H bond cleavage to form HCHO*, which desorbs into the interface and converts into $H_2COOH^-$. Meanwhile, $H_2O$* continues to dissociate, replenishing OH* and sustaining catalytic turnover. Throughout this sequence, the binding energies of oxygenated intermediates at site 1 exert a decisive influence on the reaction kinetics: moderately weak binding facilitates C–H activation, whereas excessively weak binding renders water dissociation highly unfavorable, thereby emerging as the primary limitation. Consistent with this mechanistic framework, the Bi–$Pt_3M$/C catalysts exhibit a well-defined volcano-type relationship between activity and OHBE, establishing OHBE as the critical kinetic descriptor for alkaline MOR activity.

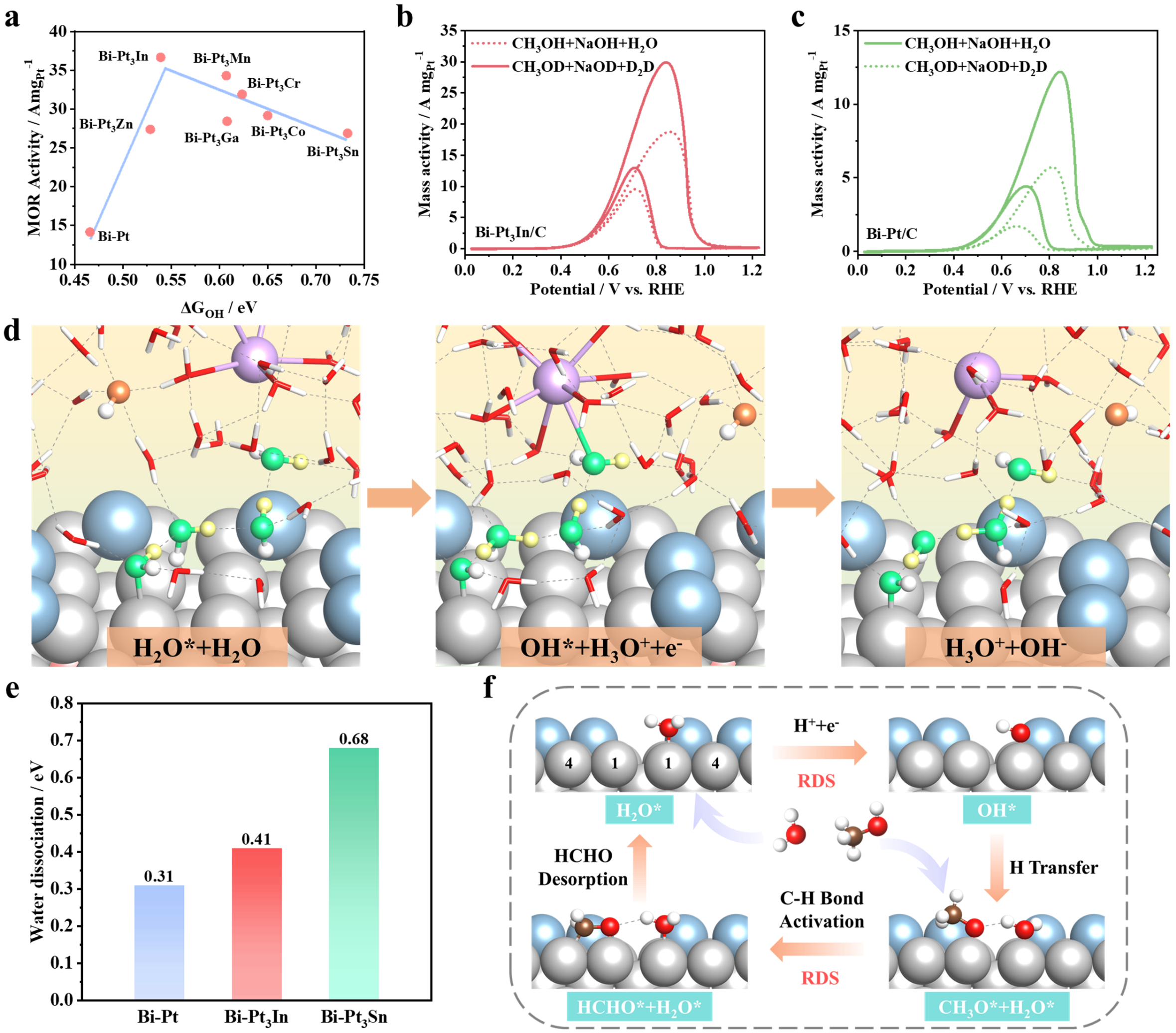

**Figure 6. Structure-to-property relationship and the RDS in Methanol Electrocatalysis Mechanism.** (a) The volcano-type trend based on the relationship between OHBE at site 1of Bi-Pt$_3$M and their MOR mass activities. A comparison of CV curves of (b) Bi-Pt$_3$In/C and (c) Pt/C between 1MCH$_3$OH+1M NaOH/H$_2$O and 1M CH$_3$OD+1M NaOD/D$_2$O. (d) The corresponding close-ups of the simulated water dissociation processes. (e) The free energy barriers of water dissociation at 0.6V$_{RHE}$. (f) Schematic illustration of kinetic control by C–H activation and water dissociation. Color code: Pt, gray; K, purple; Bi, blue; O, red; O in OH$^-$, orange; C, brown; H, white; the O involved in the proton transfer process, green; and the H involved in the proton transfer process, yellow.

## DISCUSSION

In summary, a family of Bi-Pt$_3$M/C intermetallic catalysts that follow a CO-free dominated pathway and deliver exceptional alkaline MOR activity was developed and served as an ideal materials platform for in-depth studies of kinetic origins and governing factors. To elaborate, all Bi-Pt$_3$M/C catalysts exhibit superior mass activity relative to Bi-Pt/C and Pt/C; notably, Bi-Pt$_3$In/C delivers an ultrahigh mass activity of 36.66 A mg$_{Pt}^{-1}$, surpassing state-of-the-art Pt-based MORcatalyst benchmarks. Using Bi-Pt$_3$In/C as a proof of concept, combined experimental and computational analyses revealed the origin of the enhanced reaction kinetics lies in the rapid C - H bond activation. More specifically, kinetic isotope effects tests identify that the activation barrier of the O-H bond in methanol is negligible, whereas the cleavage of the C-H bond is the rate-determining step. Assisted by AIMD stimulations, DFT calculations, and formaldehyde electrooxidation, we demonstrate that the introduction of In reduces the activation barrier of CH$_3$O* dehydrogenation step. Furthermore, the activities of Bi-Pt$_3$M/C catalysts follow a well-fitted volcano-type relationship with OHBE change. In accordance with the proposed mechanistic picture, OH* resulting from water dissociation are deemed to be indispensable for the initiation of methanol activation. Proton transfer from the methanol hydroxyl to OH* is a rapid process that occurs via hydrogen bonding. This process yields $CH_3O$* and concomitantly regenerates $H_2O$*. As a consequence, a continuous interconversion process is undergone by the OH*/$H_2O$* pair. Beyond C–H activation, the dynamics of the interconversion of OH and $H_2O$ constitute another significant factor in determining the overall kinetics and are well described by the OHBE. Hence, OHBE can serve as a reliable activity descriptor. Overall, our work provides an enlightening research paradigm to open up a hidden dimension to rationally design advanced MOR catalysts with a specific goal.

## METHODS

### Chemicals

Bismuth nitrate pentahydrate ($Bi(NO_3)_3·5H_2O$), Chromium(III) chloride hexahydrate ($CrCl_3·6H_2O$), Manganese(II) chloride tetrahydrate ($MnCl_2·4H_2O$), Cobalt chloride hexahydrate ($CoCl_2·6H_2O$), Zinc bromide ($ZnBr_2$), Indium chloride tetrahydrate ($InCl_3·4H_2O$), Gallium nitrate hydrate ($Ga(NO_3)_3·xH_2O$), Tin(II) chloride dihydrate ($SnCl_2·2H_2O$), chloroplatinic acid hydrate ($H_2PtCl_6·6H_2O$), methanol ($CH_3OH$, 99.5 %), sodium hydroxide (NaOH, 99.9%), and potassium hydroxide (KOH, 95%) were all purchased from Aladdin. Formaldehyde (HCHO, 37 %) was purchased from MACKLIN. Deuterium-substituted methanol ($CD_3OD$ and $CH_3OD$; 99.8 atom % D), sodium deuteroxide (NaOD, 40% in $D_2O$), and deuterium oxide ($D_2O$) were purchased from Energy Chemical. Sulfuric acid ($H_2SO_4$, AR) and perchloric acid ($HClO_4$, GR) were purchased from Guangzhou Chemical Reagent Factory. Ketjen Black EC300J was purchased from Suzhou Sinero Technology Co., Ltd. Commercial Pt/C was provided by was provided by TANAKA (20 wt.%). All regents were used without further purification, and all solutions were freshly prepared with ultrapure water (18.2 MΩ $cm^{-1}$).

### Preparation of $Pt_3M$/C

The synthesis of Pt-based intermetallic nanoparticles (NPs) was accomplished through a two-step method comprising freeze-drying and annealing reduction, a route that was adapted from prior work. Briefly, pretreated Ketjen Black (KB) was co-impregnated with $H_2PtCl_6·6H_2O$ and the desired metal salt solutions in stoichiometric proportions, followed by ultrasonication (1 h) to ensure uniform dispersion. The slurry was then frozen and lyophilized to afford a dry carbon–metal-salt precursor. Subsequent reductive annealing under Ar/$H_2$ (8 vol%) yielded the carbon-supported Pt intermetallics. The annealing time and temperature used are listed in Table S3.

### Preparation of Bi-Modified $Pt_3M$/C

Following our previous Bi–modified Pt/C procedure with minor adaptations, Bi was deposited onto $Pt_3M$/C by a room-temperature electrodeposition. Briefly, 2.0 mg of $Pt_3M$/C powder was dispersed in 950 μL isopropanol, 50 μL ultrapure water, and 20 μL Nafion by ultrasonication (1 h) to obtain a homogeneous ink. An 4 μL ink was dropped onto a polished glassy-carbon electrode (0.196 $cm^2$) and air-dried to serve as the working electrode. Prior to Bi deposition, the electrode was subjected to electrochemical dealloying by cyclic voltammetry in 0.1 M $HClO_4$ ($N_2$-saturated) to generate core–shell particles. The CV was conducted at 250 mV $s^{-1}$ between 0.05–1.20 V vs. RHE until the voltammogram stabilized (typically ~60 cycles). Bi electrodeposition was then performed at −0.16 V vs Ag/AgCl (sat. KCl) for 120 s in 0.5 M $H_2SO_4$ containing 1 mM $Bi^{3+}$. The electrode was rinsed thoroughly with water and used immediately for electrochemical measurements.

### Materials Characterization

Transmission electron microscopy (TEM) and energy-dispersive X-ray spectroscopy (EDS) were performed on a Talos F200X operated at 200 kV. High-resolution TEM (HRTEM) and high-angle annular dark-field scanning TEM (HAADF-STEM), and the corresponding HAADF-STEM-EDS elemental maps were acquired on a Titan Themis G2 at an accelerating voltage of 300 kV. Bulk elemental compositions were determined by ICP-OES (IRIS Intrepid II XSP, Thermo Fisher). XPS measurements were carried out on an Axis Ultra DLD spectrometer using a monochromated Al Kα source ($h\nu$ = 1486.6 eV).

## Electrochemical Measurement

All the electrochemical measurements were performed on a PINE-WaveDriver 200 electrochemical workstation with a three-electrode cell. A Pt mesh and a Hg/HgO (1 M KOH) electrode served as the counter and reference electrodes, respectively, and a glassy-carbon disk (0.196 $cm^2$) coated with catalyst ink was used as the working electrode. MOR polarization curves were recorded at a scan rate of 50 mV $s^{-1}$ in $N_2$-saturated electrolytes of 1 M KOH + 1 M $CH_3OH$. Chronoamperometry (CA) for MOR was conducted at 0.45 V vs. RHE. FOR polarization curves were collected at 50 mV $s^{-1}$ in $N_2$-saturated 1 M KOH + 1 M HCHO.

For CO-stripping experiments, measurements were performed in 1 M KOH using a graphite rod as the counter electrode. Prior to stripping, the electrolyte was bubbled with CO (99.9%) for 15 min to saturate the surface with CO at Pt active sites; the working electrode was then transferred to fresh $N_2$-saturated electrolyte, and voltammograms were acquired at 50 mV $s^{-1}$.

## Computations and Models

All density functional theory (DFT) and ab initio molecular dynamics (AIMD) calculations were carried out with the Vienna Ab initio Simulation Package (VASP) using the projector-augmented wave (PAW) formalism. Electron exchange–correlation effects were described by the revised Perdew−Burke−Ernzerhof functional (RPBE) functional within the generalized gradient approximation (GGA) framework. Unless otherwise noted, a plane-wave cutoff of 520 eV and Gaussian smearing (σ = 0.1 eV) were employed. Long-range dispersion was accounted for by the zero−damping method of Grimme (DFT-D3). Geometry optimizations were deemed converged when the total-energy change was < $1\times10^{-6}$ eV and the residual forces < $1\times10^{-2}$ eV $Å^{-1}$. Surface models employed a five-layer $Pt_3M(110)$ slab (3 × 4 surface supercell) comprising a two-layer Pt skin atop three $Pt_3M$ substrate layers; the bottom three layers were fixed during relaxation, and a 15 Å vacuum was applied along the surface normal (z). Brillouin-zone sampling used a Monkhorst–Pack 3 × 3 × 1 k-point mesh. The free energy change for each fundamental step was determined by

$$\Delta G = \Delta E + \Delta ZPE - T\Delta S$$

where ΔE is the difference in electronic energy directly obtained from the DFT simulation. ΔZPE is the contribution of variation of zero−point energy (ZPE), ΔS is the entropy (S) change, and T is the temperature, at 300 K.

AIMD calculations were performed using PAW potentials and a plane-wave basis with a kinetic-energy cutoff of 400 eV. Electronic exchange–correlation was described by the RPBE functional within the GGA framework. Gaussian smearing with a width of 0.2 eV was applied, and the DFT-D3 was included. Electronic self-consistency was converged to $1\times10^{-5}$ eV. The $Pt_3M(110)$/water interface was modeled by placing 40 $H_2O$ molecules above a five-layer 3 × 4 $Pt_3M(110)$ slab; the bottom three layers of the slab were fixed during dynamics, and a 15 Å vacuum was introduced along z. $K^+$ cations were added when constructing cation-containing interfaces. Dynamics were propagated with a 1 fs time step under the canonical (NVT) ensemble, enforced via a Nosé–Hoover thermostat at 330 K. Brillouin-zone sampling employed the Γ point only. Each system was first evolved for 10 ps to ensure adequate equilibration.

To reproduce the electrode potential in simulations, we adopted the CP-HS-DM framework for constant-potential AIMD as proposed by Liu and co-workers. In this approach, the implicit VASPsol module continuously modulates the total electron number during MD, thereby maintaining the system at the specified electrostatic potential. On the equilibrated slab–electrolyte interface, the constrained AIMD (cAIMD) simulations with a slow-growth scheme was applied, and the activation free-energy barrier was obtained by thermodynamic integration. The process was parameterized by a collective variable, ξ, which was ramped linearly from the reactant to the product basin at a fixed rate, ξ. The work required to perform the transformation from the initial to the final states can be computed as

$$W_{\text{initial-to-final}} = \int_{\xi(\text{initial})}^{\xi(\text{final})} \left(\frac{\partial F}{\partial \xi}\right) \cdot \dot{\xi} dt$$

where F is the computed free energy, which is evolving along with t, $\frac{\partial F}{\partial \xi}$ can be computed along cAIMD using the blue-moon ensemble with the SHAKE algorithm.[46] With the limit of infinitesimally small $\partial\xi$, the needed work ($W_{initial\text{-}to\text{-}final}$) corresponds to the free-energy difference between the final and initial states.

## RESOURCE AVAILABILITY

The resource availability section is required for all research articles. This section might also be required for applicable reviews and perspectives. This section must contain the following required subsections under the resource availability heading: "lead contact," "materials availability," and "data and code availability."

***Lead contact***

Requests for further information and resources should be directed to and will be fulfilled by the lead contact, Zhiming Cui (zmcui@scut.edu.cn).

***Materials availability***

This study did not generate new unique reagents.

***Data and code availability***

Data relating to the TEM, STEM, XPS data, and electrochemical results; details for DFT calculations and AIMD stimulations are available in supplemental information.

ACKNOWLEDGMENTS

This work was financially supported by the National Natural Science Foundation of China (22372062) and the key Technologies R&D Program of Guangdong Province (2023B0909060003). The computation in this work has been done using the facilities (supercomputer Ohtaka) of the Supercomputer Center, the Institute for Solid State Physics, the University of Tokyo. The authors acknowledge the support of the Supercomputing Center of Wuhan University. The authors would like to thank Prof. Yuanyue Liu for sharing the CP-VASP code and Dr. Xiaowan Bai for providing valuable insights on free energy calculations.

## AUTHOR CONTRIBUTIONS

Z.C. and L.L. conceived and supervised the work. Z.C., L.L., and H.L. wrote the manuscript. L.L. and S.Y. designed the experiments and carried out the characterization of catalysts. L.L. and H.L. performed the DFT simulations and AIMD simulations. All the other authors participated in preparing the manuscript and contributed to the discussion.

## DECLARATION OF INTERESTS

The authors declare no competing interests.

## SUPPLEMENTAL INFORMATION

**Figures S1–S39, Tables S1–S3, and supplemental reference.**

# Supporting Information

## Exceptional Alkaline Methanol Electrooxidation on Bi-modified $Pt_3M$ Intermetallics: Kinetic Origins and an OH Binding Energy Descriptor

Lecheng Liang,[†, a, c] Hengyu Li,[†, b] Shao Ye,[†, a, d] Peng Li,[c] Kaiyang Xu,[e] Jinhui Liang,[a] Binwen Zeng,[a] Mingjia Lu,[a] Bo Shen,[d] Taisuke Ozaki,[b] Zhiming Cui*[, a]

[a]Guangdong Provincial Key Laboratory of Fuel Cell Technology, School of Chemistry and Chemical Engineering, South China University of Technology, 510641 Guangzhou, China.

[b]Institute for Solid State Physics, The University of Tokyo, 277-8581 Kashiwa, Japan.

[c]College of Chemistry and Molecular Sciences, Wuhan University, 430072 Wuhan, China.

[d]Department of Materials Science and Engineering, City University of Hong Kong, Hong Kong SAR, China.

[e]Songshan Lake Materials Laboratory, Dongguan 523808, China.

* Corresponding author: zmcui@scut.edu.cn

† These authors contributed equally to this work.

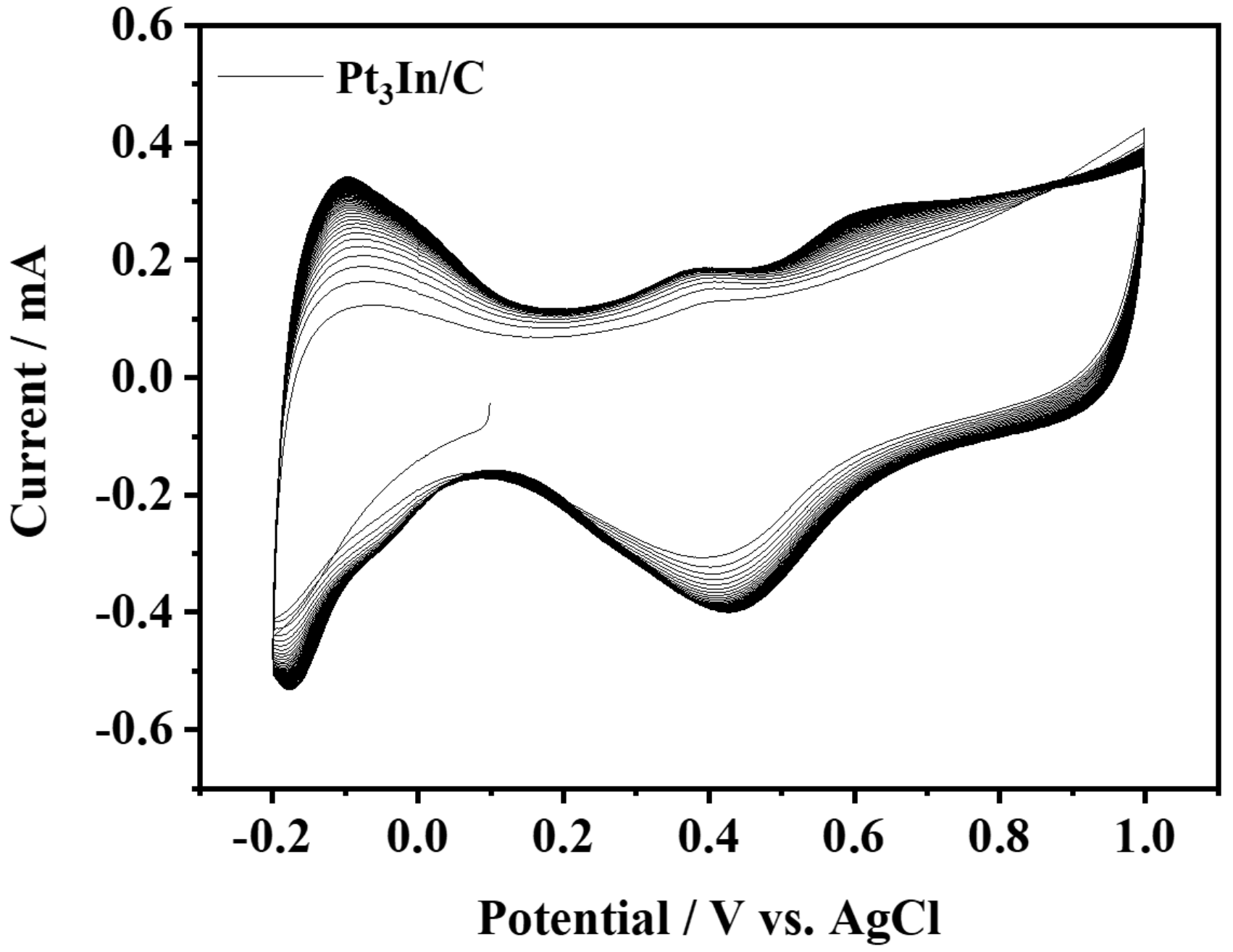

**Figure S1.** CV curves of electrochemical dealloying of $Pt_3In/C$ in $N_2$-saturated 0.1M $HClO_4$ for 60 cycles.

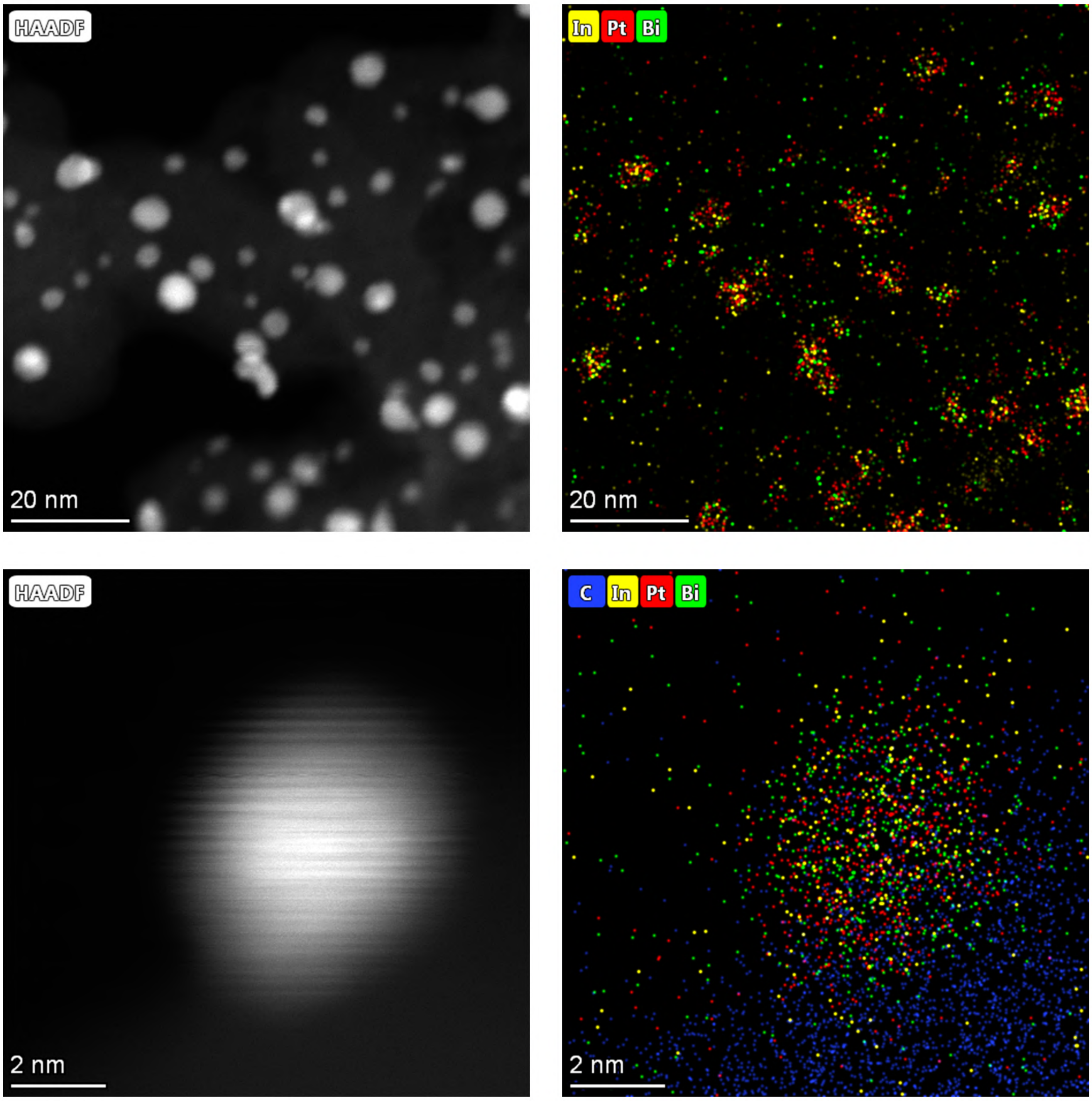

**Figure S2.** HAADF-TEM and EDS element mappings images of Bi-$Pt_3In$ NPs.

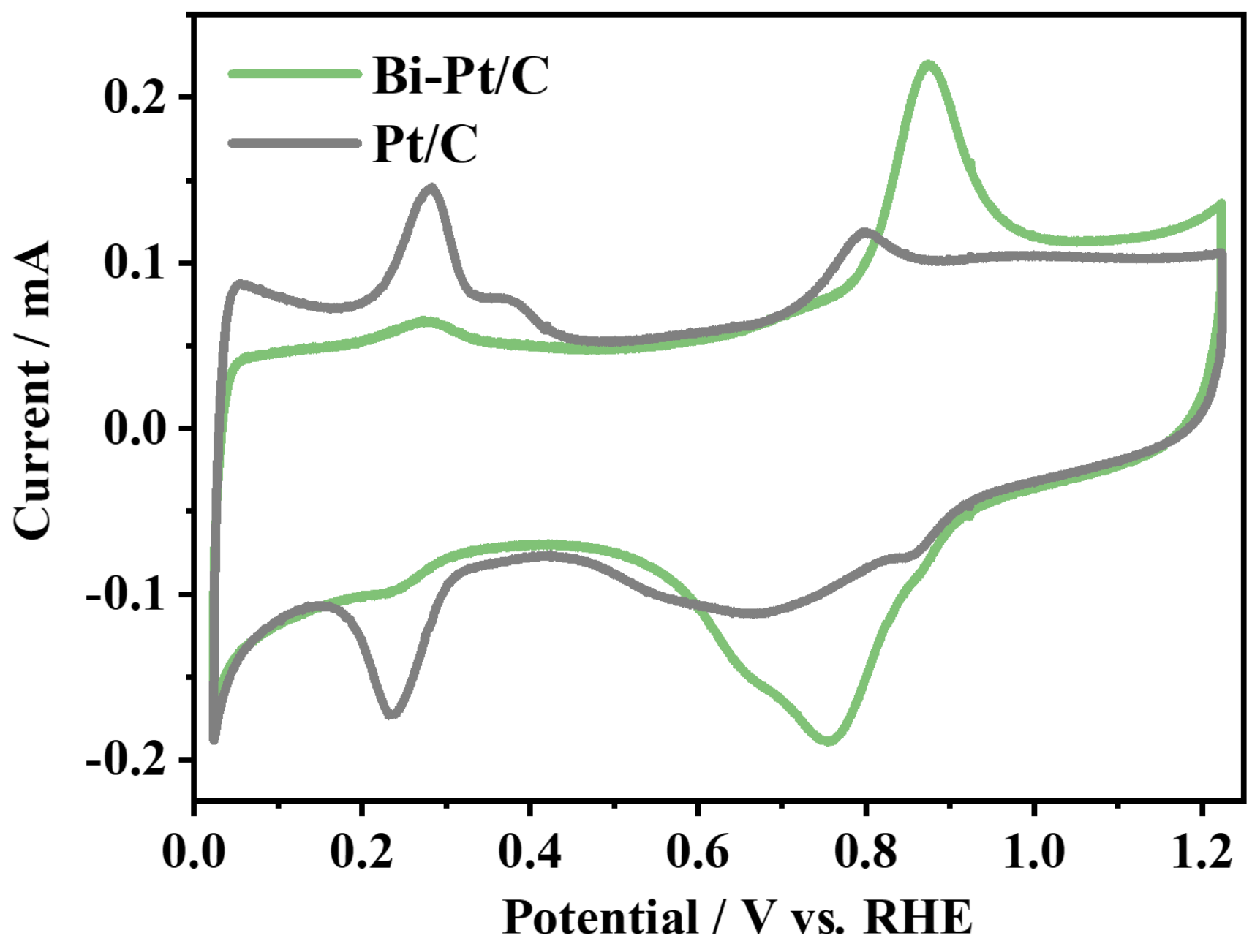

**Figure S3.** CV curves of Bi-Pt/C and Pt/C in $N_2$-satureated 1 M KOH.

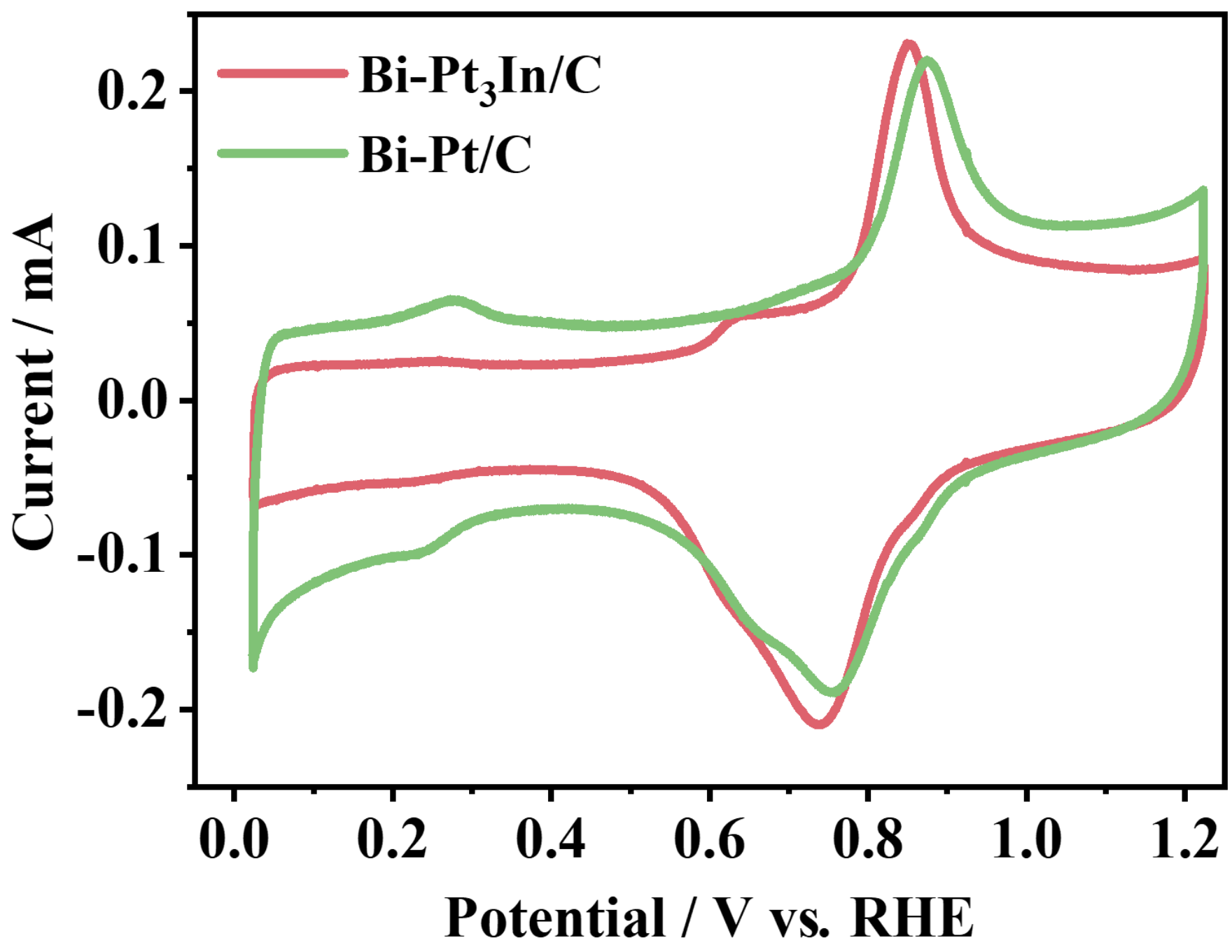

**Figure S4.** Comparison of CV curves for Bi-$Pt_3In$/C and Bi-Pt/C in $N_2$-satureated 1 M KOH.

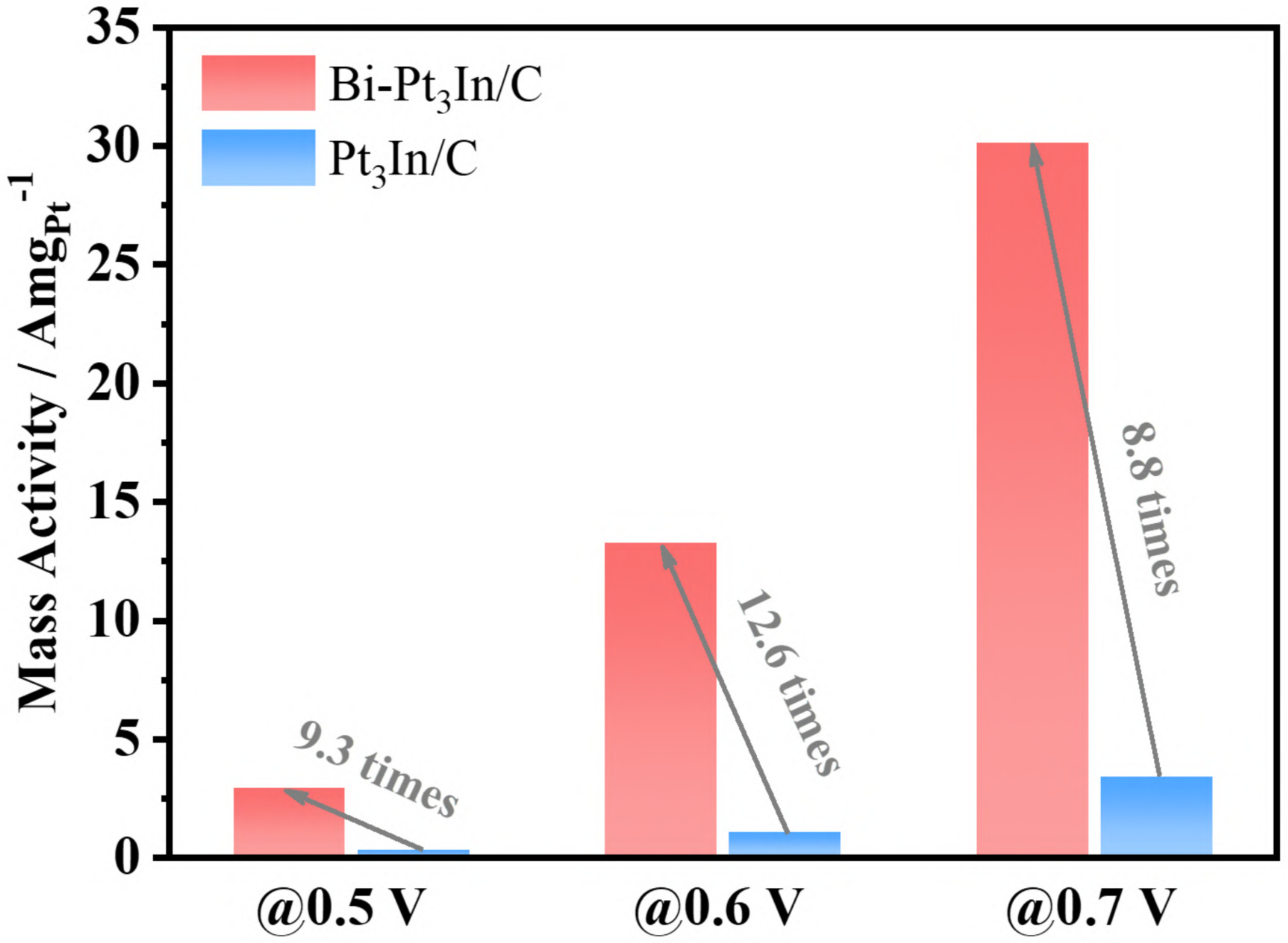

**Figure S5.** Histograms of summarized mass activity of $Pt_3In/C$ and Bi-$Pt_3In/C$ at different potentials.

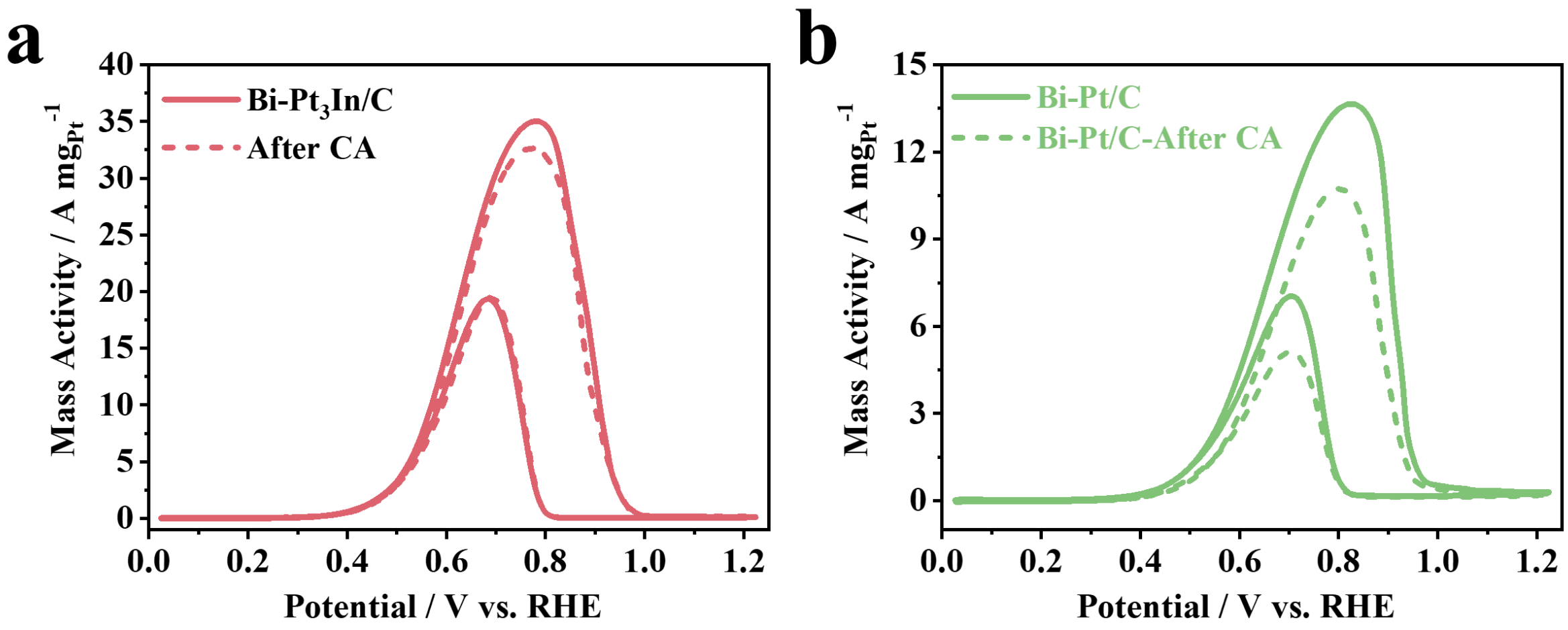

**Figure S6.** CV curves of (a) Bi-$Pt_3In$/C and (b) Bi-Pt/C in 1M $CH_3OH$/1M KOH after CA test.

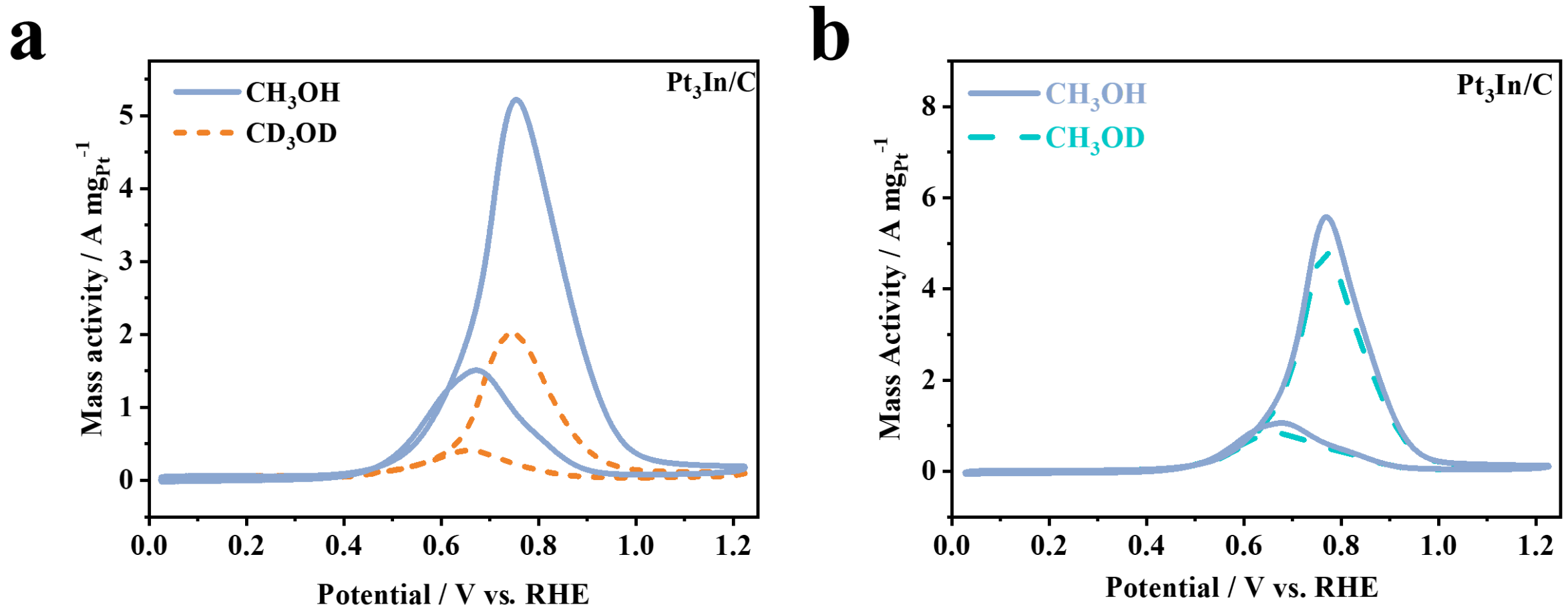

**Figure S7.** (a) A comparison of CV curves of $Pt_3In/C$ between 1M $CH_3OH$/1M KOH and 1M $CD_3OD$/1M KOH. (b) A comparison of CV curves of Bi-$Pt_3In/C$ between 1M $CH_3OH$/1M KOH and 1M $CH_3OD$/1M KOH.

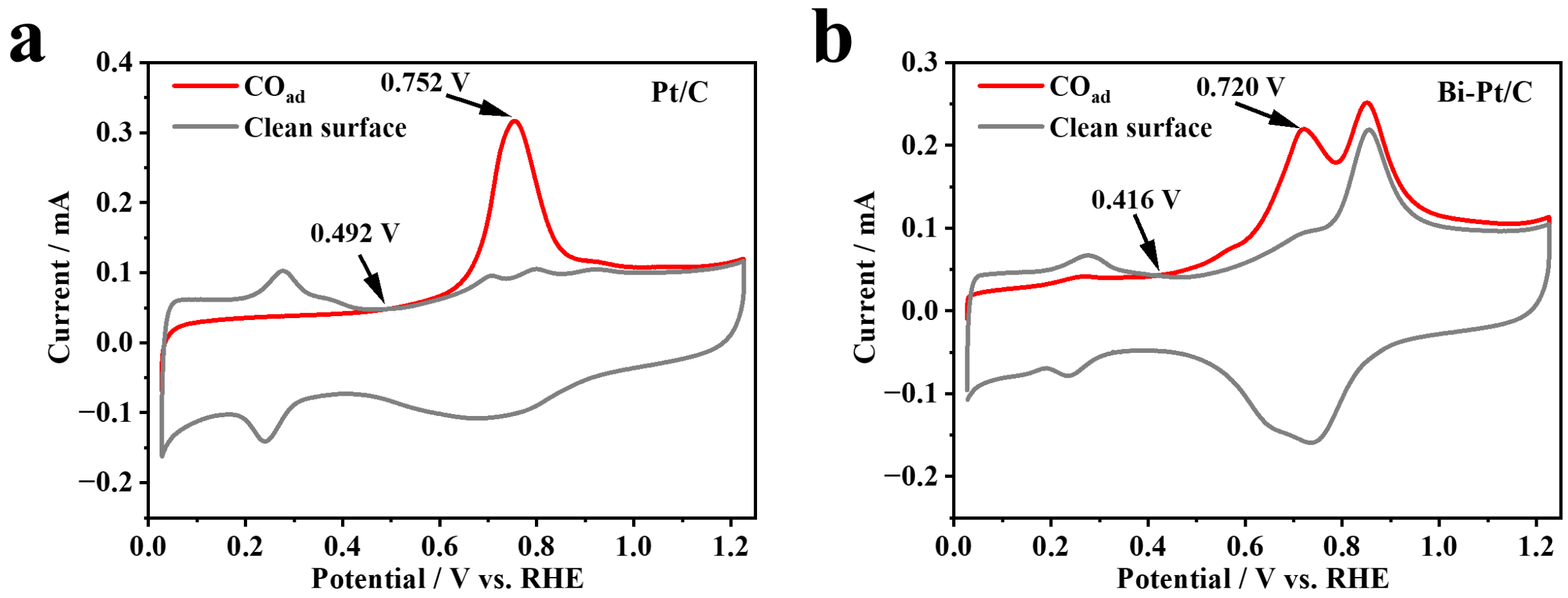

**Figure S8.** CO-stripping curves of (a) Pt/C and (b) Bi-Pt/C in 1M KOH solution.

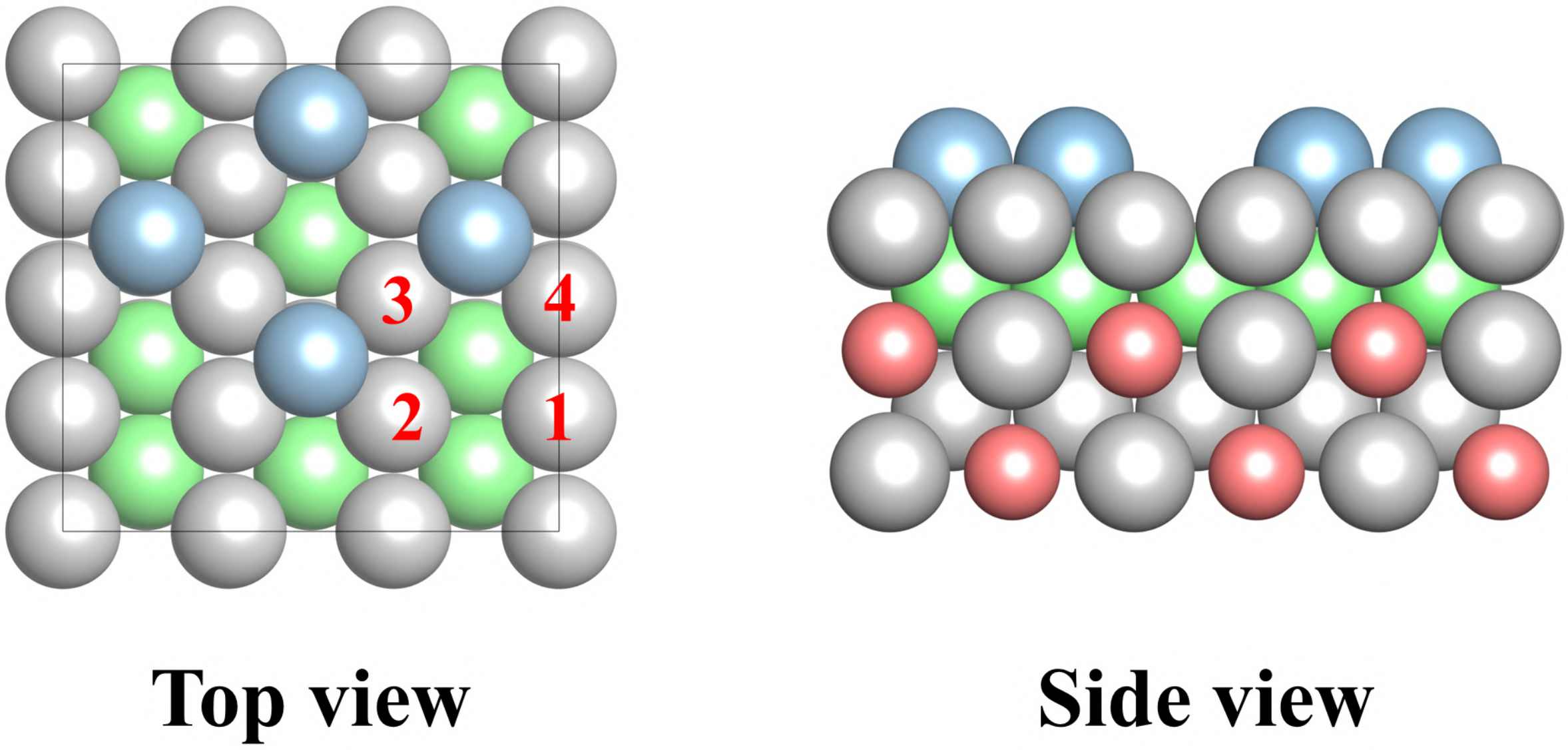

**Figure S9.** The structural model of Bi-$Pt_3In$(110) surface. Color code: outermost layer Pt, gray; subsurface Pt, green; In, light red; and Bi, blue.

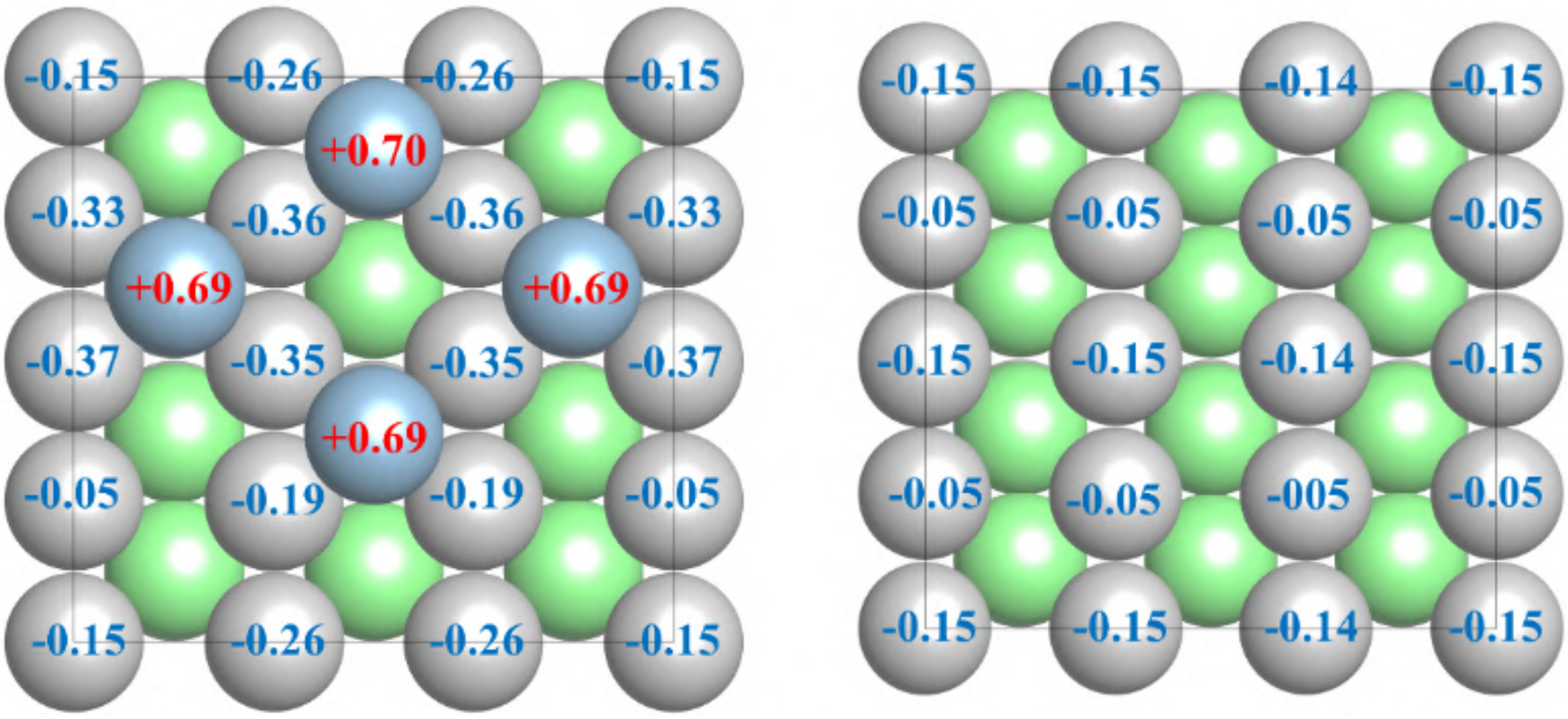

**Figure S10.** Bader charge analysis of Bi-$Pt_3In$(110) and $Pt_3In$(110).

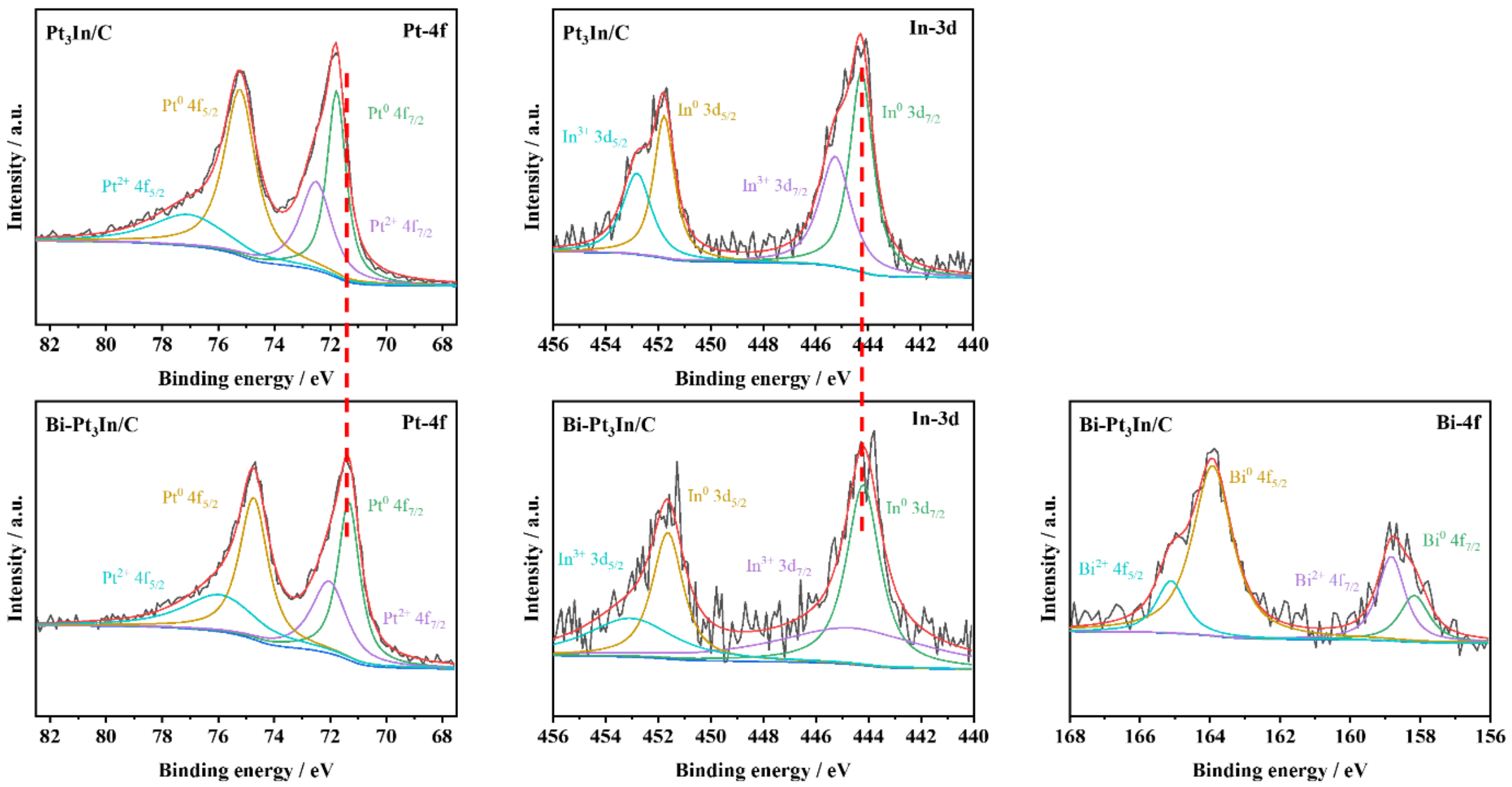

**Figure S11.** Pt 4f, In 3d, and Bi 4f XPS spectra of $Pt_3In/C$ and Bi-$Pt_3In/C$.

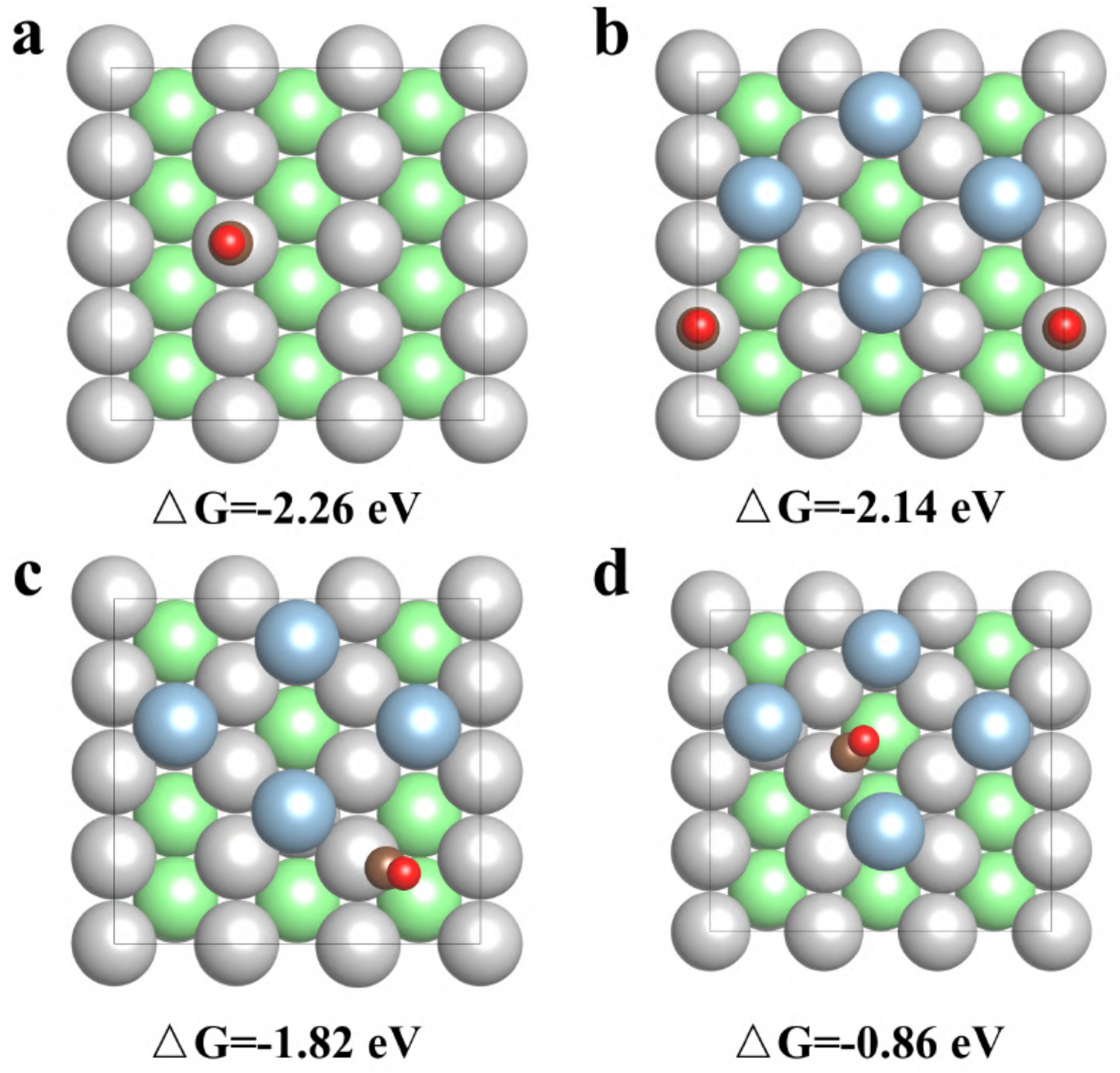

**Figure S12.** Adsorbed CO at (A) Pt top site of $Pt_3In(110)$, (B) site 1, (C) site 2, and (D) site 3 of Bi-$Pt_3In(110)$. Color code: outermost layer Pt, gray; subsurface Pt, green; Bi, blue; O, red; and C, brown.

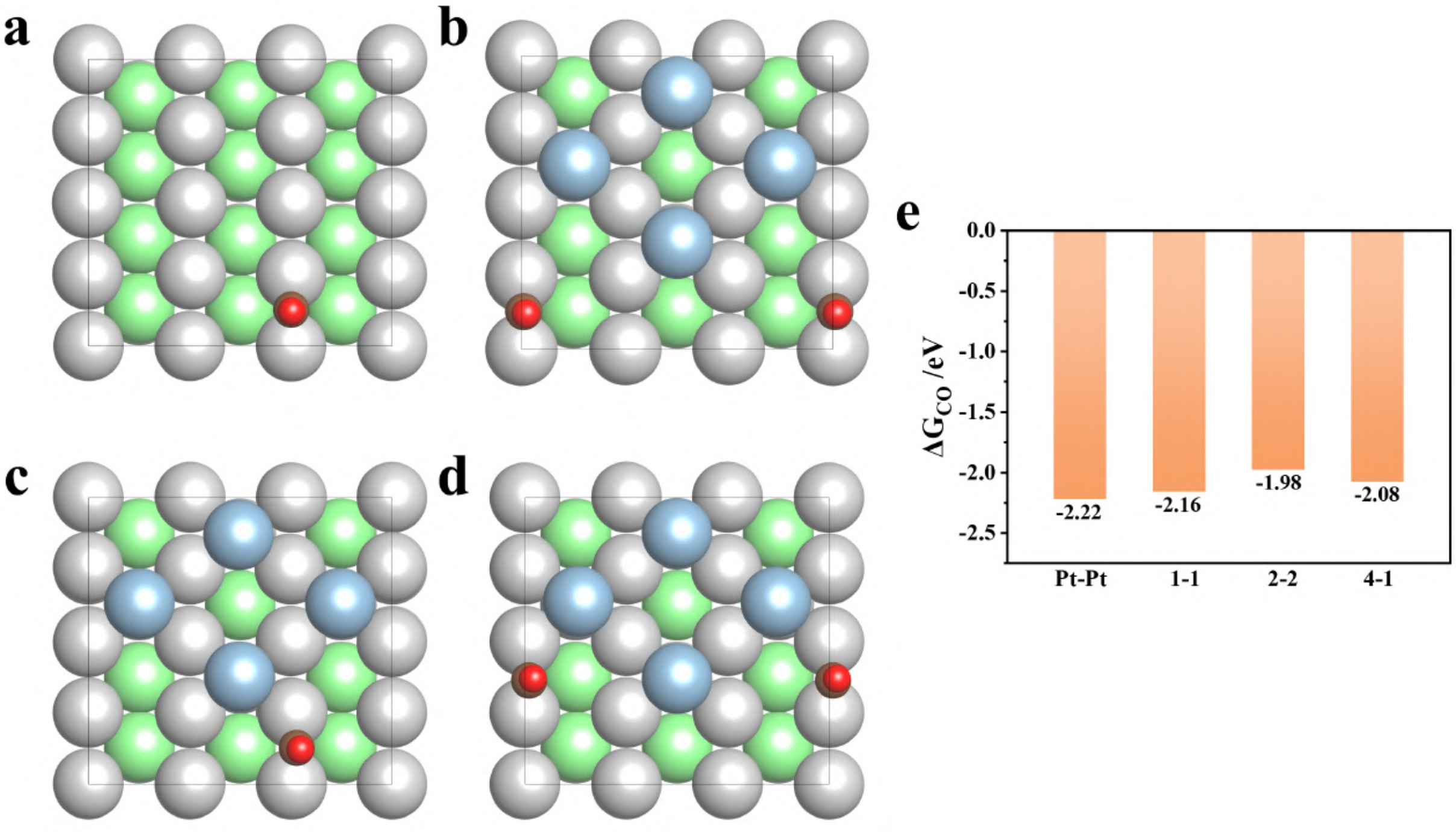

**Figure S13.** Adsorbed CO at (A) Pt-Pt bridge sites of $Pt_3In(110)$, (B) site 1-1, (C) site 2-2, and (D) site 4-1 of Bi-$Pt_3In(110)$. (E) The corresponding COBE comparison. Color code: outermost layer Pt, gray; subsurface Pt, green; Bi, blue; O, red; and C, brown.

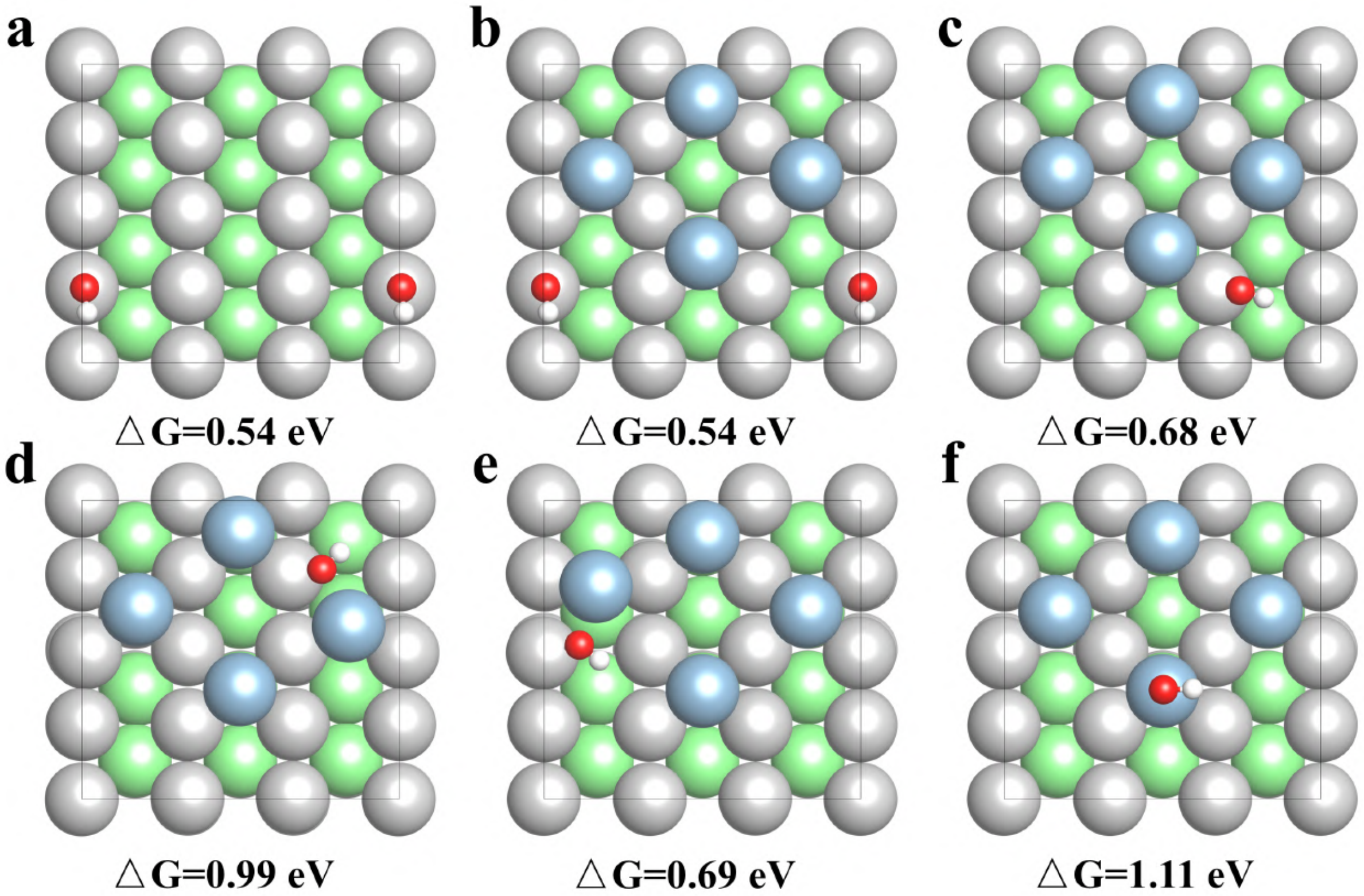

**Figure S14.** Adsorbed OH at (A) Pt top site of $Pt_3In(110)$, (B) site 1, (C) site 2, (D) site 3, (E) site 4, and (F) Bi site of Bi-$Pt_3In(110)$. Color code: outermost layer Pt, gray; subsurface Pt, green; Bi, blue; H, white; and O, red.

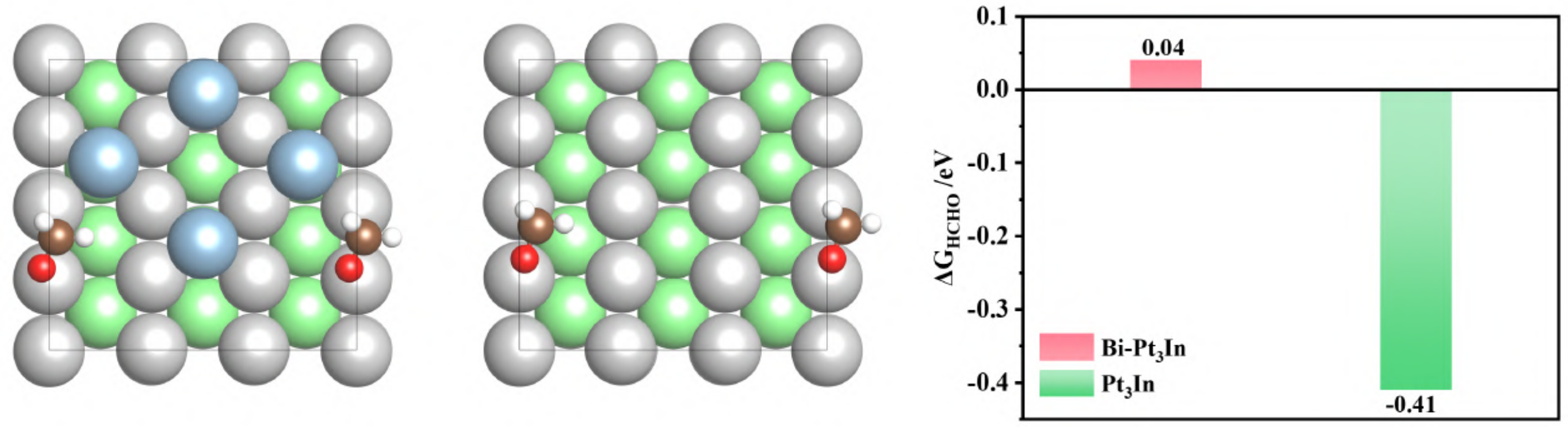

**Figure S15.** Calculated binding energy of HCHO on $Pt_3In(110)$ and Bi-$Pt_3In(110)$. Color code: outermost layer Pt, gray; subsurface Pt, green; In, light red; Bi, blue; C, brown; O, red; and H, white.

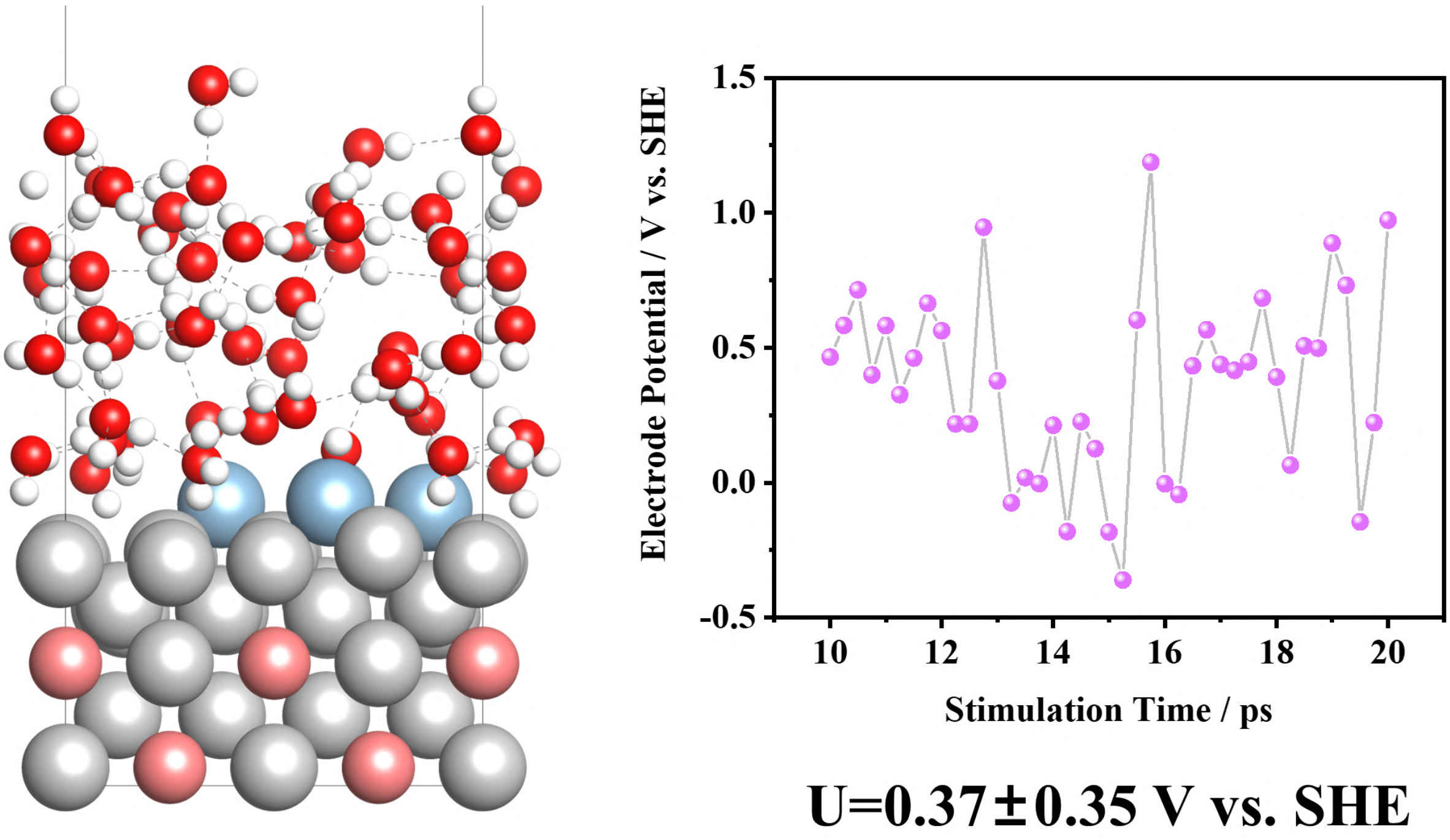

**Figure S16.** Representative snapshots of Bi- $Pt_3In(110)$/water interface and PZC evaluations. Color code: Pt, gray; Bi, blue; H, white; In, light red; and O, red.

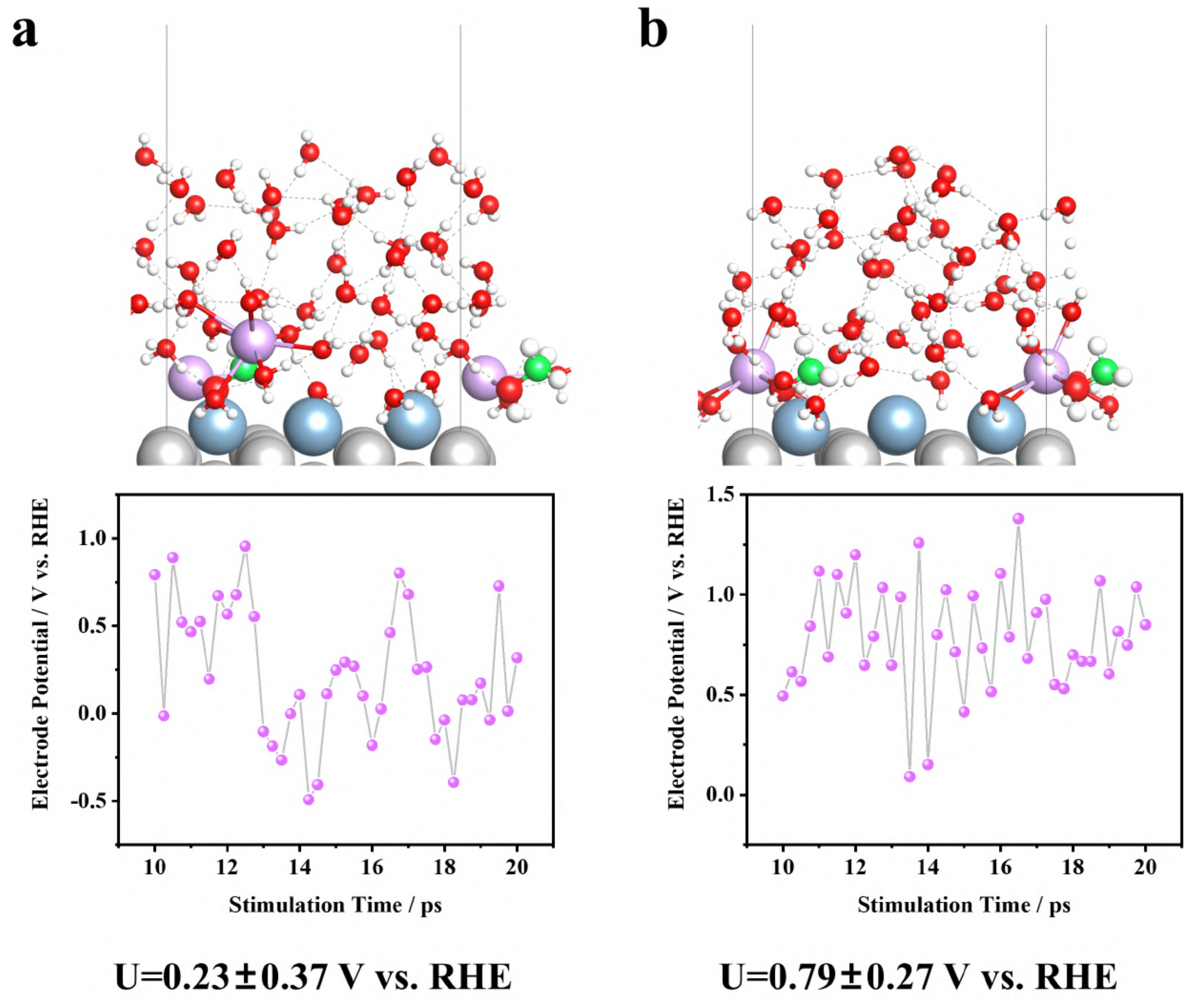

**Figure S17.** Representative snapshots of a $CH_3OH$ on (a) Bi-$Pt_3In$(110)-2$K^+$ and (b) Bi-$Pt_3In$(110)-1$K^+$ and corresponding electrode potential evaluations. Color code: Pt, gray; Bi, blue; H, white; C, green; K, purple; In, light red; and O, red.

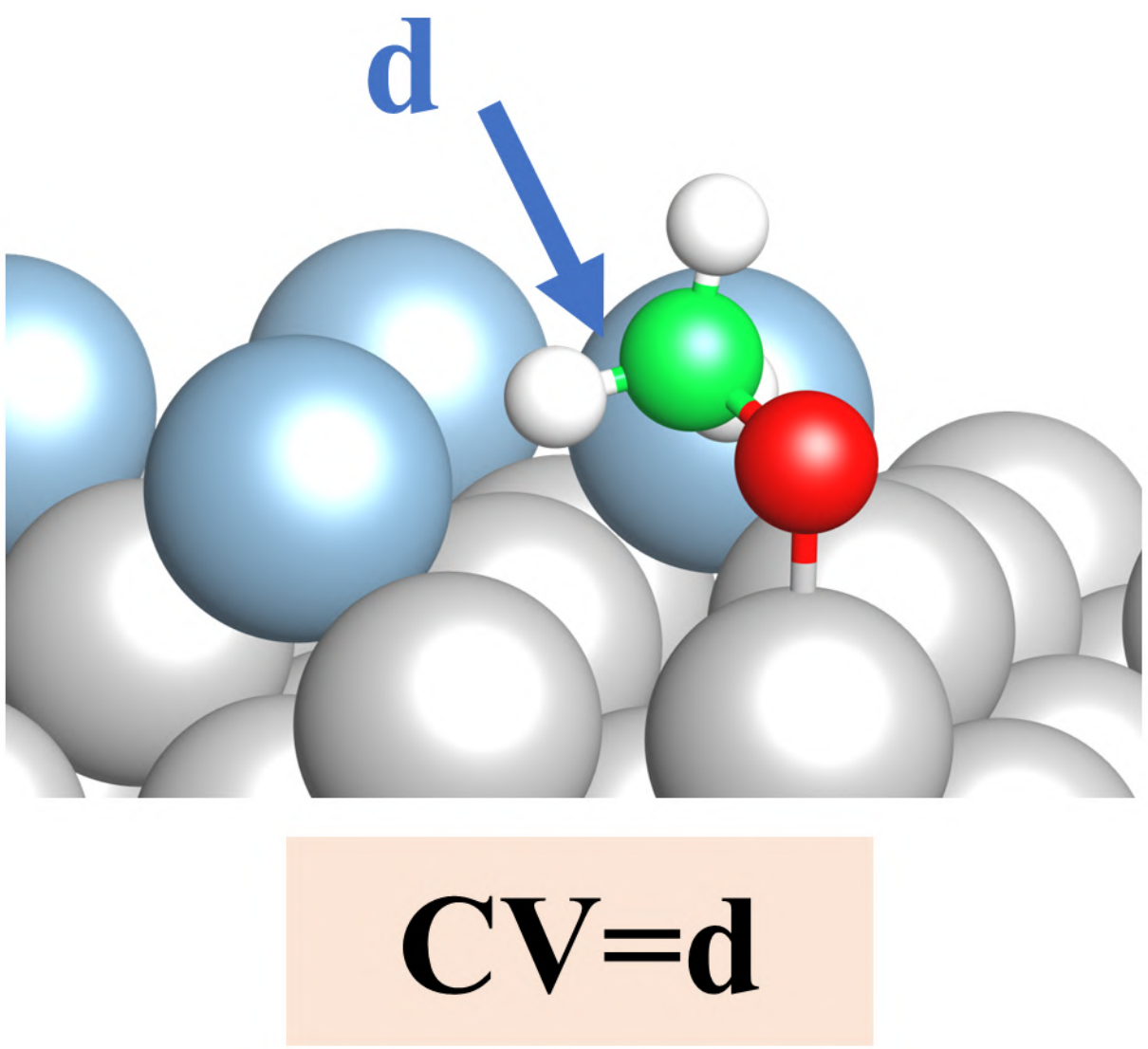

**Figure S18.** Illustration of collective variable of slow-growth approach used in cleavage of Pt-C bond in $CH_3O^*$.

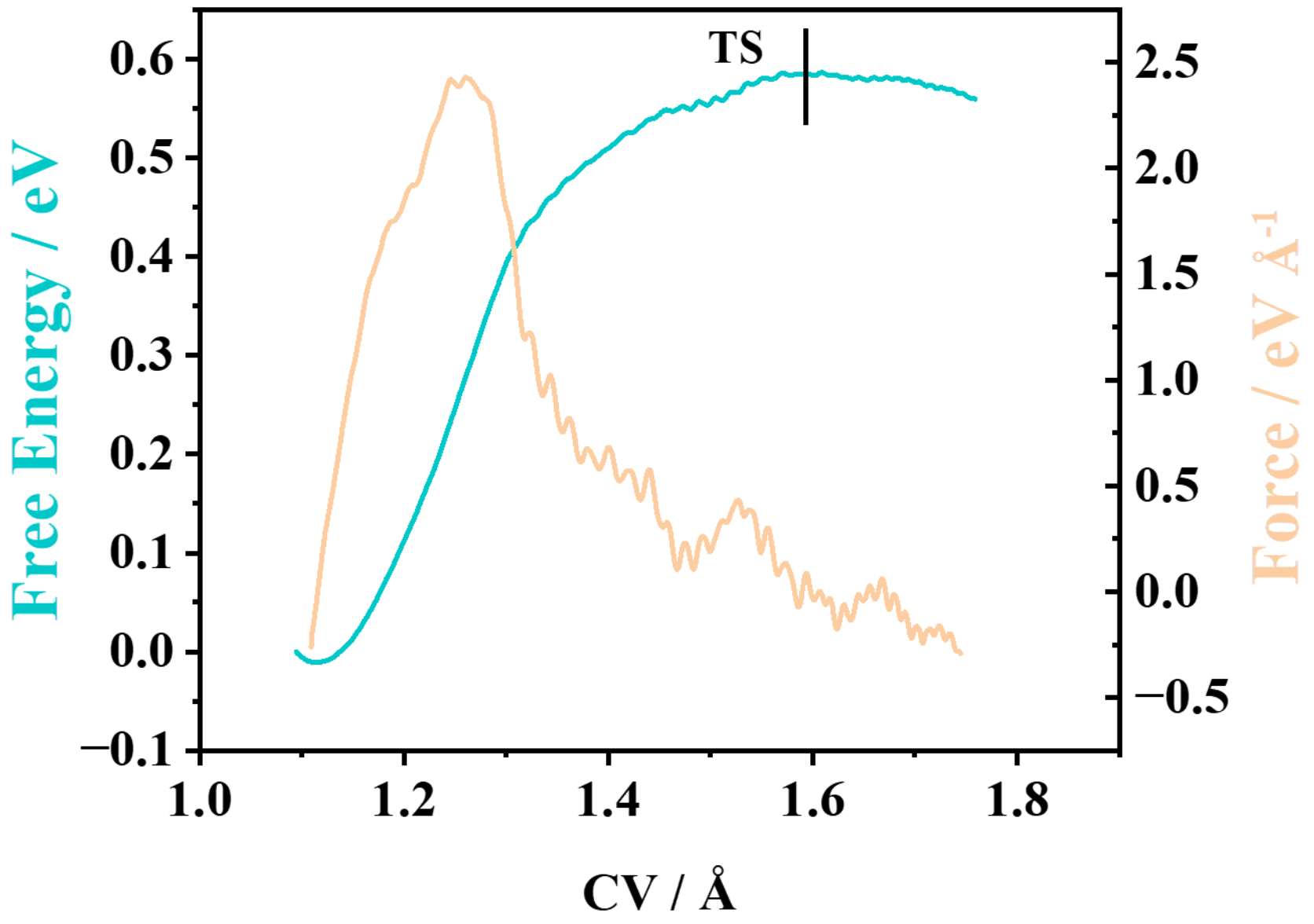

**Figure S19.** Representative potential of mean forces and corresponding free energy changes for $CH_3O^*$ dehydrogenation on Bi-$Pt_3In$(110).

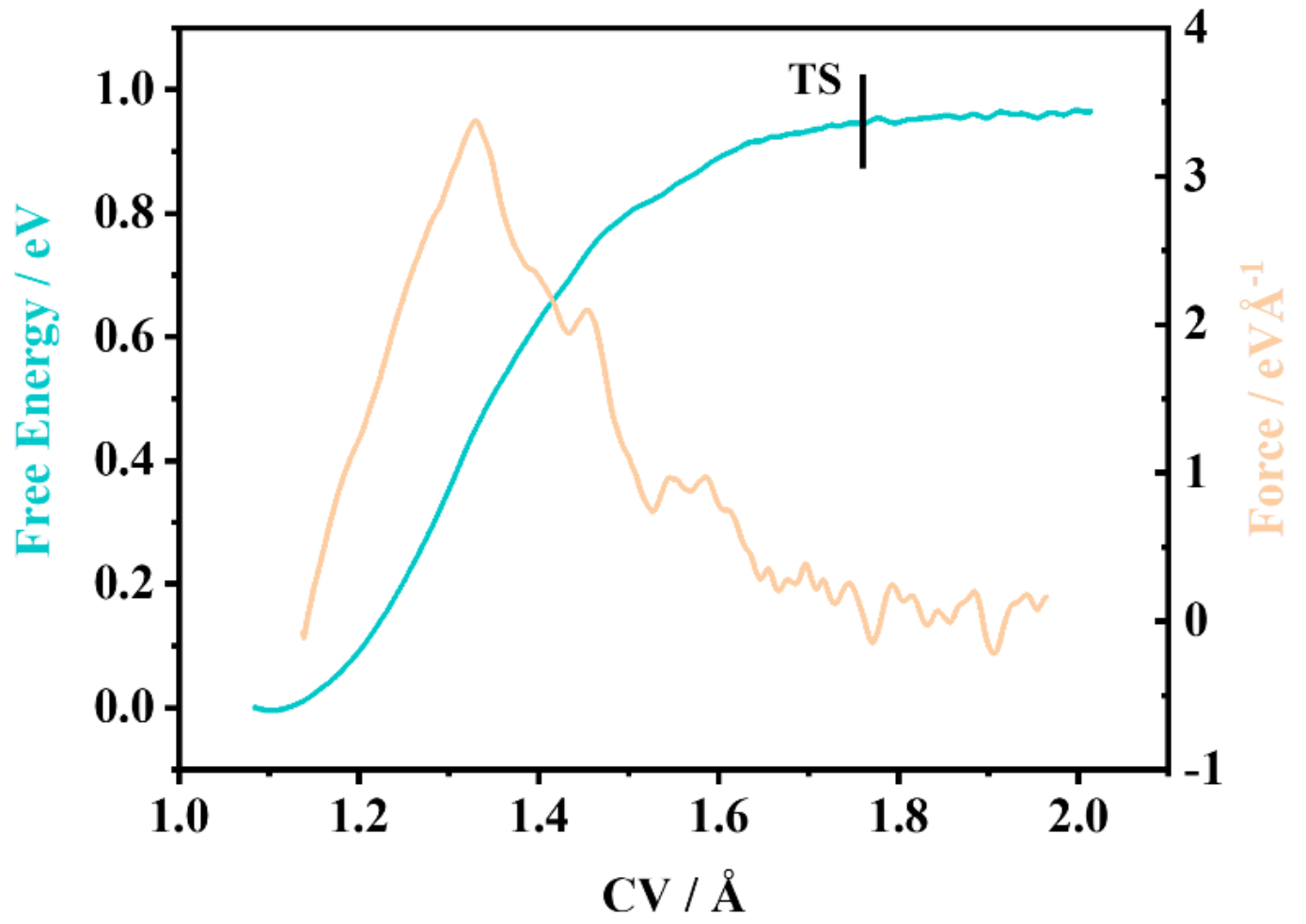

**Figure S20.** Representative potential of mean forces and corresponding free energy changes for $CH_3O$* dehydrogenation on Bi-Pt(110).

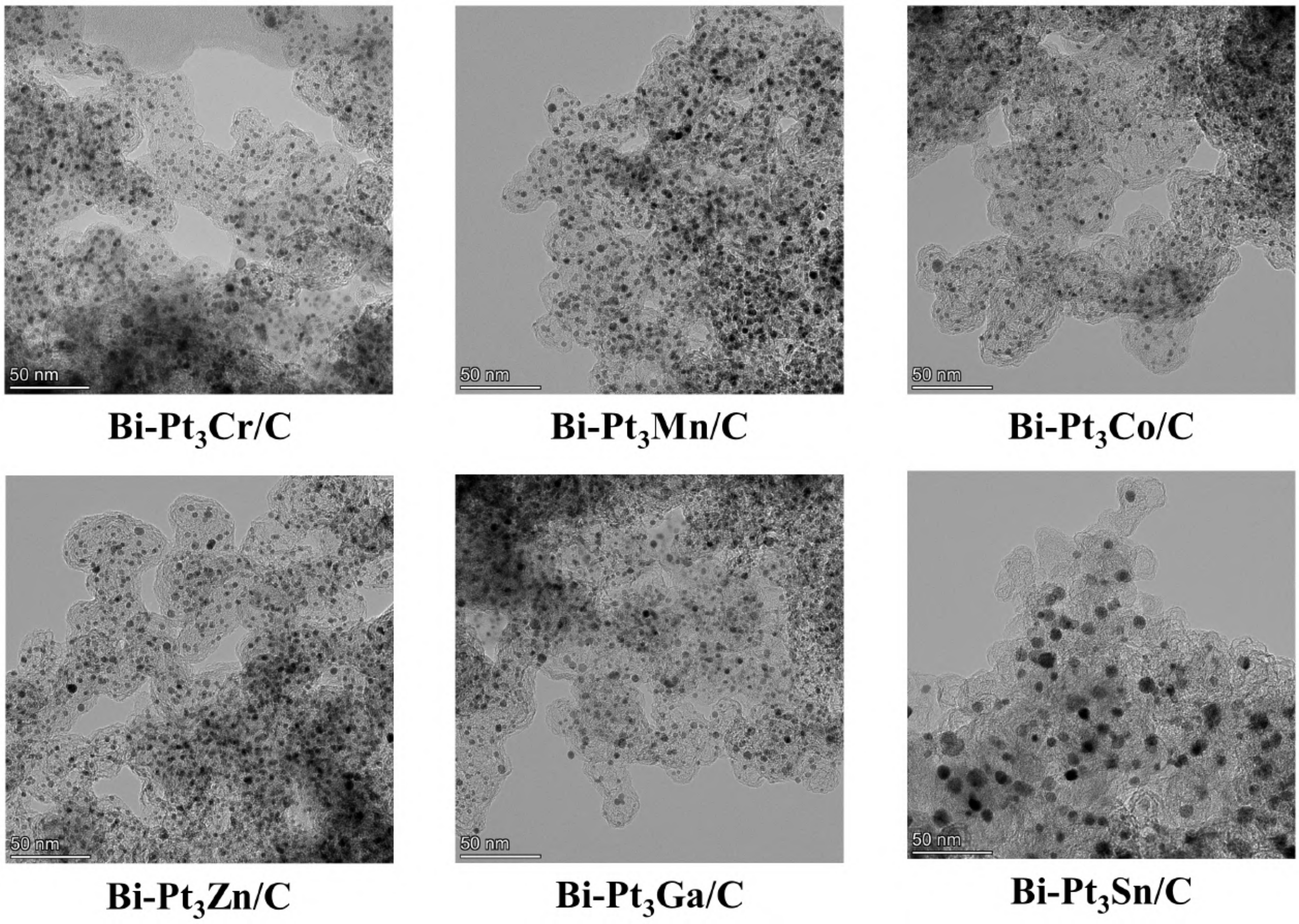

**Figure S21.** TEM images of Bi-$Pt_3M$/C (M=Cr, Mn, Co, Zn, Ga, and Sn).

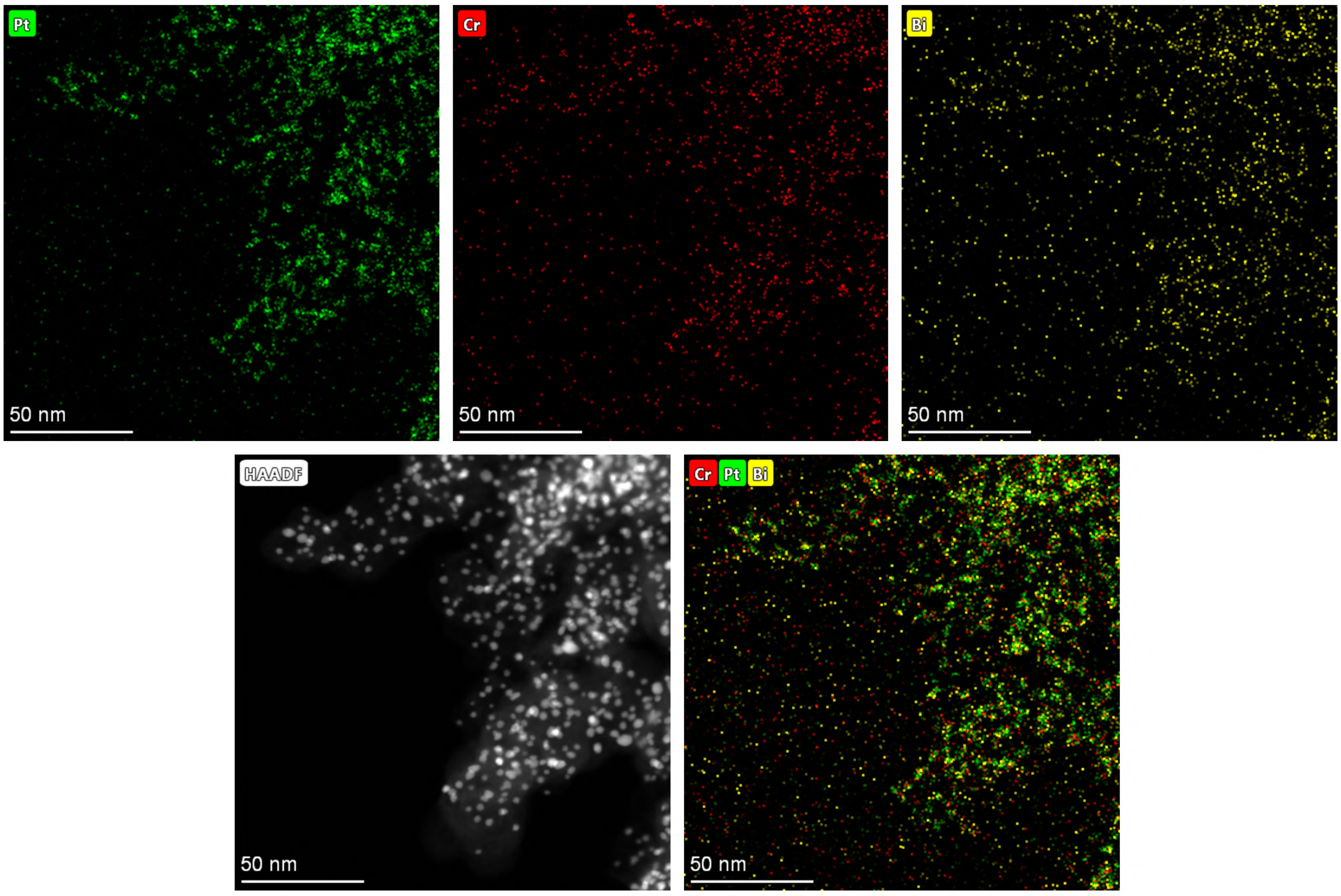

**Figure S22.** HAADF-TEM and EDS element mappings images of Bi-$Pt_3Cr$ NPs.

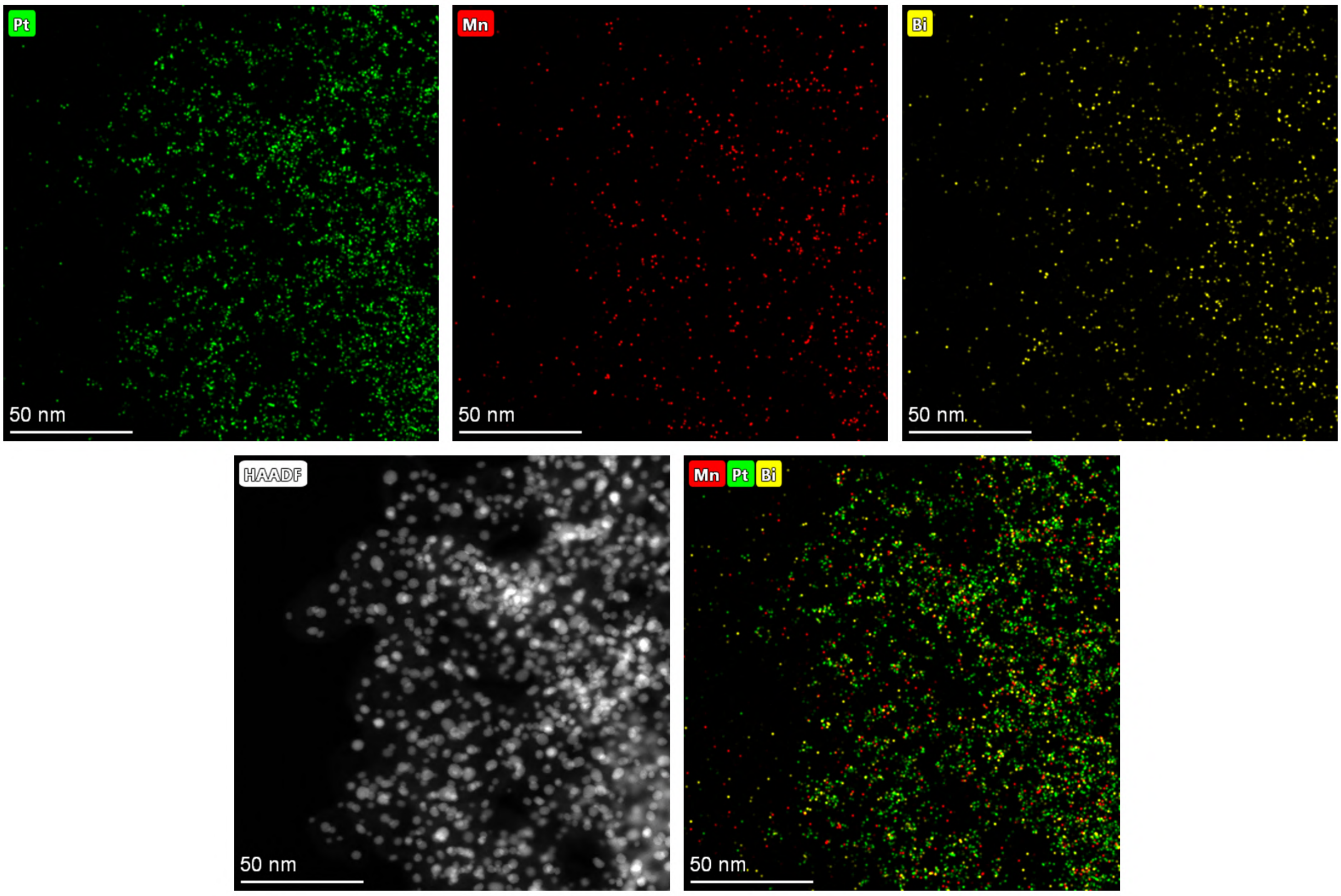

**Figure S23.** HAADF-TEM and EDS element mappings images of Bi-$Pt_3Mn$ NPs.

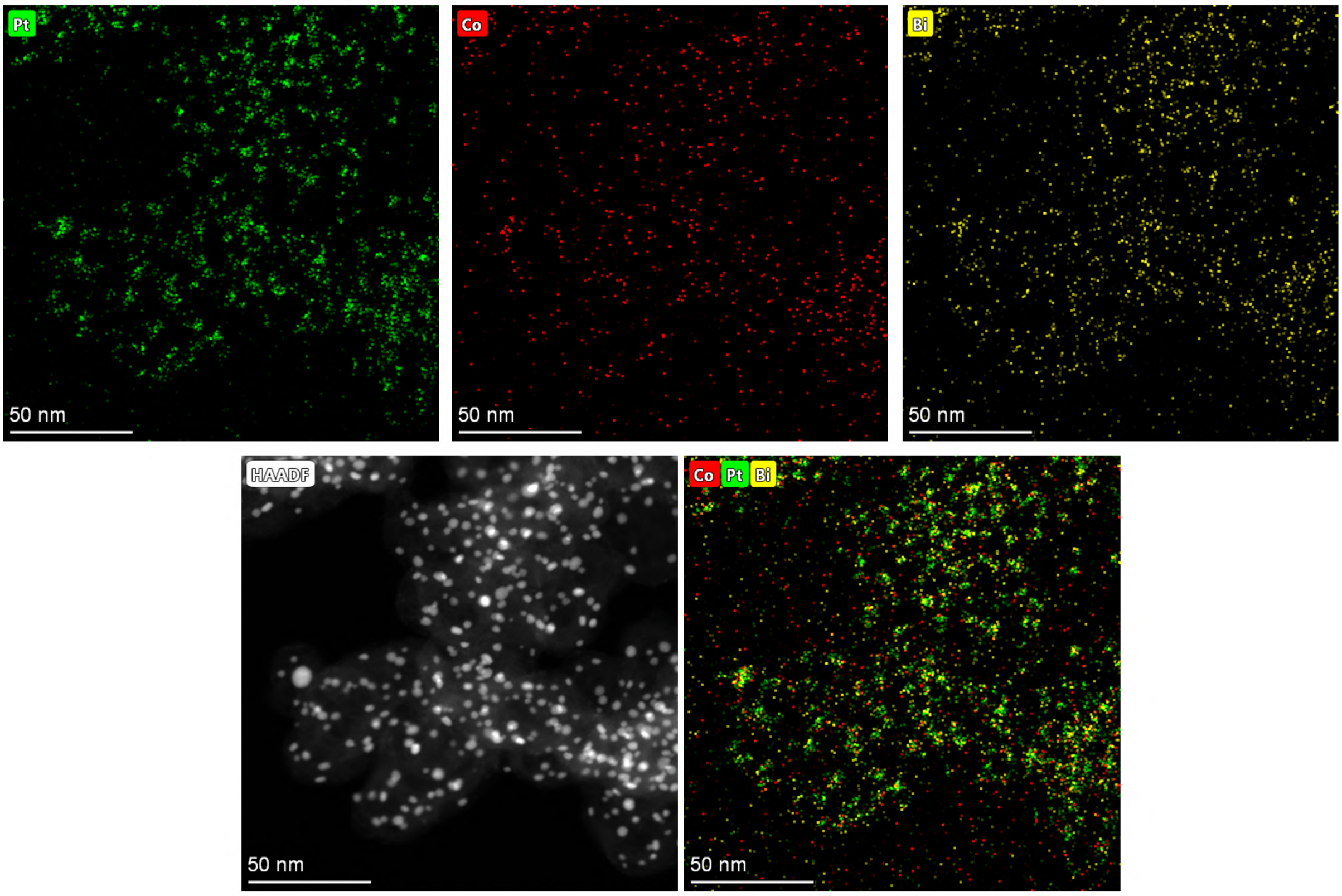

**Figure S24.** HAADF-TEM and EDS element mappings images of Bi-$Pt_3Co$ NPs.

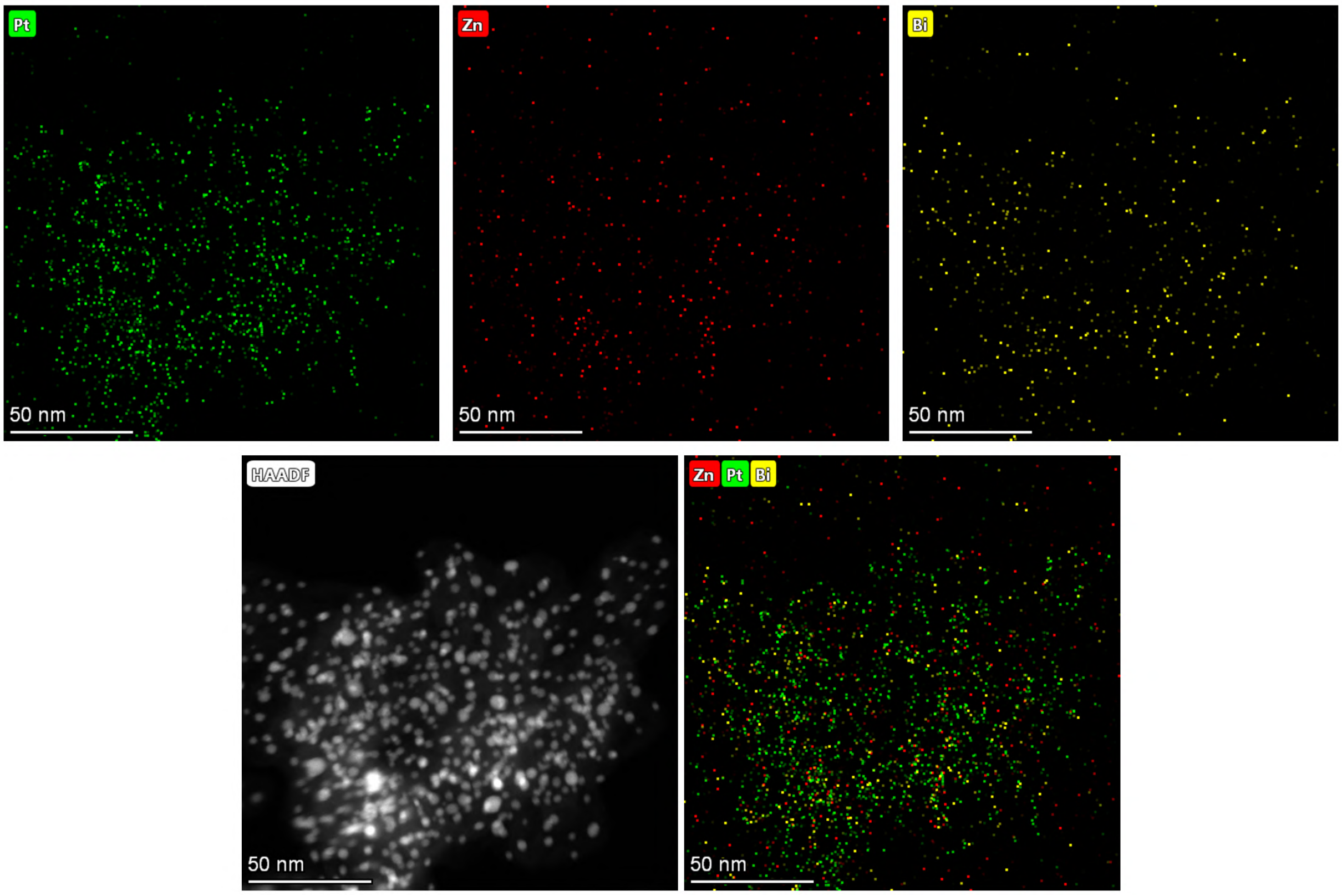

**Figure S25.** HAADF-TEM and EDS element mappings images of Bi-$Pt_3Zn$ NPs.

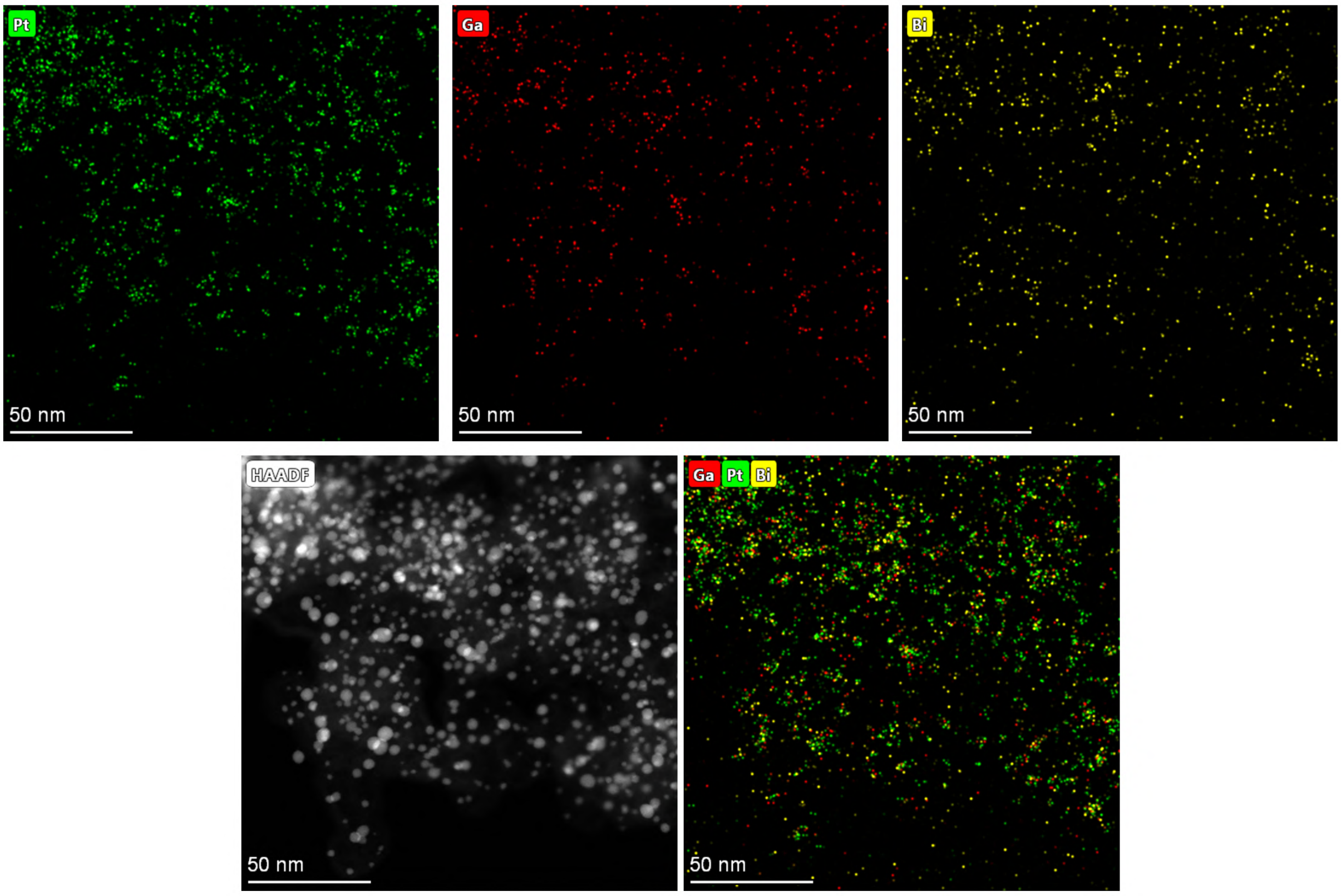

**Figure S26.** HAADF-TEM and EDS element mappings images of Bi-$Pt_3Ga$ NPs.

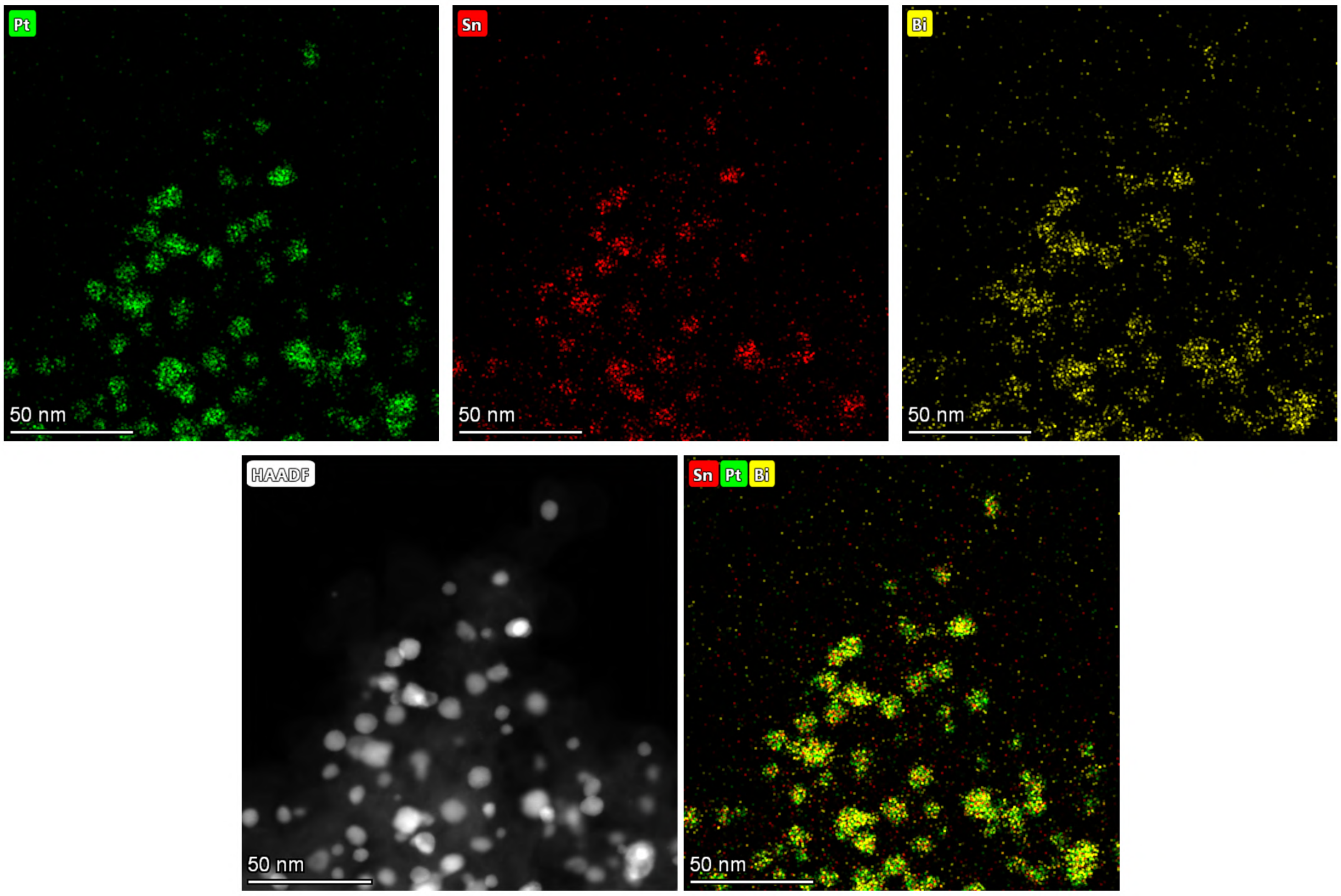

**Figure S27.** HAADF-TEM and EDS element mappings images of Bi-$Pt_3Sn$ NPs.

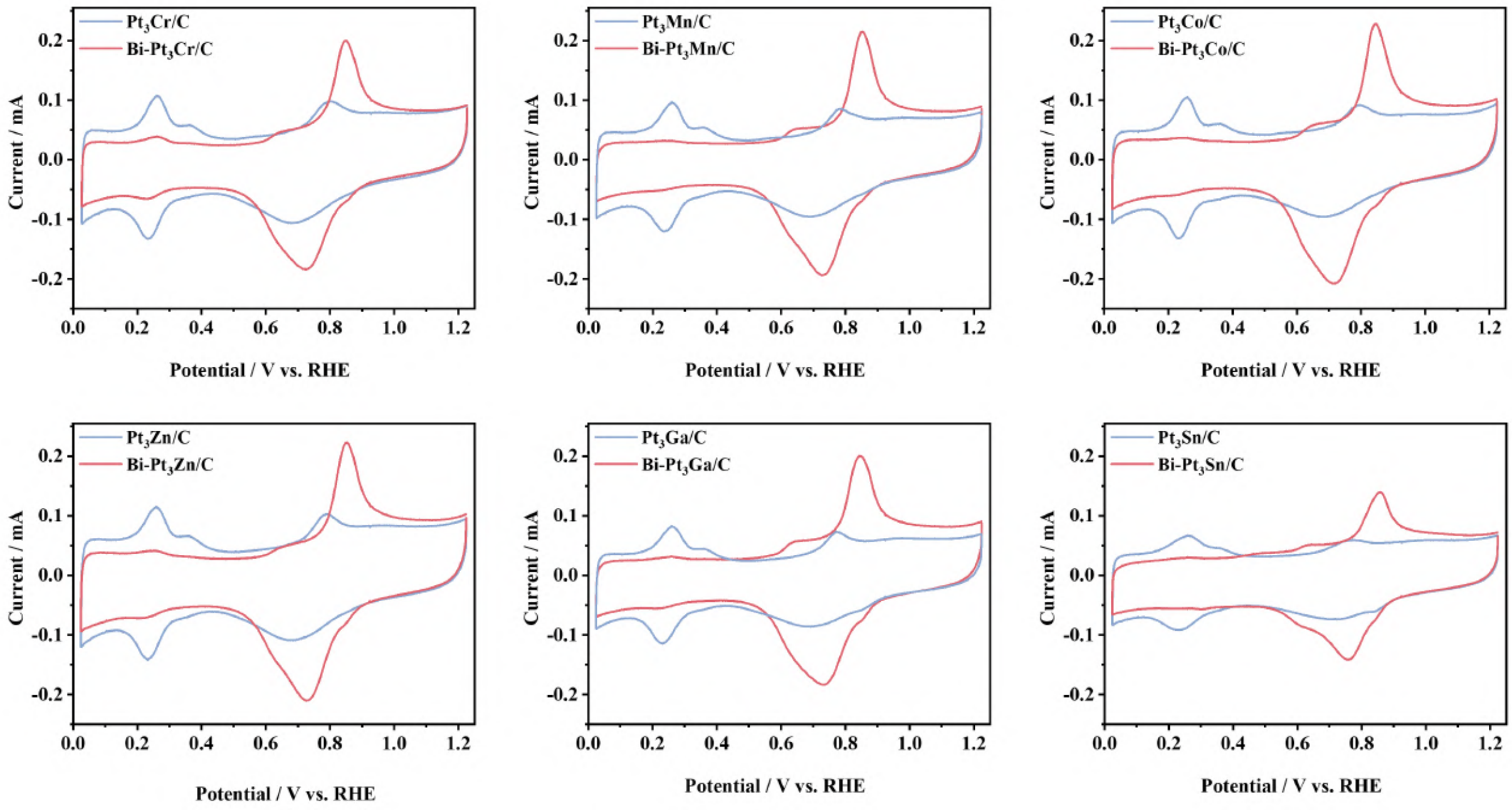

**Figure S28.** CV curves of $Pt_3M/C$ and Bi-$Pt_3M/C$ in $N_2$-saturated 1 M KOH.

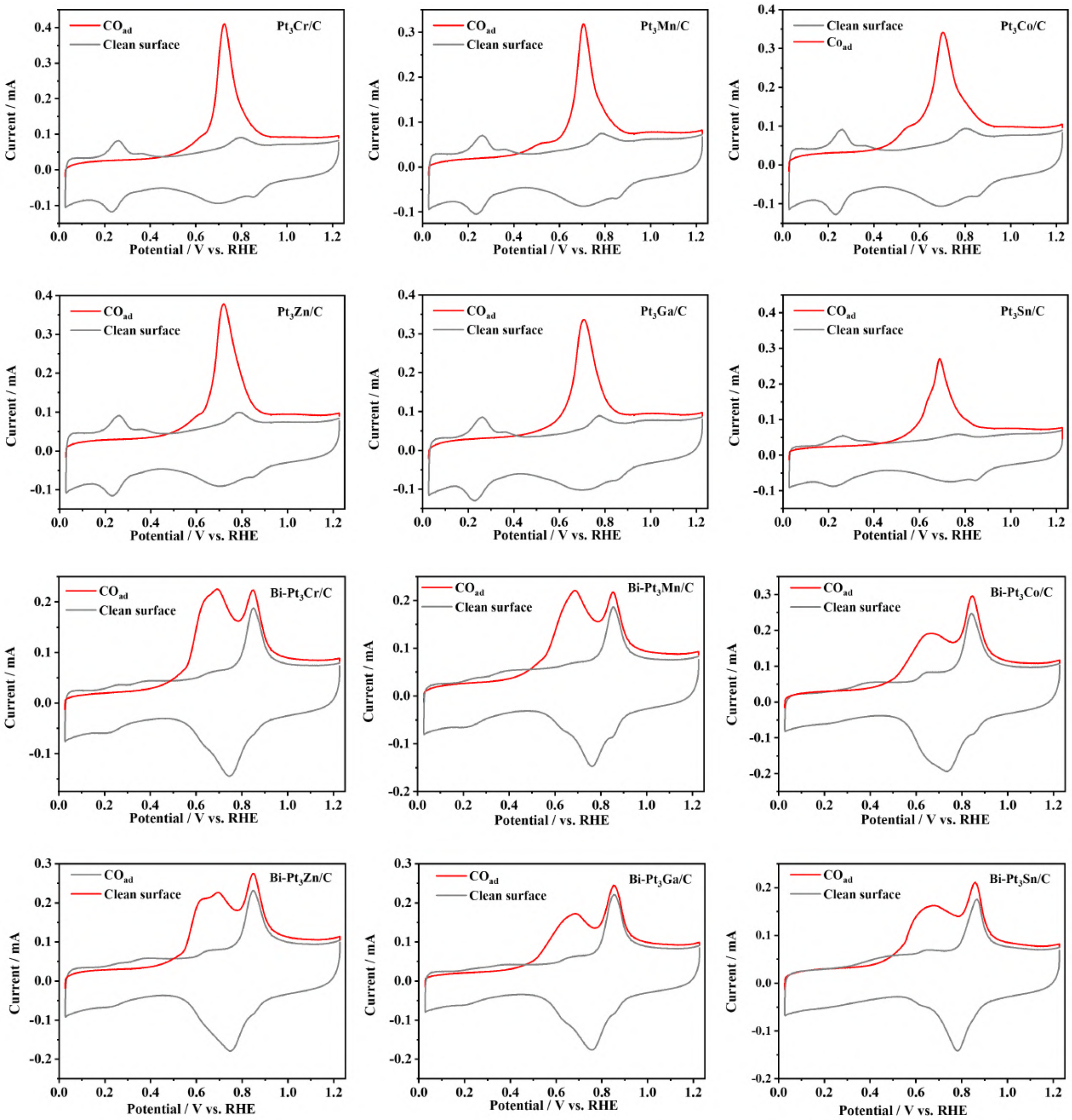

**Figure S29.** CO-stripping curves of $Pt_3M/C$ and Bi-$Pt_3M/C$ in 1 M KOH.

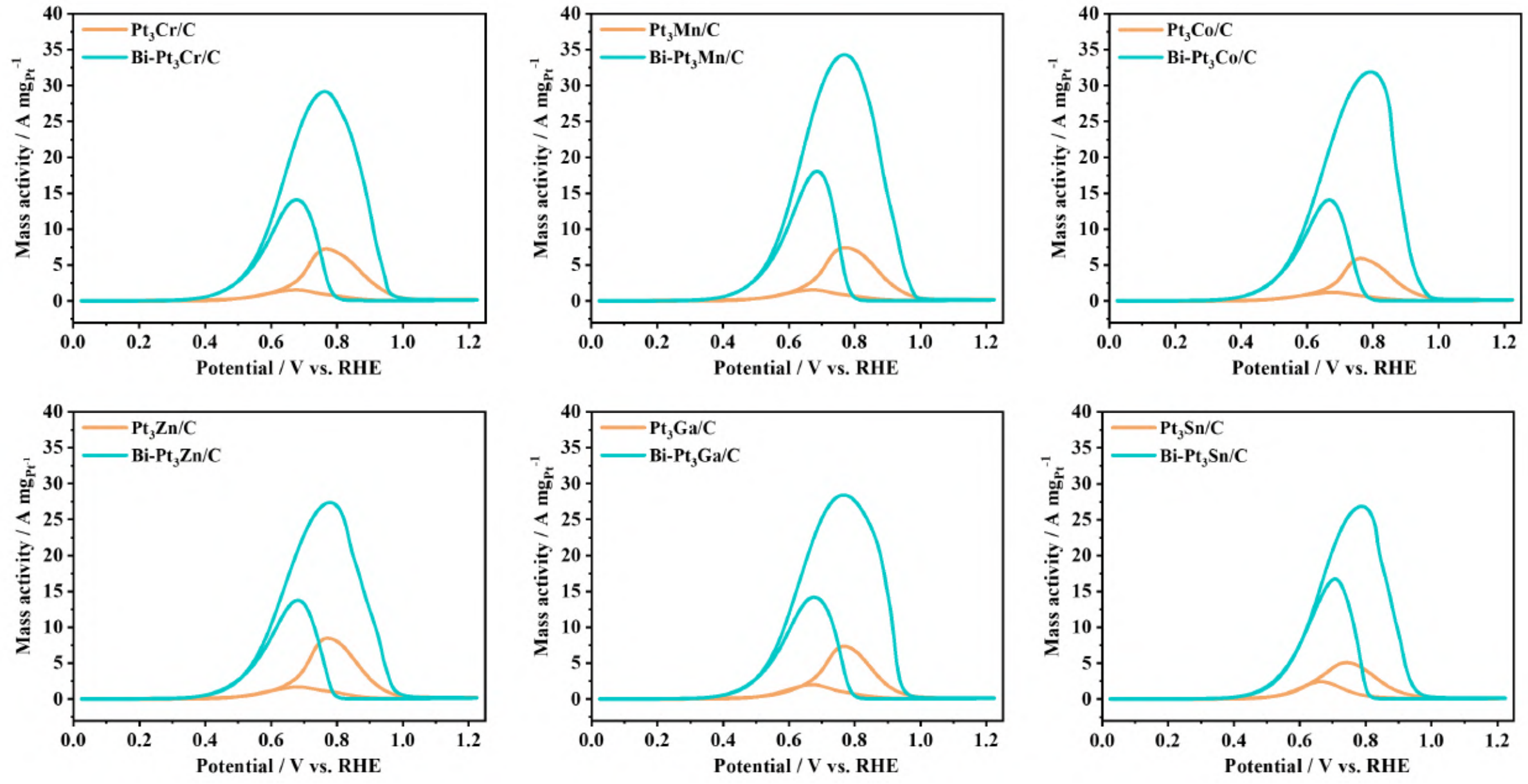

**Figure S30.** CV curves of $Pt_3M/C$ and Bi-$Pt_3M/C$ recorded at a scan rate of 50 mV $s^{-1}$ in 1 M $CH_3OH$/1 M KOH.

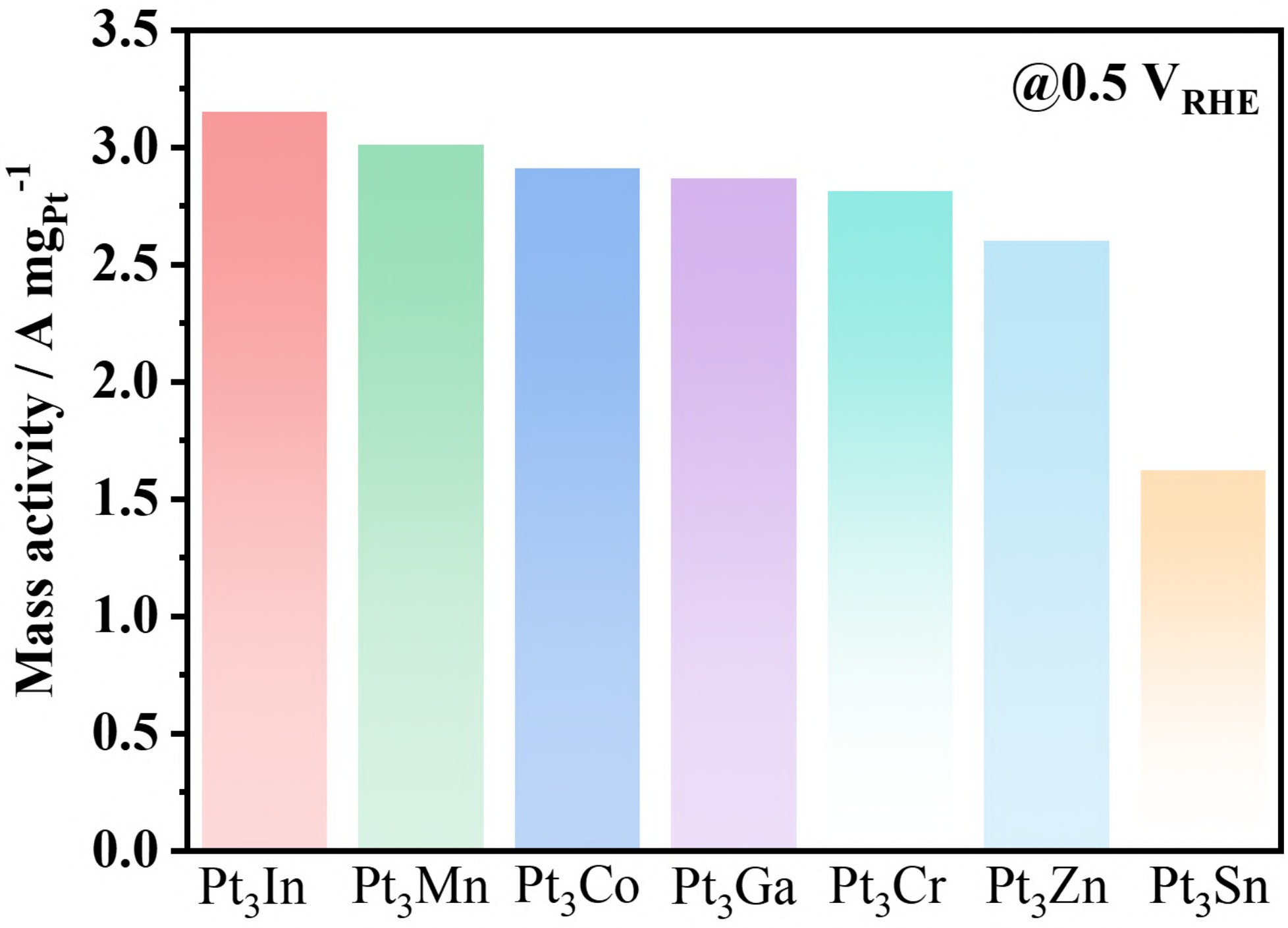

**Figure S31.** Mass activity of Bi-$Pt_3M/C$ at 0.5 $V_{RHE}$.

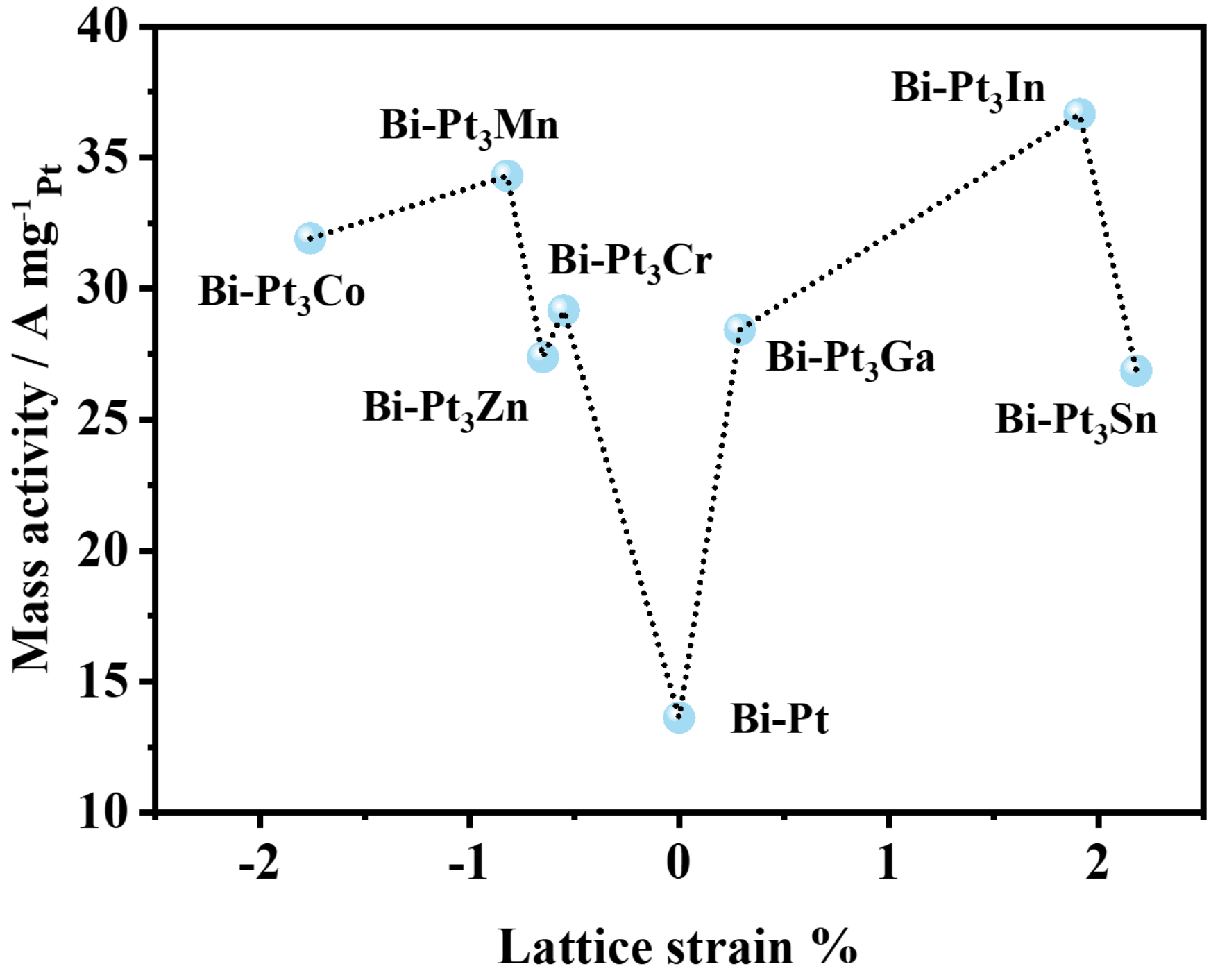

**Figure S32.** Relationship between lattice strain of $Pt_3M$ and mass activity of Bi-$Pt_3M$/C.

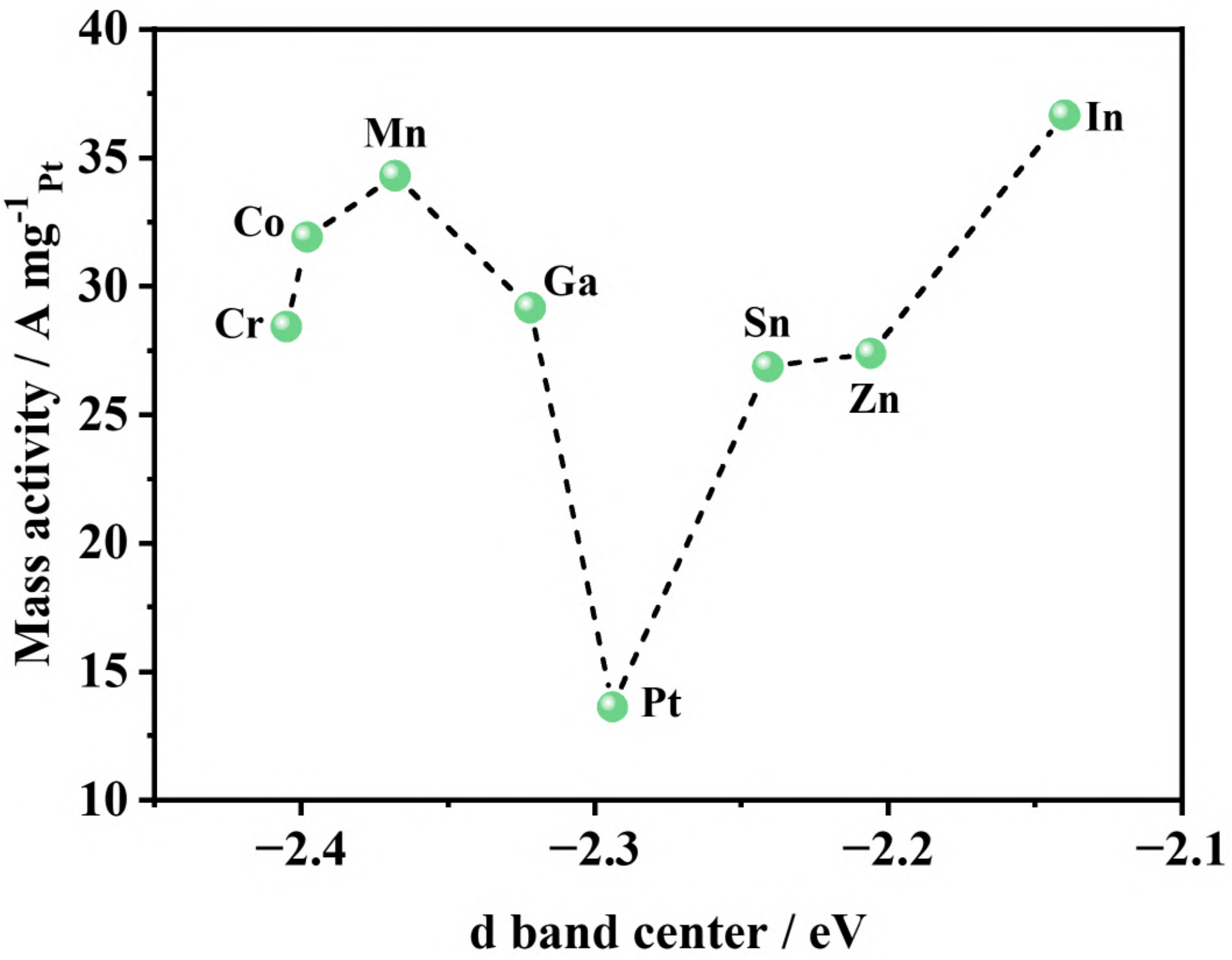

**Figure S33.** Relationship between d-band center of Bi-$Pt_3M(110)$ and their mass activities.

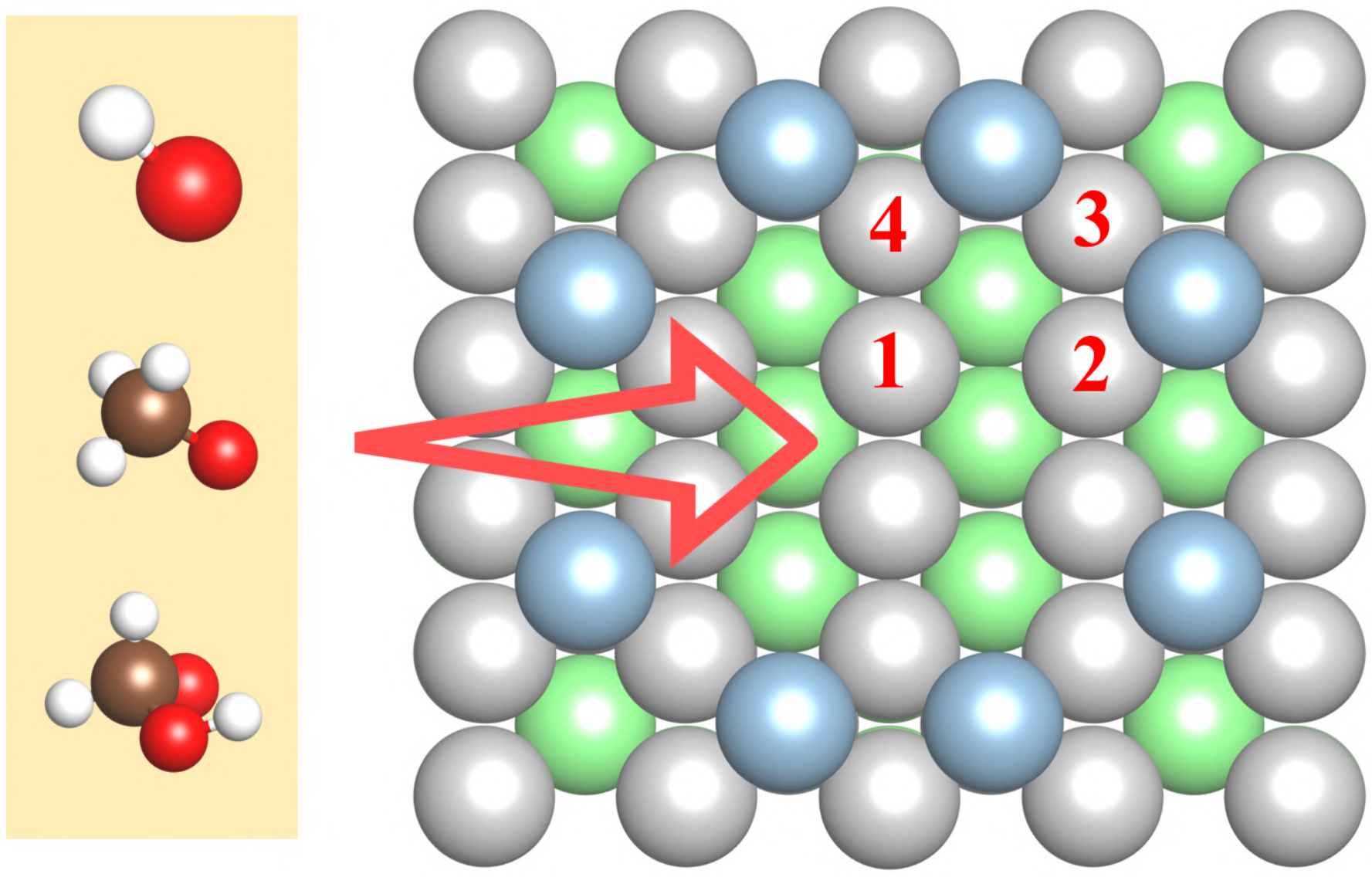

**Figure S33.** The structural model of Bi-$Pt_3M$(110) surface and key absorbed intermediates. Color code: outermost layer Pt, gray; subsurface Pt, green; Bi, blue; C, brown; O, red; and H, white.

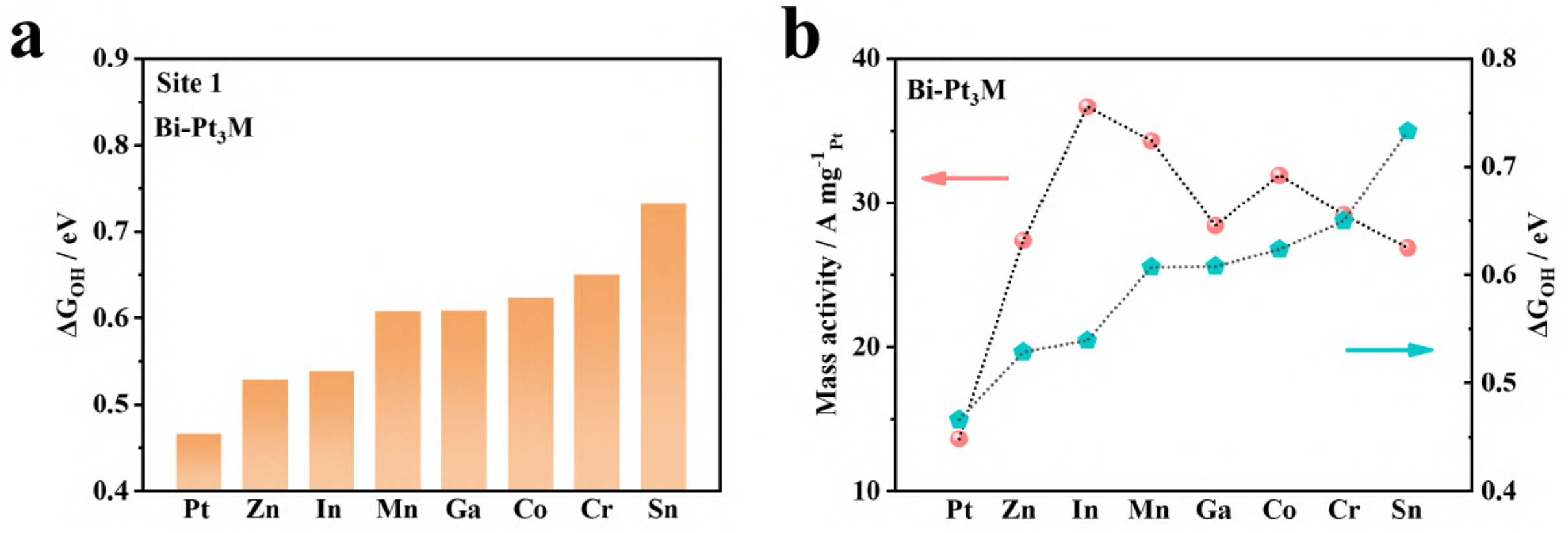

**Figure S34.** (a) OHBE at site 1 of Bi-$Pt_3M$(110); (b) Relationship between OHBE at site 1 and mass activities of Bi-$Pt_3M$/C.

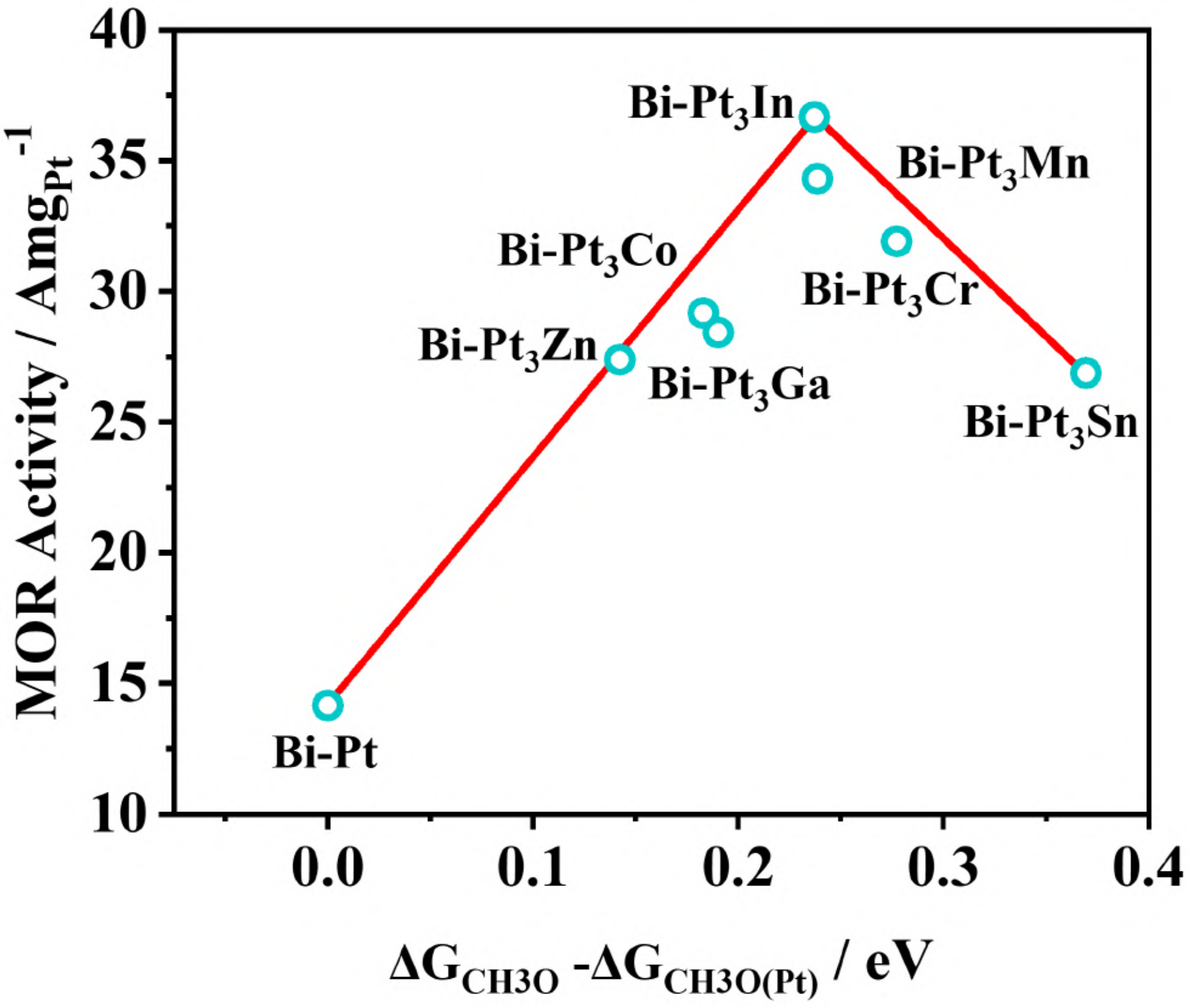

**Figure S35.** The relationship between binding energy of $CH_3O^*$ at site 1of Bi-$Pt_3M$ and their MOR mass activities. The binding energy of $CH_3O^*$ on the Bi–Pt surface is defined as 0 eV and used as the reference.

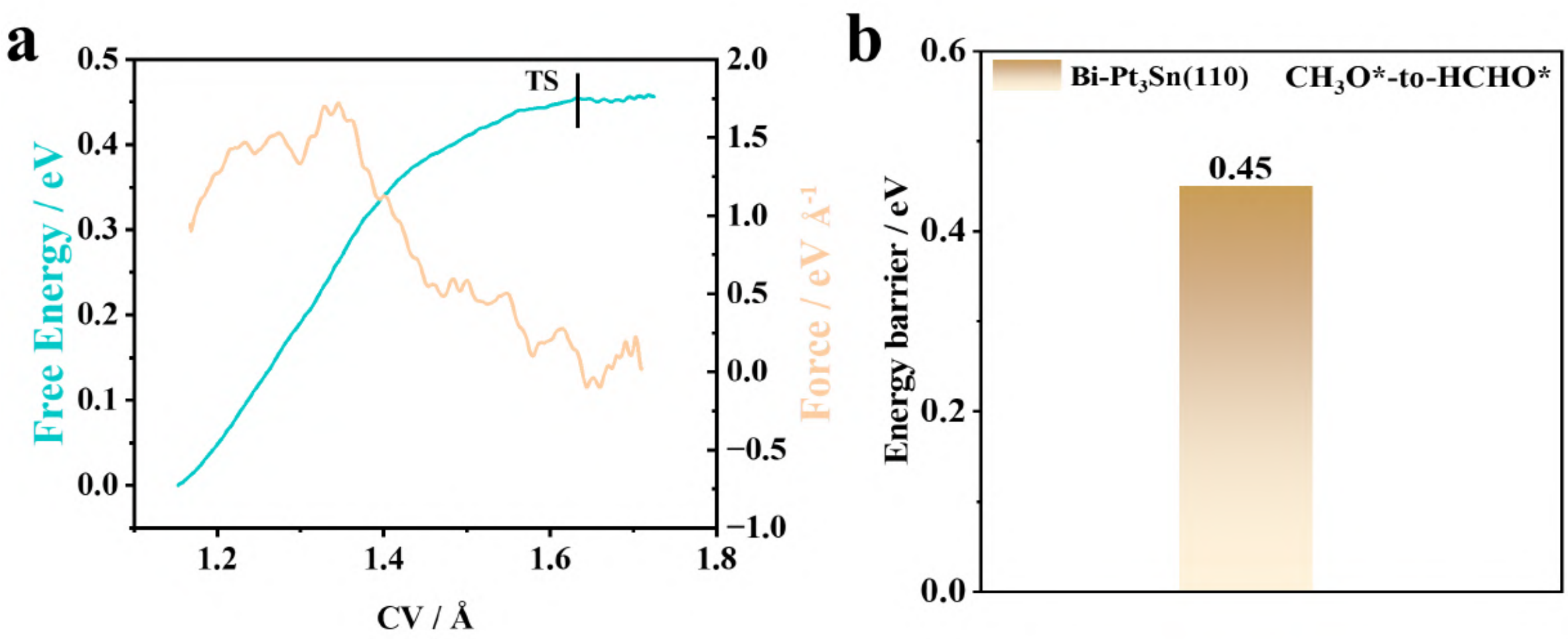

**Figure S36.** (a) The Representative potential of mean forces and corresponding free energy changes for $CH_3O^*$ dehydrogenation on Bi-$Pt_3Sn$(110). (b) Dehydrogenation barrier at 0.6 $V_{RHE}$.

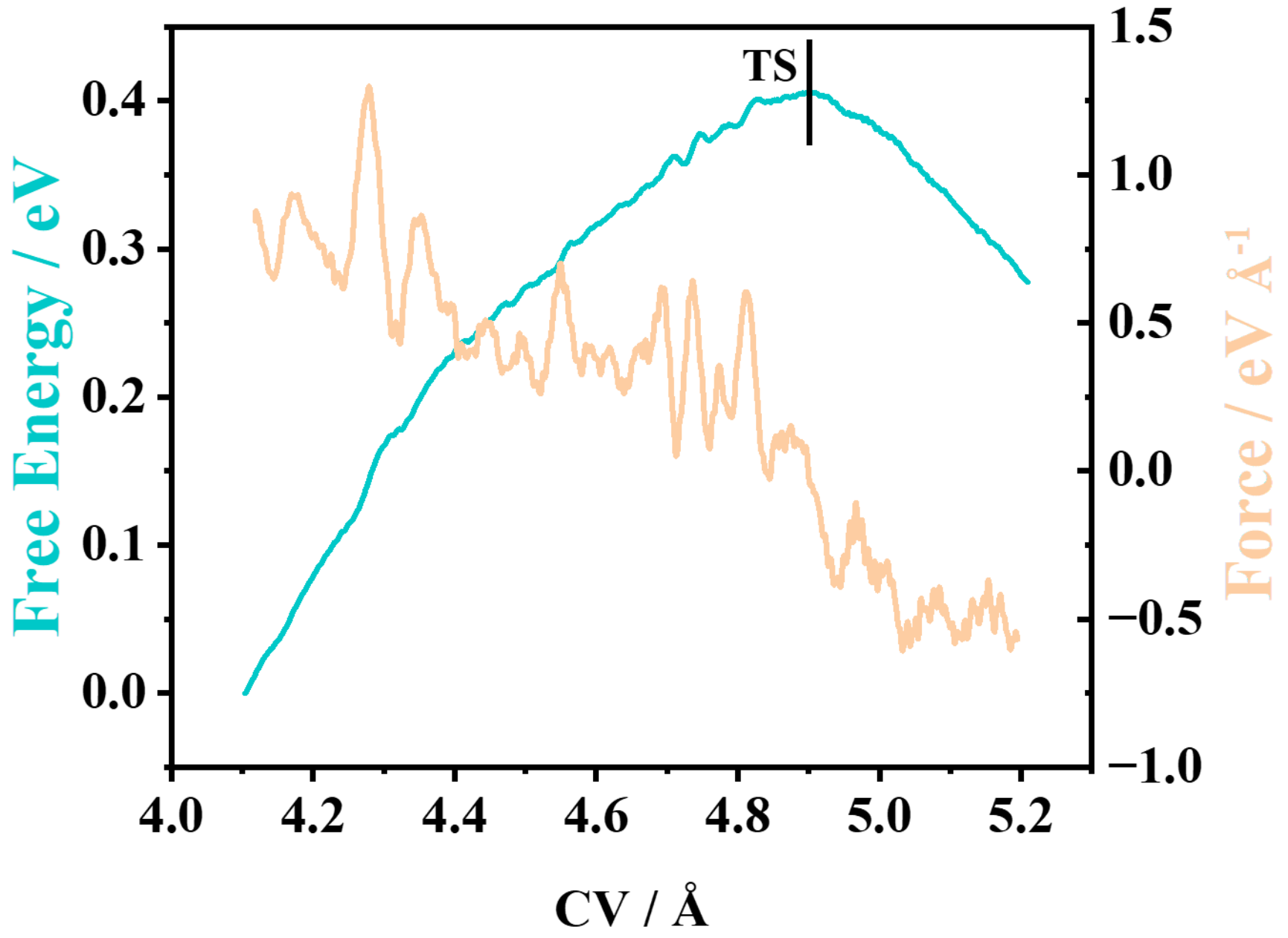

**Figure S37.** Representative potential of mean forces and corresponding free energy changes for water dissociation on Bi-$Pt_3In$(110).

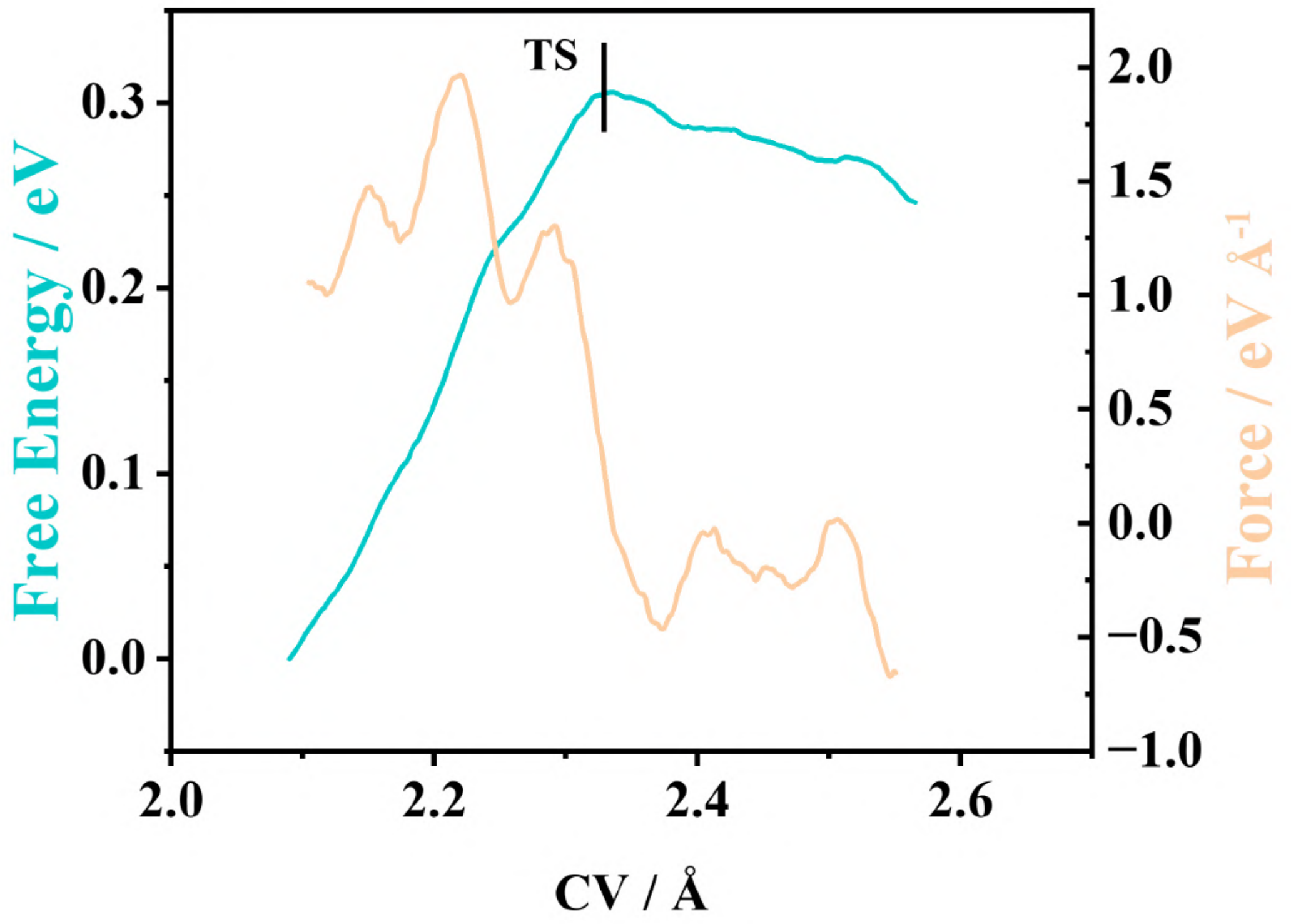

**Figure S38.** Representative potential of mean forces and corresponding free energy changes for water dissociation on Bi-Pt(110).

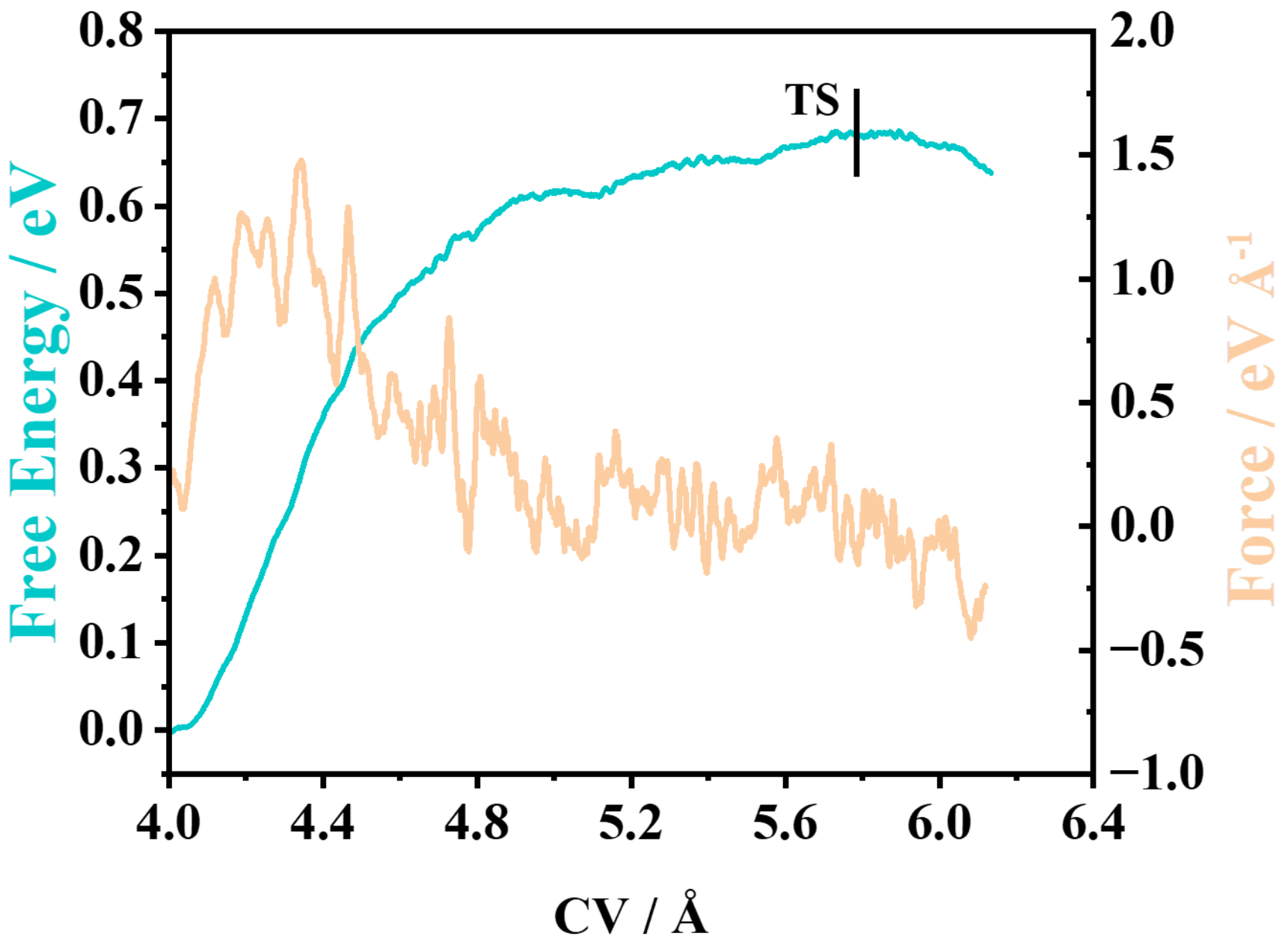

**Figure S39.** Representative potential of mean forces and corresponding free energy changes for water dissociation on Bi-$Pt_3Sn(110)$.

| $Pt_3M/C$ | Pt loading / wt% |
|---|---|
| $Pt_3Cr$ | 21.76 |
| $Pt_3Mn$ | 19.43 |
| $Pt_3Co$ | 21.64 |
| $Pt_3Zn$ | 20.57 |
| $Pt_3Ga$ | 23.08 |
| $Pt_3In$ | 20.51 |
| $Pt_3Sn$ | 21.32 |

**Table S1.** Actual content of Pt in $Pt_3M/C$ measured by ICP-OES.

| Catalysts | Mass Activity / A $mg_{Pt}^{-1}$ | Reference |
| --- | --- | --- |
| Bi-$Pt_3In$/C | 36.66 | **This work** |
| $Mo_1$-PdPtNiCuCoZn | 24.55 | Nat Commun. 2024, 15, 2290 [1] |
| SE-$Bi_1$/Pt NRs | 23.77 | Adv. Mater. 2024, 36, 2313179 [2] |
| PtRhBiSnSb | 19.53 | Adv. Mater. 2022, 34, 2206276 [3] |
| PtNiFeCoCu HEA NPs | 15.04 | Nat. Commun. 2022, 11, 5437 [4] |
| Pt ALs/$CeO_2$ | 14.87 | Angew. Chem. Int. Ed. 2024, 63, e202410545 [5] |
| HEA-NPs-(14) | 12.60 | Adv. Mater. 2023, 35, 2302499 [6] |
| A/IMC PtPbBi NSs | 11.90 | Angew. Chem. Int. Ed. 2024, 63, e202405173 [7] |
| $Pt_3Yb$ | 11.74 | ACS Nano 2024, 18, 25754-25764 [8] |
| $Pt_5Ce$ | 9.13 | Energy Environ. Sci., 2021, 14, 5911–5918 [9] |
| SANi-PtNWs | 7.93 | Nat. Catal. 2019, 2, 495–503 [10] |
| $Pt_c/Ti_3C_2T_x$ | 7.32 | J. Am. Chem. Soc. 2022, 144, 15529−15538 [11] |
| $Pt_1/RuO_2$ | 6.77 | Nat. Commun. 2021, 12, 5235 [12] |

**Table S2.** Summary of the most advanced electrocatalysts for alkaline MOR.

| $Pt_3M/C$ | Metal salt | Synthesis Condition |
| --- | --- | --- |
| $Pt_3Cr$ | $CrCl_3 \cdot 6H_2O$ | 800 °C-2h |
| $Pt_3Mn$ | $MnCl_2 \cdot 4H_2O$ | 800 °C-2h |
| $Pt_3Co$ | $CoCl_2 \cdot 6H_2O$ | 800 °C-2h |
| $Pt_3Zn$ | $ZnBr_2$ | 800 °C-2h |
| $Pt_3Ga$ | $Ga(NO_3)_3 \cdot xH_2O$ | 800 °C-2h |
| $Pt_3In$ | $InCl_3 \cdot 4H_2O$ | 850 °C-2h |
| $Pt_3Sn$ | $SnCl_2 \cdot 2H_2O$ | 950 °C-2h |

Table S3. Detail of synthesis on $Pt_3M/C$.

## Supplemental References